\documentclass[useAMS,usenatbib]{mn2e}
\usepackage{subfigure}
\usepackage{graphicx}
\usepackage{amsmath}
\usepackage{amssymb}
\usepackage{pdflscape}
\usepackage{color}

\title[Torus model properties of an ultra-hard X-ray selected sample]{Torus model properties of an ultra-hard X-ray selected sample of Seyfert galaxies}
\author[I. Garc\'ia-Bernete et al.]
{\parbox{\textwidth}{I. Garc\'ia-Bernete$^{1,2}$\thanks{E-mail: igbernete@gmail.com}, C. Ramos Almeida$^{3,4}$, A. Alonso-Herrero$^{5}$, M. J. Ward$^{6}$, J. A. Acosta-Pulido$^{3,4}$, M. Pereira-Santaella$^{7}$, A. Hern{\'a}n-Caballero$^{8}$, A. Asensio Ramos$^{3,4}$, O. Gonz\'alez-Mart\'in$^{9}$, N. A. Levenson$^{10}$,
S. Mateos$^{1}$, F. J. Carrera$^{1}$, C. Ricci$^{11,12}$, P. Roche$^{7}$, I. Marquez$^{13}$, C. Packham$^{14,15}$, J. Masegosa$^{13}$ and L. Fuller$^{14}$\\
}\vspace{0.4cm}\\
\parbox{\textwidth}{$^{1}$Instituto de F\'isica de Cantabria (CSIC-UC), Avenida de los Castros, 39005 Santander, Spain\\
$^{2}$Visiting Fellow, Centre for Extragalactic Astronomy, Durham University, South Road, Durham, DH1 3LE, UK\\
$^{3}$Instituto de Astrof\'isica de Canarias, Calle V\'ia L\'actea, s/n, E-38205, La Laguna, Tenerife, Spain\\
$^{4}$Departamento de Astrof\'isica, Universidad de La Laguna, E-38206, La Laguna, Tenerife, Spain\\
$^{5}$Centro de Astrobiolog\'ia, CSIC-INTA, ESAC Campus, E-28692, Villanueva de la Ca\~nada, Madrid, Spain\\
$^{6}$Centre for Extragalactic Astronomy, Durham University, South Road, Durham DH1 3LE, UK\\
$^{7}$Department of Physics, University of Oxford, Oxford OX1 3RH, UK\\
$^{8}$Departamento de F\'isica de la Tierra y Astrof\'isica, Facultad de CC. F\'isicas, Universidad Complutense de Madrid, E-28040 Madrid, Spain\\
$^{9}$Instituto de Radioastronom\'ia y Astrof\'isica (IRyA), 3-72 (Xangari), 8701, Morelia, Mexico\\
$^{10}$Space Telescope Science Institute, 3700 San Martin Drive, Baltimore, MD, 21218, USA\\
$^{11}$ N\'ucleo de Astronom\'ıa de la Facultad de Ingenier\'ıa, Universidad Diego Portales, Av. Ej\'ercito Libertador 441, Santiago, Chile\\
$^{12}$ Kavli Institute for Astronomy and Astrophysics, Peking University, Beijing 100871, China\\
$^{13}$Instituto de Astrof\'isica de Andaluc\'ia, CSIC, Glorieta de la Astronom\'ia s/n, 18008, Granada, Spain\\
$^{14}$Department of Physics and Astronomy, University of Texas at San Antonio, One UTSA Circle, San Antonio, TX 78249, USA\\
$^{15}$National Astronomical Observatory of Japan, 2-21-1 Osawa, Mitaka, Tokyo 181-8588, Japan\\
}
}

\begin{document}
\date{}
\pagerange{\pageref{firstpage}--\pageref{lastpage}} \pubyear{2019}
\maketitle
\label{firstpage}
\begin{abstract} We characterize for the first time the torus properties of an ultra-hard X-ray (14--195~keV) volume-limited (D$_L<$40\,Mpc) sample of 24 Seyfert (Sy) galaxies (BCS$_{40}$ sample). The sample was selected from the {\textit{Swift/BAT}} nine month catalog. We use high angular resolution nuclear infrared (IR) photometry and N-band spectroscopy, the CLUMPY torus models and a Bayesian tool to characterize the properties of the nuclear dust. In the case of the Sy1s we estimate the accretion disk contribution to the subarcsecond resolution nuclear IR SEDs ($\sim$0.4$\arcsec$) which is, on average, 46$\pm$28, 23$\pm$13 and 11$\pm$5\% in the J-, H- and K-bands, respectively. This indicates that the accretion disk templates that assume a steep fall for longer wavelengths than 1~$\mu$m might underestimate its contribution to the near-IR emission. Using both optical (broad vs narrow lines) and X-ray (unabsorbed vs absorbed) classifications, we compare the global posterior distribution of the torus model parameters. We confirm that Sy2s have larger values of the torus covering factor (C$_T\sim$0.95) than Sy1s (C$_T\sim$0.65) in our volume-limited Seyfert sample. These findings are independent of whether we use an optical or X-ray classification. We find that the torus covering factor remains essentially constant within the errors in our luminosity range and there is no clear dependence with the Eddington ratio. Finally, we find tentative evidence that even an ultra hard X-ray selection is missing a significant fraction of highly absorbed type 2 sources with very high covering factor tori.
\end{abstract}

\begin{keywords}
galaxies: active -- galaxies: nuclei -- galaxies: photometry -- galaxies: spectroscopy.
\\
\end{keywords}

\section{Introduction}
\label{intro}
Active galactic nuclei (AGNs) are powered by accretion of material onto supermassive black holes (SMBHs), which release energy in the form of radiation and/or mechanical outflows to the host galaxy interstellar medium. Although they comprise just a small fraction of the galaxies in the local universe ($\sim$10\%), AGNs are now considered to be a short but recurrent phase in the overall lifetime of galaxies. Accordingly, galaxies are observed as AGN during an active phase when their SMBHs are accreting material at a relatively high rate (e.g. \citealt{Bennert11}). Several studies found a correlation between the SMBH and host galaxy bulge mass (e.g. \citealt{Kormendy13} and references therein) which is interpreted as a sign of co--evolution of AGNs and their host galaxies. However, the study of the AGN--host galaxy connection is difficult due to the very different spatial and temporal scales involved. Therefore it is of great importance to investigate the innermost regions of AGN to better understand this connection (see \citealt{Ramos17} and references therein). 

The key piece of the AGN unified model \citep{Antonucci1993} is a dusty molecular torus that obscures the central engines of type 2, and allows a direct view in the case of type 1 sources. This dusty torus absorbs part of the AGN radiation and reprocesses it to emerge in the infrared (IR).

To correctly separate the nuclear emission from the foreground galaxy emission and be able to characterize the properties of the nuclear obscurer the highest possible spatial resolution is required. Since Seyfert (Sy) galaxies are intermediate-luminosity AGNs, and, in general, are relatively nearby, they are one of the best astrophysical laboratories to study the inner regions of active galaxies. 

The torus radius has been constrained to be compact ($\sim$0.1-10~pc) in the mid-IR (MIR; $\sim$5--30~$\mu$m). For example, using MIR direct imaging, \citet{Packham05} and \citet{Radomski2008} found for Circinus and Centaurus\,A that the MIR size of the torus is less than $\sim$4~pc (diameter). The modelling of MIR interferometric data shows a relatively compact torus of r$<10$~pc (e.g. \citealt{Jaffe04,Tristram07,Tristram09,Burtscher09,Raban09,Burtscher13,Lopez-Gonzaga16}). Atacama Large Millimeter/submillimeter Array (ALMA) observations of the archetypal Seyfert 2 galaxy NGC\,1068 have spatially resolved for the first time the submillimeter (sub-mm) counterpart of the putative torus \citep{Garcia-Burillo16,Gallimore16,Imanishi18}. This is a disk of $\sim$7-10~pc diameter. More recently, \citet{Herrero18} and \citet{Combes18} have found even larger nuclear molecular disks for other Seyfert galaxies and low-luminosity AGNs. Thus, as theoretically predicted (e.g. \citealt{Schartmann08,Stalevski2012}), the radii measured in the sub-mm for the dusty and molecular torus are found to be larger than those inferred from IR observations. Therefore, to constrain the properties of the warm dust, we still need to compare torus models to the observed SEDs.
 
Torus models can be broadly grouped in two categories: physical (e.g. \citealt{Wada02,Schartmann08,Wada12}) and geometrical (\textit{ad-hoc}; e.g. \citealt{Pier92,Efstathiou95,Nenkova08a,Nenkova08b,Honig10,Stalevski2012,Siebenmorgen2015,Honig17}). 
Some of the geometrical models also include a polar component in the MIR range (e.g. \citealt{Honig17}). However, this polar emission has been detected so far in six Seyfert galaxies of the 23 observed using IR interferometry \citep{Lopez-Gonzaga16, Leftley18} and, therefore, more observations are needed in order to study whether this is a common feature in AGNs. The physical models are more realistic since they include important processes, such as supernovae and AGN feedback. However, they require large computational times and therefore it is more difficult to compare with observations. On the other hand, geometrical torus models are more degenerate, but they can be easily compared with the observations, assuming various geometries and compositions of the dust (see \citealt{Ramos17} for a review). 

Recent studies reported good fits to the nuclear IR SED of nearby AGNs assuming a clumpy distribution of dust surrounding the central engine (e.g., \citealt{Mason06,Mason09,Nikutta09}; \citealt{Ramos09}; hereafter RA09; \citealt{Ramos11b}; hereafter RA11; \citealt{Honig10}; \citealt{Herrero11}; hereafter AH11; \citealt{Sales11,Lira13,Ichikawa15,Bernete2015,
Siebenmorgen2015,Fuller16,Audibert17, Garcia-Gonzalez17}). Although the torus properties of nearby Seyfert galaxies have been extensively studied in the literature, to date there have been no studies based on an ultra-hard X-ray (14--195~keV) selected sample of these galaxies using high angular resolution IR data. 

In this work we use the \citet{Nenkova08a,Nenkova08b} clumpy torus models, known as CLUMPY, and the Bayesian tool \textsc{BayesClumpy} \citep{Asensio09,Asensio13} to fit the nuclear IR emission of an ultra-hard X-ray selected sample of Seyfert galaxies. Our aim is to study the torus properties that are driving the Seyfert type classification, the difference in the dusty torus of the various Seyfert types and how they vary with the central engine properties.

The paper is organized as follows. Section \ref{sample} and \ref{observations} describe the sample selection, the observations and data compilation, respectively. The nuclear IR SED construction and modelling are presented in Section \ref{nuclear_sed}. In Section \ref{Comparison_parameter}, we compare the torus properties for the different Seyfert subgroups. Finally, in Section \ref{Conclusions} we summarize the main conclusions of this work. Throughout this paper, we assumed a cosmology with H$_0$=73 km~s$^{-1}$~Mpc$^{-1}$, $\Omega_m$=0.27, and $\Omega_{\Lambda}$=0.73, and a velocity-field corrected using the \citet{Mould00} model, which includes the influence of the Virgo cluster, the Great Attractor, and the Shapley supercluster.

\begin{table*}
\tiny 
\centering
\begin{tabular}{lccccccc}
\hline
Name &	R.A.&	Dec.&D$_{L}$ &Spatial&Seyfert&b/a&Foreground\\
 	&	(J2000)&	(J2000)&(Mpc) &scale& type& &extinction  (A$_{V}^{for}$) \\
 	&				&& &(pc arcsec$^{-1}$)& && (mag)\\
 (1)&(2)&(3)&(4)&(5)&(6)&(7)&(8)\\	
\hline
ESO\,005-G004		&	06h05m41.6s&	-86d37m55s&24.1&116&2.0	&0.21& $\cdots$	\\
MCG-05-23-016		&	09h47m40.1s&	-30h56m55s&35.8&171&2.0	&0.45& $>$6$^a$\\
MCG-06-30-015		&	13h35m53.7s&	-34d17m44s&26.8&128&1.2	&0.60&	$\sim$1.8-3.0$^b$\\
NGC\,1365		&	03h33m36.4s&	-36d08m25s&21.5&103&1.8		&0.55&	$<$5$^c$\\
NGC\,2110		&	05h52m11.4s&	-07d27m22s&32.4&155&2.0		&0.76&	5.0$^d$\\
NGC\,2992		&	09h45m42.0s&	-14d19m35s&34.4&164&1.9		&0.31&	3.8$^e$\\
NGC\,3081		&	09h59m29.5s&	-22d49m35s&34.5&164&2.0		&0.76&	$\cdots$\\
NGC\,3227		&	10h23m30.6s&	+19d51m54s&20.4&98&1.5		&0.67&	2.3$^e$\\
NGC\,3783		&	11h39m01.7s&	-37d44m19s&36.4&173&1.2		&0.89&	0.8$^f$\\
NGC\,4051		&	12h03m09.6s&	+44d31m53s&12.9&62&1.2		&0.75&	1.0$^g$\\
NGC\,4138		&	12h09m29.8s&	+43d41m07s&17.7&85&1.9		&0.65&	$\cdots$\\
NGC\,4151		&	12h10m32.6s&	+39d24m21s&20.0&96&1.5		&0.71&	1.0$^e$\\
NGC\,4388*		&	12h25m46.7s&	+12d39m44s&17.0&82&2.0		&0.19&	5.9$^h$\\
NGC\,4395		&	12h25m48.8s&	+33d32m49s&3.84&19&1.8		&0.83&	0.4$^i$\\
NGC\,4945		&	13h05m27.5s&	-49d28m06s&4.36&21&2.0		&0.19&	$\cdots$\\
NGC\,5128 (CenA)	&	13h25m27.6s&	-43d01m09s&4.28&21&2.0	&0.78&	$\sim$7-8$^j,k$\\
NGC\,5506		&	14h13m14.9s&	-03d12m27s&30.1&144&1.9		&0.24&	$\geq$11$^l$\\
NGC\,6300		&	17h16m59.5s&	-62d49m14s&14.0&68&2.0		&0.67&	$\cdots$\\
NGC\,6814		&	19h42n40.6s&	-10d19m25s&25.8&123&1.5		&0.93&	$\cdots$\\
NGC\,7172		&	22h02m01.9s&	-31d52m11s&37.9&180&2.0		&0.56&	$\cdots$\\
NGC\,7213		&	22h09m16.3s&	-47d10m00s&25.1&120&1.5		&0.90&	0.6$^m$\\
NGC\,7314		&	22h35m46.2s&	-26d03m02s&20.9&100&1.9		&0.46&	$\cdots$\\
NGC\,7582		&	23h18m23.5s&	-42d22m14s&22.1&106&2.0		&0.42&	$\sim$8-13$^n$\\
UGC\,6728		&	11h45m16.0s&	+79d40m53s&32.1&153&1.2		&0.63&	$\cdots$\\

\hline
\end{tabular}						 
\caption{BCS$_{40}$ sample. Right ascension (R.A.), declination (Dec.), Seyfert type and galaxy inclination (b/a) were taken from the NASA/IPAC Extragalactic Database (NED). *This galaxy is part of the Virgo Cluster \citep{Binggeli85}. A$_{V}^{for}$ corresponds to the foreground extinction due to the host galaxy. References: a) \citet{Veilleux97}; b) \citet{Reynolds97}; c) \citet{Alloin81}; d) \citet{Storchi-Bergmann99}; e) \citet{Ward87b}; f) \citet{Ward84}; g) \citet{Contini99}; h) \citet{Ardila17}; i) \citet{Lira99}; j) \citet{Packham96}; k) \citet{Marconi00};  l) \citet{Goodrich94}; m) \citet{Halpern84}; n) \citet{Winge00}.} 
\label{tab1}
\end{table*}

\section{Sample Selection}
\label{sample}

The sample studied here consists of 24 Seyfert galaxies previously presented in Garc\'ia-Bernete et al. (2016; hereafter GB16). It was drawn from the {\textit{Swift/BAT}} nine month catalog \citep{Tueller2008}. The ultra hard 14-195 keV band of the parent sample is far less sensitive to the effects of obscuration than optical or softer X-ray wavelengths, making this AGN selection one of the least biased for N$_H$ $<$10$^{24}$~cm$^{-2}$ to date (see e.g. \citealt{Winter2009,Winter2010,Weaver2010,Ichikawa2012,Ricci15,Ueda15}). 

We selected all the Seyfert galaxies in the nine month catalog with luminosity distances D$_L<$40\,Mpc. We used this distance limit to ensure a resolution element of $\leqslant$50\,pc in the MIR, considering the average angular resolution of 8-10~m-class ground-based telescopes ($\sim$0.3\arcsec ~at 10~$\mu$m). This constraint provides us with a sample of 24 local Seyfert galaxies (hereafter BCS$_{40}$ sample; GB16) containing 8 Sy1 (Sy1, Sy1.2 and Sy1.5), 6 Sy1.8/1.9 and 10 Sy2 galaxies. This sample covers an AGN luminosity range log(L${_{\textrm{int}}^{2-10~\textrm{keV}}}$)$\sim$40.5--43.4~erg~s$^{-1}$. See GB16 for further details on the sample selection. The properties of the BCS$_{40}$ sample are shown in Table \ref{tab1}.

\section{Observations}
\label{observations}
Our aim is to construct high angular resolution IR SEDs for the whole sample. In the following we describe the new and archival MIR and near-IR (NIR; $\sim$1--5~$\mu$m) observations used in this work.

\subsection{New Observations}
\label{new_observations}
\subsubsection{Gran Telescopio CANARIAS/CanariCam}
\label{CC}

We obtained subarcsecond resolution N-band spectra (7.5-13~$\mu$m) of two Seyfert galaxies (NGC\,4138 and UGC\,6728) using the low spectral resolution (R$\sim$175) grating available in the instrument CanariCam (CC; \citealt{Telesco03,Packham2005}), on the 10.4~m~Gran Telescopio CANARIAS (GTC). CC is a MIR (7.5-25~$\mu$m) imager with spectroscopic, coronagraphic and polarimetric capabilities. It uses a Si:As detector, which covers a field of view (FOV) of 26$\times$19 arcsec$^2$ on the sky and it has a pixel scale of 0.08 arcsec. NGC\,4138 and UGC\,6728 were observed in 2016 March and the slit, of width 0.52 arcsec, was oriented at PA= 145 and 150 degrees, respectively. The total on-source integration times were 1061 and 1415\,s, respectively. In both cases, the standard MIR chopping-nodding technique was used with chop and nod throws of 15 arcsec (see Table \ref{tab2}). The data were taken on 2016 March 14 and 15 as part of a Director's Discretionary Time program (GTC04-15B DDT; PI: I. Garc\'ia-Bernete). 
Using the acquisition images of the standard stars used for NGC\,4138 (HD\,95121) and UGC\,6728 (HD\,105943), we measured for the standard stars full width at half-maximum (FWHM) values of  0.28\arcsec (at $\lambda$=10.3~$\mu$m) and 0.34\arcsec (at $\lambda$=8.7~$\mu$m), respectively.

\begin{landscape}
\begin{table}
\scriptsize
\centering
\begin{tabular}{lccccccccccc}
\hline
Name		& \multicolumn{5}{|c|}{NIR Flux Density (mJy)}&  & & \multicolumn{2}{|c|}{N-band Spectroscopy} \\
			&  \multicolumn{5}{|c|}{------------------------------------------------------------------------------------------------}& &&\multicolumn{2}{|c|}{------------------------------------} \\
			& J-band&H-band & K-band& L-band& M-band&Filters &Ref.&  Slit Width& P.A. & Ref.\\ 
						&  &  &  &  &  &  & &  (arcsec) & (deg) &\\
 (1)&(2)&(3)&(4)&(5)&(6)&(7)&(8)&(9)&(10)&(11)\\	
						
\hline
ESO\,005-G004& $<$14.3		& $<$28		& $<$39.6		& $\cdots$  & $\cdots$  &2MASS/J, H, K & a &$\cdots$&$\cdots$&$\cdots$\\
MCG-05-23-016& 1.1$\pm$0.2	& 3.7$\pm$0.6	& 10.7$\pm$1.6	& $<$79.5	& $\cdots$  &UKIRT/J, H, K, L				& b &0.75& 50& o\\
MCG-06-30-015& 6.5$\pm$1.0	& 11.8$\pm$1.8	& 31.3$\pm$4.7	& $\cdots$	& $\cdots$	&UKIRT/J, H, K					& c&0.75& 80& o\\
NGC\,1365	& $\cdots$		& 8.3$\pm$0.8	& $<$24.2		& $\cdots$  & $\cdots$	&HST/F160W, AAT/K				& d, e& 0.35& 15& p\\
NGC\,2110	& 6.6$\pm$1.0	& 8.7$\pm$1.3	& 12.2$\pm$1.8	& $<$58.1	& $\cdots$	&CIRCE/J, H, K, UKIRT/L				& {\bf{f}}, g &0.75& 55& o\\
NGC\,2992	& 1.8$\pm$0.3	& 2.3$\pm$0.3   & 5.4$\pm$0.8	& $<$22.7	& $\cdots$	&CIRCE/J, H, K, UKIRT/L				& {\bf{f}}, g&0.52& 30& q, r\\
NGC\,3081	& $\cdots$		& 0.22$\pm$0.02	& $<$1.8		& $<$11.3	& $\cdots$	&HST/F160W, SINFONI/K, UKIRT/L	& h, i, g& 0.65 & 0 & p\\
NGC\,3227	& $\cdots$		& 7.8$\pm$0.8	& 16.6$\pm$1.7	& $<$46.7	& $\cdots$	&HST/F160W, F222M, NFSCam/L 	& j, k & 0.52 & 0 & r\\
NGC\,3783	& 23$\pm$3	& $\cdots$		& 73$\pm$11	& $<$170	& $\cdots$	&NACO/J, K, L					& l&0.75& 315& o\\
NGC\,4051	& $\cdots$		& 12.8$\pm$1.9	& 15.1$\pm$2.3	& $<$73.5	& $\cdots$	&WHT/H, SINFONI/K, NFSCam/L		& e, i, k & 0.52 & 310 & r\\
NGC\,4138	& $\cdots$		& $\cdots$		& 2.3$\pm$0.4	& $\cdots$	& $\cdots$	&HST/F190N						& {\bf{f}} &0.52& 145 & {\bf{f}} \\
NGC\,4151	& 69$\pm$7	& 103.6$\pm$10.4& 177.5$\pm$17.8& $<$325	& $<$449	&HST/F110W, F160W, F222M, KPNO/L, M	& k, b & 0.36 & 60 &s\\
NGC\,4388	& 0.06$\pm$0.01	& 0.7$\pm$0.1	& $<$7.5		& $<$40	& $\cdots$	&HST/F110W, F160W, SINFONI/K, UKIRT/L & k, i &0.52&90&r\\
NGC\,4395	& $\cdots$		& 0.9$\pm$0.1	& $<$1.1		& $\cdots$	& $\cdots$	&HST/F160W, NFSCam/K			& k &$\cdots$&$\cdots$&$\cdots$\\
NGC\,4945	& $\cdots$		& $<$0.15		& $<$25.8		& $\cdots$	& $\cdots$	&HST/F160W, AAT/K				& h, e& 0.65&45&p\\
NGC\,5128	& 1.3$\pm$0.2	& 5.8$\pm$0.6	&30$\pm$4	& 200$\pm$30	& $\cdots$	&NACO/J, HST/F160W, NACO/K, L & l, h &0.65&0 &p\\
NGC\,5506	& 13$\pm$2	& 53.1$\pm$5.3	&80$\pm$12	&290$\pm$44		& $<$530	&NACO/J, HST/F160W, NACO/K, L, UKIRT/M & l, h, b &0.35& 60& p\\
NGC\,6300	&$\cdots$		& 2.0$\pm$0.2	& $<$5.4		& $\cdots$	& $\cdots$	&HST/F160W, SINFONI/K			& h, i &$\cdots$&$\cdots$&$\cdots$\\
NGC\,6814   & $\cdots$		& 6.2$\pm$0.6   & 6.3$\pm$1.0   & $\cdots$  & $\cdots$  &HST/F160W, SINFONI/K           & h, i &$\cdots$&$\cdots$&$\cdots$\\
NGC\,7172	&1.3$\pm$0.2	&8.8$\pm$1.3	&15$\pm$2	& $<$30	& $<$61.4	&CIRCE/J, H, K, UKIRT/L, M			& {\bf{f}}, b&0.35&60&p\\
NGC\,7213   & 1.8$\pm$0.3   & 5.2$\pm$0.8   & 14.7$\pm$2.2  & $\cdots$  & $\cdots$  &SOFI/J, H, K                   & m &0.75& 300& o\\
NGC\,7314	&1.9$\pm$0.3	&3.6$\pm$0.5	&7.1$\pm$1.1	& $<$21.9	& $\cdots$	&CIRCE/J, H, K, UKIRT/L				& {\bf{f}}, n&$\cdots$&$\cdots$&$\cdots$\\
NGC\,7582	&$\cdots$		& $<$11		& $<$18		&$\cdots$	& $\cdots$	&HST/F160W, NACO/K					& l&0.70&0&p\\
UGC\,6728	& $<$ 9.2       & $<$10.8       & $<$11.8       & $\cdots$  & $\cdots$  &2MASS/J, H, K                  & a &0.52& 150 & {\bf{f}}\\
\hline
\end{tabular}						 
\caption{Summary of the NIR fluxes and MIR spectroscopy employed in this work. Columns from 2 to 6 list the J-, H-, K-, L- and M-band fluxes available. Columns 7 and 8 correspond to their corresponding instruments/filters and references, respectively. Columns 9, 10 and 11 list the N-band spectroscopy slit widths, position angles (P.A.) and references. References: a) \citet{Skrutskie06}; b) \citet{Alonso-Herrero01}; c) \citet{Kotilainen92b}; d) \citet{Carollo02}; e) \citet{Sosa-Brito01}; {\bf{f) This Work}}; g) \citet{Alonso-Herrero98}; h) \citet{Quillen01}; i) \citet{Burtscher15}; j) \citet{Kishimoto07}; k) \citet{Alonso-Herrero03}; l)  \citet{Prieto10}; m) \citet{Prieto02b}; n) \citet{Ward87a}; o) \citet{Honig10}; p) \citet{Gonzalez-Martin2013}; q) \citet{Bernete2015}; r) \citet{Herrero16}; s)\citet{Herrero11}.}
\label{tab2}
\end{table}
\end{landscape}

The data reduction was carried out with the \textit{RedCan} pipeline \citep{Gonzalez-Martin2013}, which performs sky subtraction, stacking of individual observation, rejection of the bad frames (due to excess array of sky noise), wavelength and flux calibration, trace determination and spectral extraction. We extracted the nuclear spectra as a point source for both galaxies. Note that for point source extraction, \textit{RedCan} uses an aperture that increases with wavelength to take into account the decreasing angular resolution, and it also performs a correction account for slit loses (see \citealt{Gonzalez-Martin2013} for further details on CC data reduction).

\subsubsection{Gran Telescopio CANARIAS/CIRCE}
\label{CIRCE}

We obtained NIR imaging data (J, H \& K bands) with the Canarias InfraRed Camera Experiment (CIRCE; \citealt{Garner14}) on the 10.4-m GTC. The instrument was equipped with an engineering grade Hawaii2RG detector with a total FOV of 3.4$\times$3.4 arcmin$^2$ and a plate scale of 0.1 arcsec pixel$^{-1}$. Note that all the observations were taken using a 5 dither pattern. See Table \ref{tab3} for observation details.

We performed the data reduction by using the IDL (Interactive Data Language) routines employed in \citet{DAmmando17}. The first step in the data processing includes the subtraction of dark current frames. From twilight sky exposures, we obtained an illumination correction to compensate a decrease of about 40 per cent from the centre to the border of the FOV. At this point, we introduced a correction to remove a pattern of inclined stripes related to reading amplifiers. Once this pattern was removed, the images corresponding to each dither cycle were median combined to form a sky frame, which was subtracted for each frame of the cycle. We then combined all sky-subtracted images with the commonly used shift-and-add technique. During the combination of these frames, we applied a bad-pixel mask, which includes the two vertical bands corresponding to non-functional amplifiers. Finally, we obtained the photometric calibrations relative to photometric standard PSF stars using their Two-Micron All-Sky Survey (2MASS) photometry.

To estimate the NIR nuclear fluxes in the J, H \& K bands we used the PSF subtraction method (see GB16 and references therein), which consists of subtracting the PSF star from the galaxy profiles. This method has been widely used in ground-based IR images (e.g. \citealt{Soifer00,Radomski2002,Radomski2003,Bernete2015}). 
\begin{table}
\scriptsize 
\centering
\begin{tabular}{lccccc}
\hline
Name		&Filter &Obs.	  &	Total  &PSF&	FWHM\\
			&name    &Date	  &on-source& star	&PSF\\
			&    &	          &time (s)& name	& \\
 (1)&(2)&(3)&(4)&(5)&(6)\\	
			
\hline
NGC\,2110	&J	&11/10/2016& 75&		AS05\,0 &	0.89\arcsec	\\
            &H	&11/10/2016& 75&			    &	0.75\arcsec	\\
            &Ks	&11/10/2016& 75&		        &	0.76\arcsec	\\
NGC\,2992	&J	&05/02/2017& 125&		S708\,D &	0.98\arcsec	\\
            &H	&05/02/2017& 125&				&	0.78\arcsec	\\
            &Ks	&05/02/2017& 125&				&	0.76\arcsec	\\
NGC\,7172	&J	&16/10/2016& 125&		AS31\,1 &	0.71\arcsec	\\
            &H	&16/10/2016& 125&				&	0.70\arcsec	\\
            &Ks	&16/10/2016& 125&				&	0.63\arcsec	\\
NGC\,7314	&J	&05/02/2017& 150&		AS05\,0 &	0.50\arcsec	\\
            &H	&05/02/2017& 150&				&	0.54\arcsec	\\
            &Ks	&05/02/2017& 150&				&	0.53\arcsec	\\
\hline
\end{tabular}						 
\caption{Summary of the GTC/CIRCE NIR imaging observations.} 
\label{tab3}
\end{table}

\subsection{Archival Data}
\label{archival_data}

We downloaded the fully reduced NIR imaging data of NGC\,4138 (unpublished, to our knowledge) from the ESA Hubble Legacy Archive\footnote{http://archives.esac.esa.int/hst/}. This Seyfert 1.9 galaxy was observed in February 2008 with the Near Infrared Camera and Multi-Object Spectrometer (NICMOS) and the narrow F190N filter ($\lambda_c$=1.9~$\mu$m). This observation was taken using the NIC3 camera, which has a FOV 51.2$\times$51.2 arcsec$^2$ on the sky and a pixel scale of 0.2\arcsec. This image was taken as part of the Hubble programs GO11080 (cycle:15, PI: D. Calzetti) and the exposure time was 13474 s. 

In order to accurately subtract the unresolved AGN component, first, we generated a theoretical Tiny Tim\footnote{http://tinytim.stsci.edu/cgi-bin/tinytimweb.cgi} PSF \citep{Krist95,Krist11} for the NIC3 camera F190N filter and, then, we used the PSF subtraction method.

\subsection{Literature High Angular Resolution IR Observations}
\label{nuclear_fluxes}

We compiled the highest angular resolution IR ($\sim$1-30~$\mu$m) nuclear fluxes available from the literature for our sample. The compiled nuclear NIR fluxes are from both ground- and space-based (i.e. Hubble Space Telescope; HST) data (see Table \ref{tab2}). In the case of the MIR nuclear fluxes, we used the measurements of the unresolved MIR emission (angular resolutions ranging from 0.2\arcsec to 0.6\arcsec) calculated in GB16, where we employed the PSF subtraction method on high angular resolution MIR images from 8-10 m-class ground-based telescopes (GTC/CanariCam, VLT/VISIR, Gemini/T-ReCS and MICHELLE; see Table 2 of GB16).

We retrieved 31.5~$\mu$m high angular resolution (FWHM$\sim$3.1\arcsec) nuclear fluxes of six Seyfert galaxies (see Table \ref{tab4}), which were observed with the long-wavelength camera (LWC;  $\lambda >$ 25~$\mu$m) within the Faint Object Infrared Camera for the SOFIA Telescope (FORCAST; \citealt{Herter12}) on the 2.5~m SOFIA telescope. These observations were obtained using the 31.5~$\mu$m filter ($\Delta\lambda$= 5.7~$\mu$m). See \citet{Fuller16} for further details on the observations, data reduction and obtention of unresolved nuclear fluxes.

\begin{table*}
\scriptsize 
\centering
\begin{tabular}{lcccccc}
\hline
Name		& \multicolumn{6}{|c|}{Flux Density (mJy)} \\
			&  \multicolumn{6}{|c|}{-----------------------------------------------------------------------------------------------------------------------------} \\
			& 4.5~$\mu$m&5.5~$\mu$m & 18.0~$\mu$m & 25.0~$\mu$m& 30.0~$\mu$m&  31.5~$\mu$m\\
			& & & & & &SOFIA/FORCAST\\
 (1)&(2)&(3)&(4)&(5)&(6)&(7)\\	
			
\hline
ESO\,005-G004& 2.9$\pm$0.4		& 3.8$\pm$0.6		& $\cdots$  			& 141$\pm$28	& 163$\pm$33				& $\cdots$ \\
MCG-05-23-016& $\cdots$			& 101.1$\pm$15.2	& $\cdots$  			& $<$1762.1			& $<$1898						&$<$1640\\
MCG-06-30-015& 57.3$\pm$8.6		& 69.2$\pm$10.4		& 308.5$\pm$61.7		& 352.1$\pm$70.4	& $<$519.8						& $\cdots$ \\
NGC\,1365	& 39.3$\pm$5.9		& 49.8$\pm$7.5		& $\cdots$  			& 514.5$\pm$102.9	& 554.2$\pm$110.8 $^{\dagger}$	& $\cdots$ \\
NGC\,2110	& 112.8$\pm$16.9	& 125.3$\pm$18.8	& 508.9$\pm$101.8		& 598.9$\pm$119.8	& $<$858.1		  				&$<$860 \\
NGC\,2992	& 16.1$\pm$2.4		& 28.8$\pm$4.3		& $\cdots$  			& 773.2$\pm$154.6	& 965$\pm$193			&$<$810\\
NGC\,3081	& $\cdots$			& 18.8$\pm$2.8		& $\cdots$  			& 452.2$\pm$90.4	& 520$\pm$104				&$<$800\\
NGC\,3227	& $\cdots$			& 47$\pm$7		& 839.7$\pm$167.9		& 947$\pm$189	& 1018.8$\pm$203.8				&$<$1300\\
NGC\,3783	& $\cdots$			& 133.8$\pm$20.1	& $\cdots$  			& 1022.6$\pm$204.5	& 1182.8$\pm$236.6				& $\cdots$ \\
NGC\,4051	& 76.5$\pm$11.5		& 97$\pm$15		& 661.3$\pm$132.3 		& 1001.1$\pm$200.2	& $<$1354.7						& $\cdots$ \\
NGC\,4138	& 5.8$\pm$0.9		& 6.9$\pm$1.0		& 30.9$\pm$6.2  		& 35.3$\pm$7.1		& 37$\pm$7 $^{\dagger}$		& $\cdots$ \\
NGC\,4151	& $\cdots$			& 404$\pm$61	& $\cdots$  			& 3187.4$\pm$637.5	& 2965.2$\pm$593.0				& $\cdots$ \\
NGC\,4388	& 23.5$\pm$3.5		& 30.2$\pm$4.5		& 788.6$\pm$157.7		& 1127.1$\pm$225.4	& 1305.7$\pm$261.1				&$<$2040\\
NGC\,4395	& $\cdots$			& 1.4$\pm$0.2		& 19.8$\pm$4.0  		& 25.6$\pm$5.1		& 30.3$\pm$6.1					& $\cdots$ \\
NGC\,4945	& $\cdots$			& 4.9$\pm$0.7		& $\cdots$  			& 194.4$\pm$38.9	& 252.5$\pm$50.5				& $\cdots$ \\
NGC\,5128	& $\cdots$			& 372.2$\pm$55.8	& $\cdots$  			& 3526.2$\pm$705.2	& 4095.3$\pm$819.1				& $\cdots$ \\
NGC\,5506	& $\cdots$			& 490.6$\pm$73.6	& $\cdots$  			& $<$3273.3			& $<$3960.9						&$<$3660\\
NGC\,6300	& 11.4$\pm$1.7		& 28.1$\pm$4.2		& 614.3$\pm$112.9 		& 1831.1$\pm$366.2	& $<$2694.1						& $\cdots$ \\
NGC\,6814   & 12.2$\pm$1.8		& 15.5$\pm$2.3		& $\cdots$  			& 160$\pm$32	& $<$249.7						& $\cdots$ \\
NGC\,7172	& $\cdots$			& 41.8$\pm$6.3		& $\cdots$  			& 146.6$\pm$29.3	& 166.3$\pm$33.3				& $\cdots$ \\
NGC\,7213   & $\cdots$			& 19.2$\pm$2.9		& $\cdots$  			& $<$386.6			& $<$389.7						& $\cdots$ \\
NGC\,7314	& $\cdots$			& $<$21.5			& $\cdots$  			& 180.8$\pm$36.2	& 203.8$\pm$40.8				& $\cdots$ \\
NGC\,7582	& $\cdots$			& 27.3$\pm$4.1		& $\cdots$  			& 649.7$\pm$129.9	& 923.8$\pm$184.8				& $\cdots$ \\
UGC\,6728	& 12.1$\pm$1.8		& 14.7$\pm$2.2		& 57.2$\pm$11.4			& 54.5$\pm$10.9		& 51.7$\pm$10.3					& $\cdots$ \\
\hline
\end{tabular}						 
\caption{Summary of the nuclear MIR emission derived from the AGN contribution based on spectral decomposition of Spitzer/IRS spectra and the SOFIA/FORCAST 31.5~$\mu$m fluxes. Column 1 corresponds to the galaxy name. Columns from 2 to 6 list the 4.5, 5.5, 18.0, 25.0 and 30.0 fluxes, respectively. The final column 7, corresponds to the SOFIA/FORCAST 31.5~$\mu$m fluxes, reported by \citet{Fuller16}. Note that $\dagger$ corresponds to nuclear fluxes calculated at 28~$\mu$m instead of 30~$\mu$m. See Section \ref{sed_construction} for further details.}
\label{tab4}
\end{table*}

Finally, we compiled N-band spectra (7.5--13~$\mu$m) for the majority of the sample (17/24 sources), which were obtained with different instruments (GTC/CC, VLT/VISIR, Gemini/T-ReCS and MICHELLE). Details on these observations are given in Table \ref{tab2}, and we used the fully reduced and flux calibrated spectra noted.

\section{Nuclear IR SEDs}
\label{nuclear_sed}
\subsection{SED construction}
\label{sed_construction}

To construct the entire nuclear IR SEDs sampling similar physical scales, we use NIR nuclear fluxes from our own GTC/CIRCE observations, HST archival data, or the highest angular resolution nuclear IR fluxes available in the literature. For those cases in which the angular resolution available is greater than 1\arcsec ~or there is evidence of a possible extra contribution from the host galaxy we used them as upper limits (see Table \ref{tab2}).

When available, we used the subarcsecond nuclear spectra extracted as a point source, resampling them to 50 points, following the same methodology as in previous works using N-band nuclear spectra and clumpy torus models (e.g. AH11; \citealt{Ramos14,Bernete2015}). In general, there is a good agreement between the flux calibration of the nuclear spectra and the N-band nuclear fluxes. However, for consistency, we systematically scaled the spectra to the N-band nuclear fluxes, unless there is any evidence to discard them due to the possible contribution of either emission lines or PAH features in the specific spectral window of the filters (e.g. NGC\,7582). We estimated a $\sim$15\% total uncertainty for the nuclear spectra by quadratically adding the errors in the flux calibration and point source extraction.

In addition, we estimated the AGN contribution at 5.5, 25 and 30 $\mu$m for all the galaxies based on spectral decomposition of Spitzer/IRS galaxies (see Table \ref{tab4})
\footnote{Note that when the derived rest-frame AGN component does not extend as far as 30 $\mu$m, we calculated 28 $\mu$m fluxes (e.g. NGC\,1365 and NGC\,4138). If that is not possible, we used the Spitzer/IRS spectra to estimate the 30~$\mu$m fluxes and considered the IRS fluxes as upper limits, due to the low angular resolution of Spitzer. The latter also applies to the 25~$\mu$m fluxes. We note that the 25 and 30 $\mu$m fluxes could have some contribution from the host galaxy.}. To do so, we first scaled the AGN component to the N-band fluxes and then calculated homogeneous nuclear fluxes at 5.5, 25 and 30 $\mu$m using a 1 $\mu$m window in the scaled AGN component, using the same method as in GB16. We remark that when a specific rest-frame AGN template extends down to $\sim$4 $\mu$m, which occurs for roughly half of our sample (11/24 sources), we also derived the 4.5 $\mu$m nuclear fluxes (see Table \ref{tab4}). Finally, for those sources without Q-band (17--25~$\mu$m) photometry (e.g. NGC\,4388), we calculated the 18 $\mu$m fluxes using the same methodology.

Five sources lack high angular resolution nuclear spectra (NGC\,4395, NGC\,6300, NGC\,6814, NGC\,7314 and ESO\,005-G004). Nevertheless, we have high angular resolution photometry in the N- and Q-bands, and we then used the scaled AGN components derived from the IRS spectra to obtain N-band ``pseudo-nuclear'' spectra (e.g. \citealt{Hernan-caballero2015}). For consistency with the other 19 nuclear IR SEDs, we restricted the scaled AGN component to have the same wavelength range (7.5--13~$\mu$m) as the ground-based spectra and resampled to 50 points. Note that we also use the ``pseudo-nuclear'' spectra for NGC\,4138, NGC\,4945, NGC\,7172 and UGC\,6728. In the case of NGC\,4945 and NGC\,7172, their nuclear spectra show a strong contribution from the host galaxy, while those of NGC\,4138 and UGC\,6728 are very noisy and practically identical in spectral shape to the AGN component.

\subsection{SED observational properties}
\label{sed_properties}

\begin{figure*}
\centering
\includegraphics[width=8.5cm]{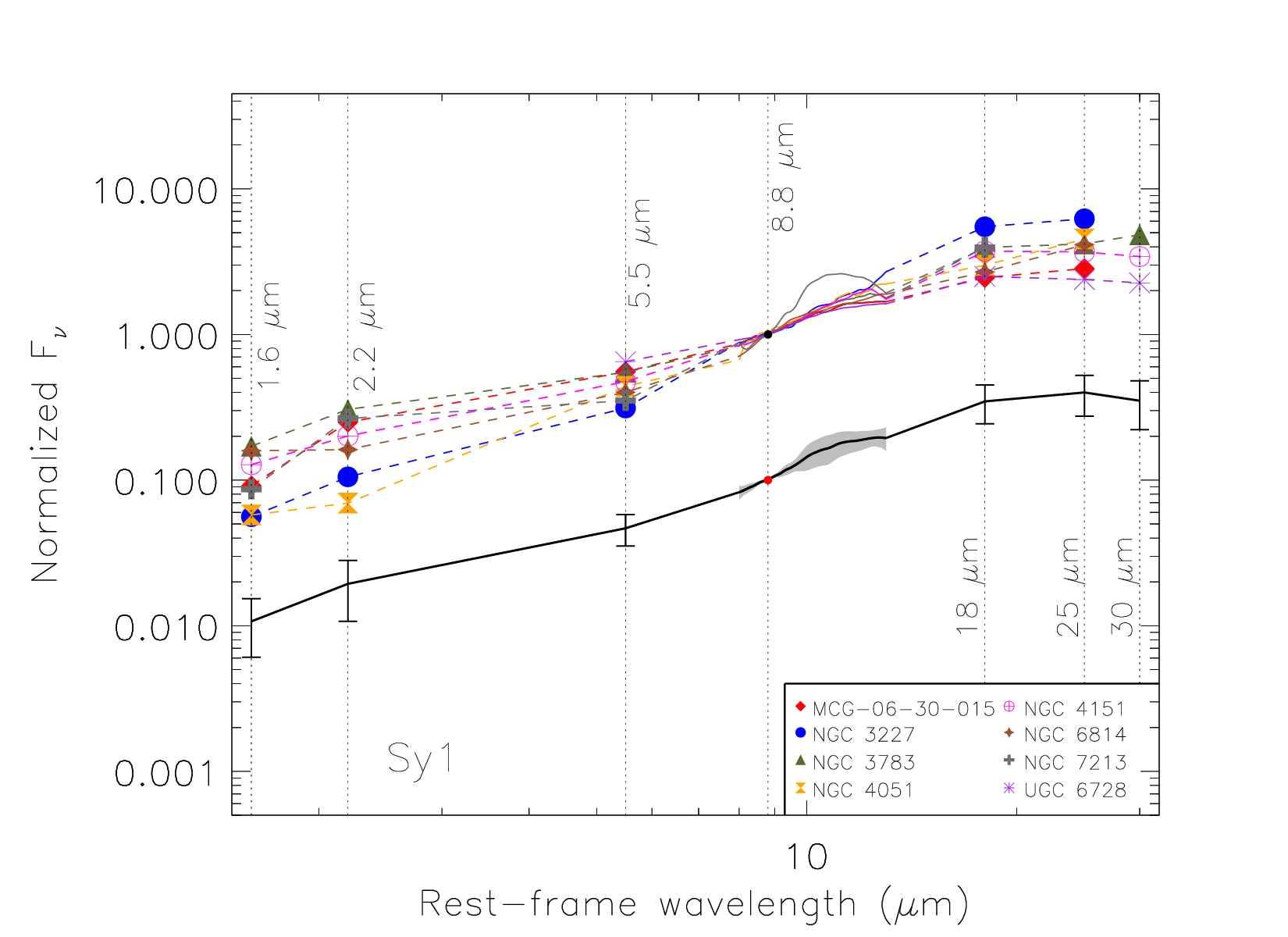}
\includegraphics[width=8.5cm]{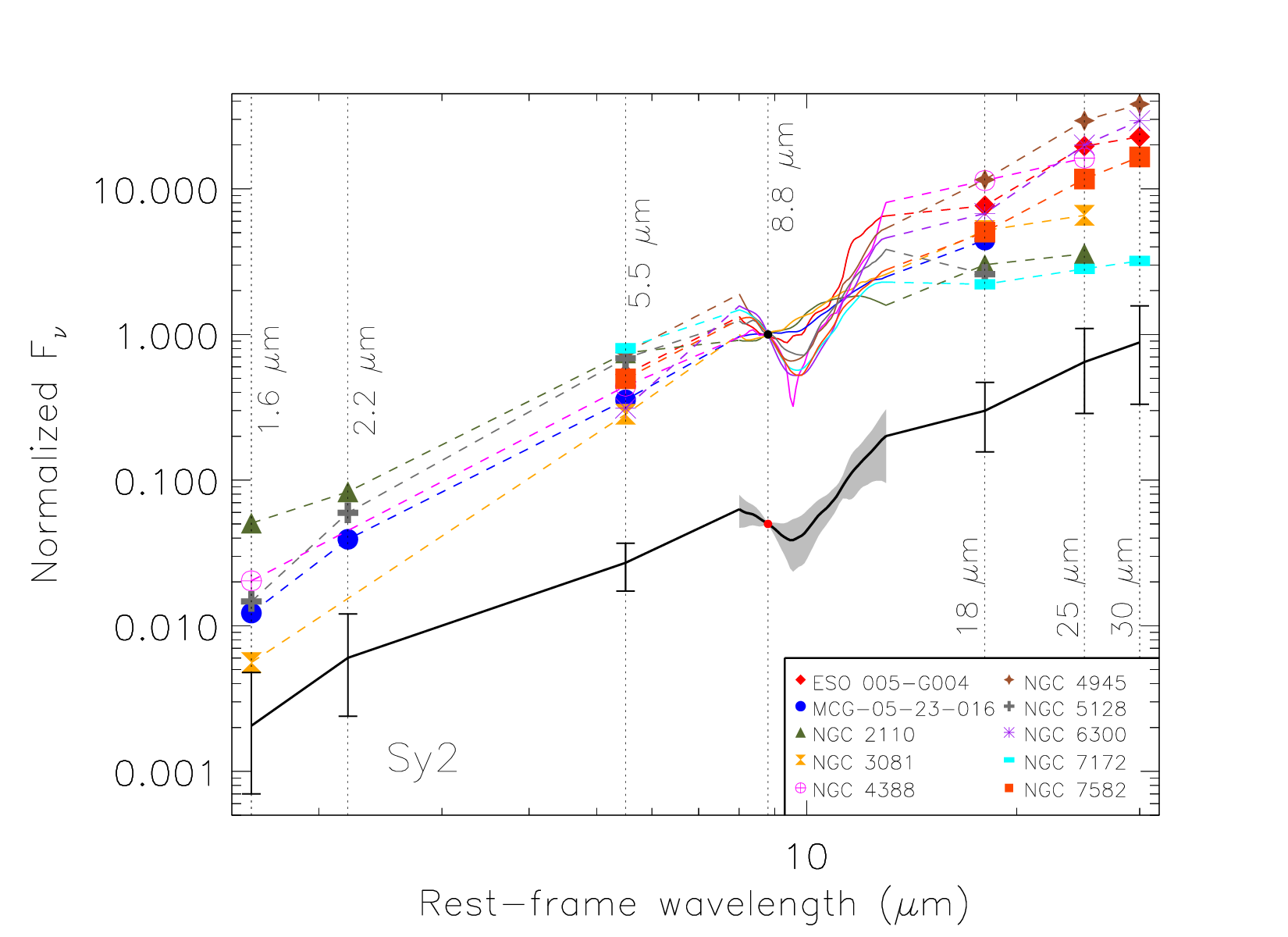}
\includegraphics[width=8.5cm]{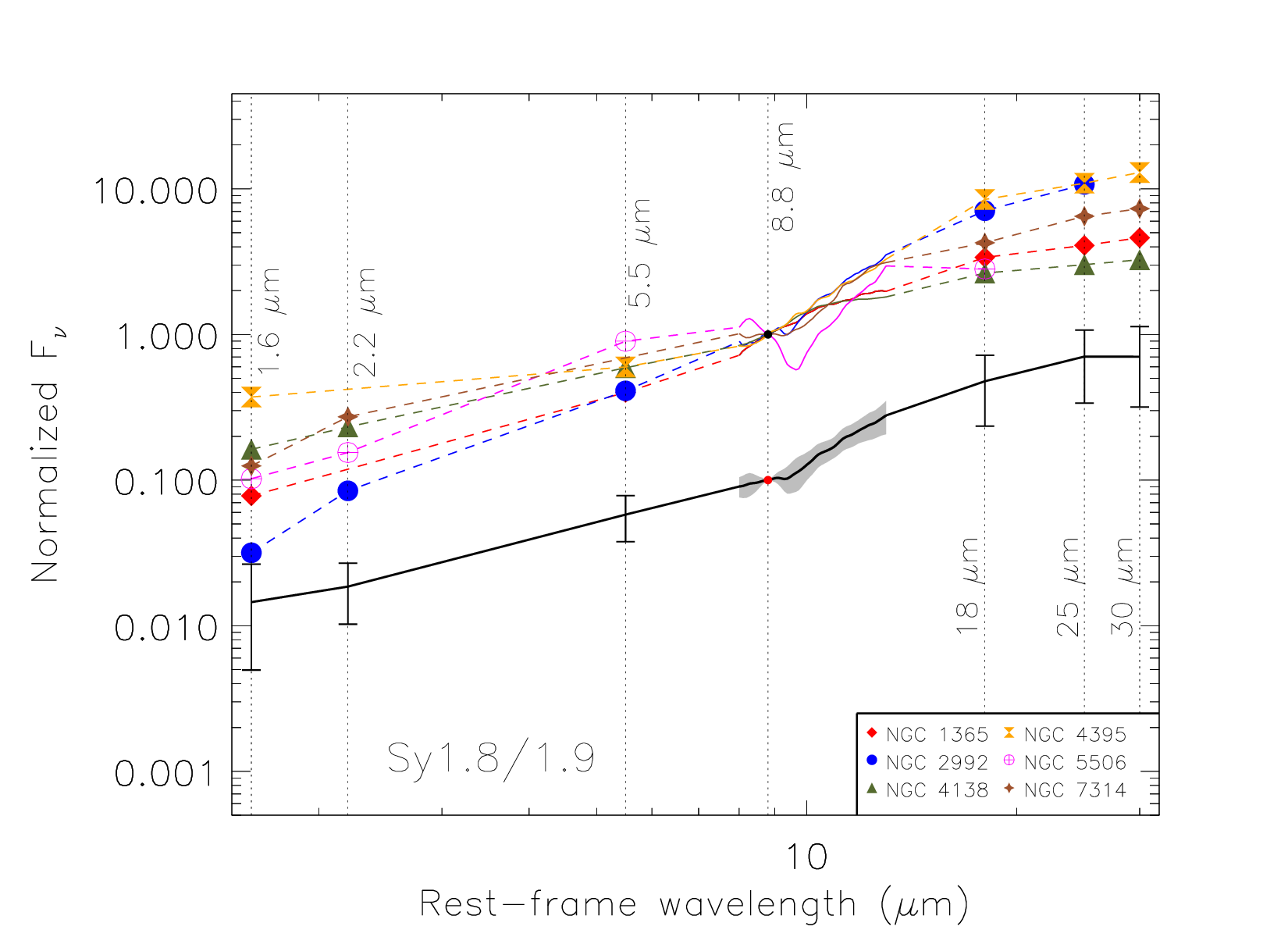}
\includegraphics[width=8.5cm]{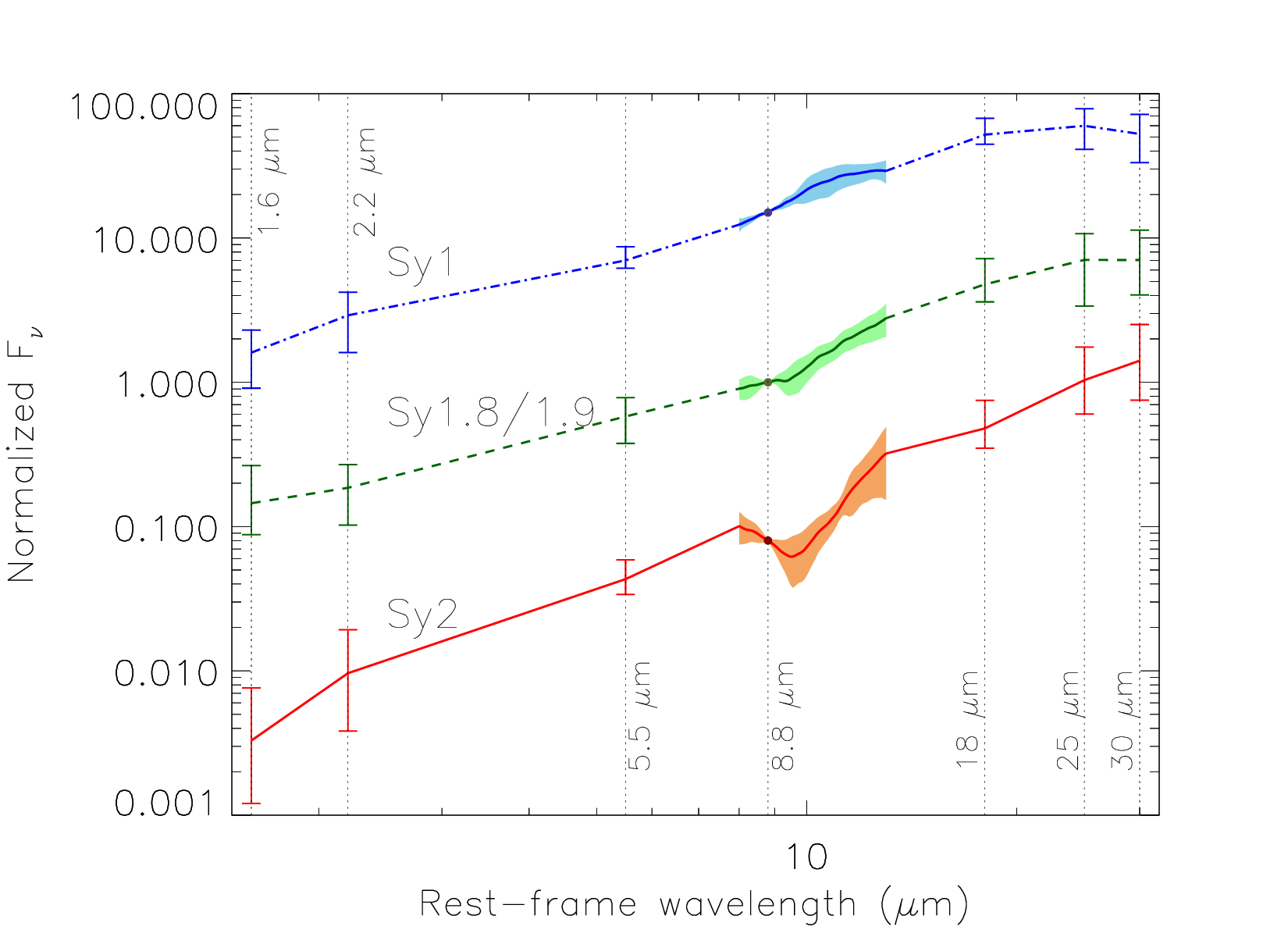}
\caption{Observed nuclear IR SEDs for the Sy1, Sy1.8/1.9 and Sy2 galaxies in the BCS$_{40}$ sample. Note that different colours and symbols correspond to the galaxies labelled in each panel. All SEDs have been normalized at 8.8~$\mu$m, and the average Sy1, Sy1.8/1.9 and Sy2 (black dashed line) have been shifted in the Y-axis for clarity. The error bars correspond to the standard deviation of each averaged point. Bottom-right panel: average Sy1, Sy1.8/1.9 and Sy2 nuclear IR SEDs. Blue dot-dashed, green dashed and red solid lines correspond to the Sy1, Sy1.8/1.9 and Sy2 templates, respectively. The average SEDs have been shifted in the Y-axis.}
\label{fig1}
\end{figure*}

In Fig. \ref{fig1} we present the nuclear IR SEDs ($\sim$1--30~$\mu$m) of our sample of Seyfert galaxies. In these plots we compare the spectral shapes and the average nuclear IR SEDs for the different Seyfert types considered in this study (Sy1, Sy1.8/1.9 and Sy2 galaxies). The average nuclear IR Sy1, Sy1.8/1.9 and Sy2 templates were constructed using the nuclear IR SEDs described in Section \ref{sed_construction}, but excluding the lowest angular resolution data (i.e. upper limits). For consistency, we used the same wavelength grid for all the photometry (1.6, 2.2, 5.5, 8.8, 18.0, 25.0, 30.0~$\mu$m). To do so, we performed a quadratic interpolation of nearby measurements for each galaxy. In this process, we avoid using L- and M-bands due to the large number of upper limits at these wavelengths. Note that we computed the interpolated fluxes for the sole purpose of deriving the average Seyfert templates. In addition, we used N-band spectra, either the subarcsecond angular resolution or the ``pseudo-nuclear'' spectra (see Section \ref{sed_construction}).

We measured the NIR (1.6--8~$\mu$m), MIR (8--18~$\mu$m) and total IR (1.6--25~$\mu$m) spectral indices ($f_\nu$~$\alpha$~$\nu^\alpha$), the H/N and N/Q flux ratios, and the strength of the silicate feature (9.7~$\mu$m) for each galaxy in the sample. We also repeated these measurements in the derived Sy1, Sy1.8/1.9 and Sy2 templates, which are representative of each group of SEDs (see Table \ref{tab5}). We find steeper IR slopes for Sy2 than for Sy1, and the Sy1.8/1.9 and Sy1 slopes are very similar. Steeper IR slopes for type-2 AGN have been previously reported in the literature for Seyfert galaxies (e.g. \citealt{Alonso-Herrero03}, RA11 and references therein) and more luminous AGNs (e.g. \citealt{Mateos16}).
In addition, we measured practically the same MIR slopes for the three groups ($\alpha_{MIR}\sim$-2) within the errors, in good agreement with the results reported by RA11. Following the same methodology as in RA09 and RA11, we also compare the spectral shapes of the different Seyfert types using the H/N and N/Q flux ratios. In agreement with the values reported by the latter authors, we found similar N/Q flux ratios ($\sim$0.3-0.2). On the other hand, we found that Sy1 (0.11$\pm$0.05) and Sy1.8/1.9 (0.15$\pm$0.12) galaxies have slightly larger values of the H/N flux ratio than those of Sy2 (0.04$\pm$0.05), but the values are consistent within the errors.

\begin{table*}
\centering
\begin{tabular}{lccccccc}
\hline
    & H/N & N/Q & $\alpha_{IR}$& $\alpha_{NIR}$ & $\alpha_{MIR}$ & S$_{Sil}$ \\ 
		& 1.6/8.8~$\mu$m &8.8/18~$\mu$m &1.6--25~$\mu$m& 1.6--8.0~$\mu$m& 8--18~$\mu$m& 9.7~$\mu$m\\
		 (1)&(2)&(3)&(4)&(5)&(6)&(7)\\	
\hline
Average Sy1& 0.11$\pm$0.05 &		0.31$\pm$0.08 & -1.4$\pm$0.2 &-1.3$\pm$0.3 & -1.7$\pm$0.3 & 0.07$\pm$0.14 \\
Average Sy1.8/1.9& 0.15$\pm$0.12 &		0.25$\pm$0.11  &		-1.5$\pm$0.4	 &	  -1.3$\pm$0.5 &		-1.9$\pm$0.7 & 	-0.33$\pm$0.45 \\
Average Sy2 & 0.04$\pm$0.05 & 0.22$\pm$0.13 & -2.0$\pm$0.6 & -2.4$\pm$0.6 & -1.8$\pm$0.7 & -1.01$\pm$0.65 \\
\hline
\end{tabular}					 
\caption{Spectral shape information of the nuclear IR SEDs. The strength of the 9.7~$\mu$m silicate feature is computed as S$_{Sil}$= ln (f$_{cont}$/f$_{9.7}$), where f$_{cont}$ and f$_{9.7}$ are the flux densities of the continuum and the feature, which we measured at 9.7~$\mu$m.}
\label{tab5}
\end{table*}

Taking advantage of the spectroscopy data we compare the strength of the silicate feature (9.7~$\mu$m) for the different Seyfert types (see Table \ref{tab5}). The latter is computed as S$_{Sil}$= ln (f$_{cont}$/f$_{9.7}$), where f$_{cont}$ and f$_{9.7}$ are the flux densities of the continuum and the feature, which we measured at 9.7~$\mu$m. As can be seen from the top-left panel of Fig. \ref{fig1}, the majority of the Sy1 galaxies show weak or moderate emission (S$_{Sil}>$0; the only exception is NGC\,3227, which has a value of -0.2 and it could be related to the emission of PAHs), whereas Sy1.8/1.9 and Sy2 galaxies have relatively deep silicate features (S$_{Sil}$=-0.3 and -1.0, respectively). This feature is normally observed in weak emission or absent in Sy1 and in shallow absorption in type 2 Seyfert galaxies when observed in subarcsecond resolution data (e.g. \citealt{Herrero16}, \citealt{Garcia-Gonzalez17} and references therein). 

\subsection{Accretion disk fitting}
\label{accretion_disk_section}

The NIR emission of AGN is mainly produced by the emission of very hot dust and the direct emission from the AGN (i.e. accretion disk) in the case of type 1s, although another important contribution can be stellar emission from the host galaxy. The contribution from the accretion disk declines with increasing wavelength. According to both theoretical models (e.g. \citealt{Hubeny01}) and polarized light observations (e.g. \citealt{Kishimoto01}) the NIR emission of the accretion disk can be explained by a power-law extension of the optical/UV spectrum to the NIR range. This power-law extrapolation is commonly used to fit the AGN direct emission in Seyfert 1 galaxies (e.g. \citealt{Stalevski2012}). However, the clumpy torus models of \citet{Nenkova08a,Nenkova08b} assume a steep fall of the disk spectrum for wavelengths longer than 1~$\mu$m. We note that the CLUMPY models cannot reproduce the NIR bumps observed in the SEDs of some Sy1s (e.g. \citealt{Mor09}; RA11; AH11). For example, \citet{Mateos16} successfully reproduced the IR SEDs of a sample of X-ray selected quasars using a non-truncated disk component and the CLUMPY torus models.

In order to quantify the contribution from the accretion disk to the nuclear NIR emission, we follow the same procedure as described in \citet{Hernan-caballero2016} using optical, NIR and MIR photometry (see Tables \ref{tab2}, \ref{tab4} and \ref{tab6}) to fit the accretion disk emission for all Sy1 galaxies\footnote{Since Sy1.8/1.9 tend to have relatively high values of foreground extinction (see Table \ref{tab1}), we did not consider the direct AGN contribution like in the case of Sy2 galaxies (see also RA09 and RA11).} in our sample. This method used a semi-empirical model consisting of a single template for the accretion disk and two blackbodies for the dust emission. 

\begin{table}
\centering
\begin{tabular}{lcccc}
\hline
Name            &F$_{J}$&F$_{H}$&F$_{K}$& Ref.\\ 
 (1)&(2)&(3)&(4)&(5)\\	

\hline
MCG-06-30-015	&0.17	&0.09	& 0.03	& a\\    
NGC\,3227   	&0.52	&0.25	& 0.10	& b,c\\
NGC\,3783   	&0.27	&0.14	& 0.07	& d\\
NGC\,4051   	&0.21	&0.11	& 0.08	& c\\
NGC\,4151   	&0.83	&0.40	& 0.19	& c\\
NGC\,6814   	&0.41	&0.24	& 0.20	& e\\
NGC\,7213   	&0.49	&0.15	& 0.04	& f\\
UGC\,6728   	&0.22$^\dagger$	&0.18$^\dagger$	& 0.14$^\dagger$	& h\\
Average Sy1		&0.46$\pm$0.28	 &0.23$\pm$0.13	&0.11$\pm$0.05	&...\\
\hline
\end{tabular}					 
\caption{Accretion disk measurements derived from the fitting of Sy1s. 
Columns 2, 3 and 4 list the fractional contribution of the accretion disk component to the J-, H- and K-band emission, respectively. References for the optical photometry: a) \citet{Bentz16a}; b) \citet{Munoz-Marin07}; c) \citet{Hoandpeng01}; d) \citet{Prieto10}; e) \citet{Bentz13}; f) \citet{Lauer05}; g) \citet{Bentz16b}. Note that for the average values we used only sources with subarcsecond resolution data.}
$\dagger$ Derived from NIR upper limits. 
\label{tab6}
\end{table}

\begin{figure*}
\centering
\par{
\includegraphics[width=8.0cm]{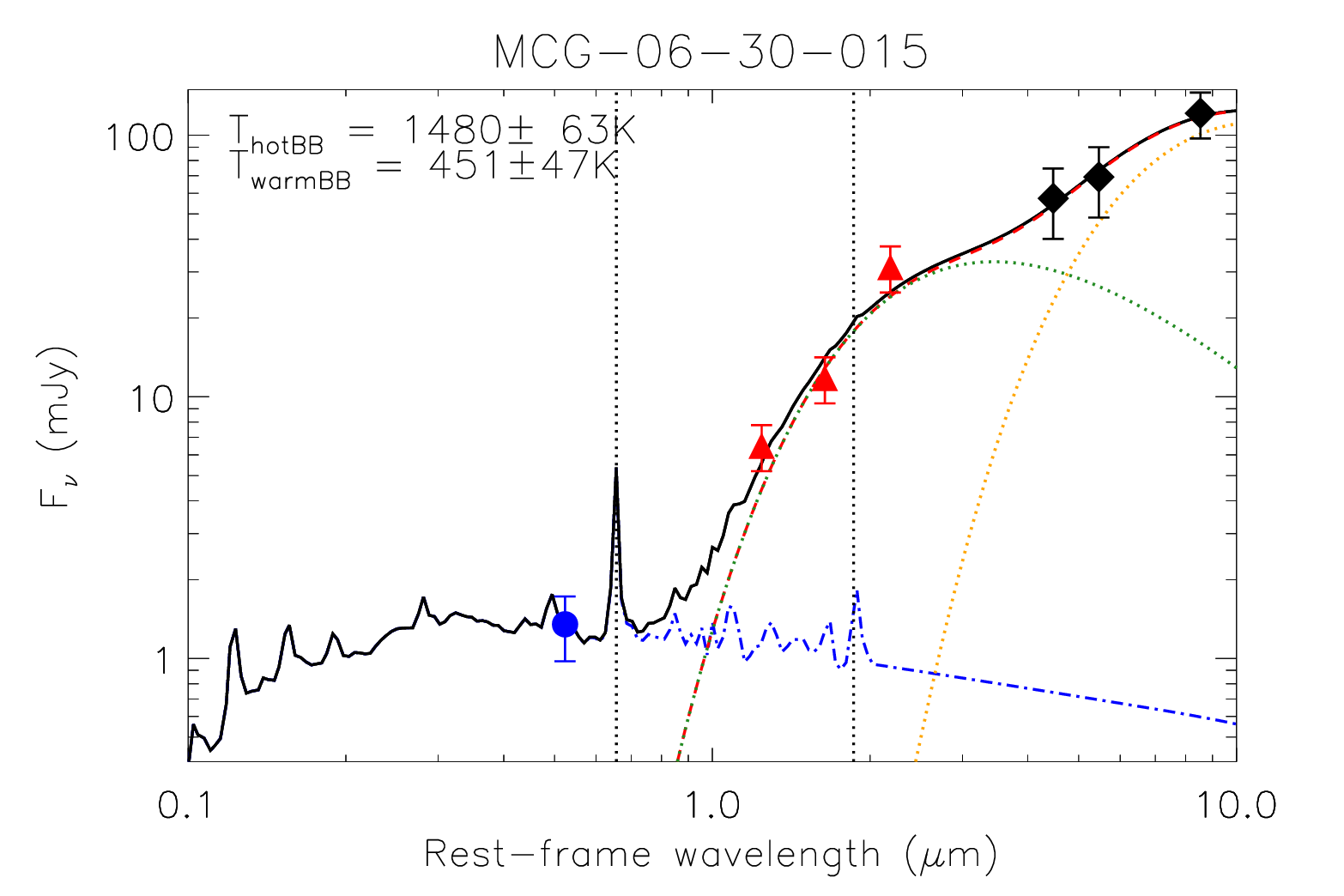}
\includegraphics[width=8.0cm]{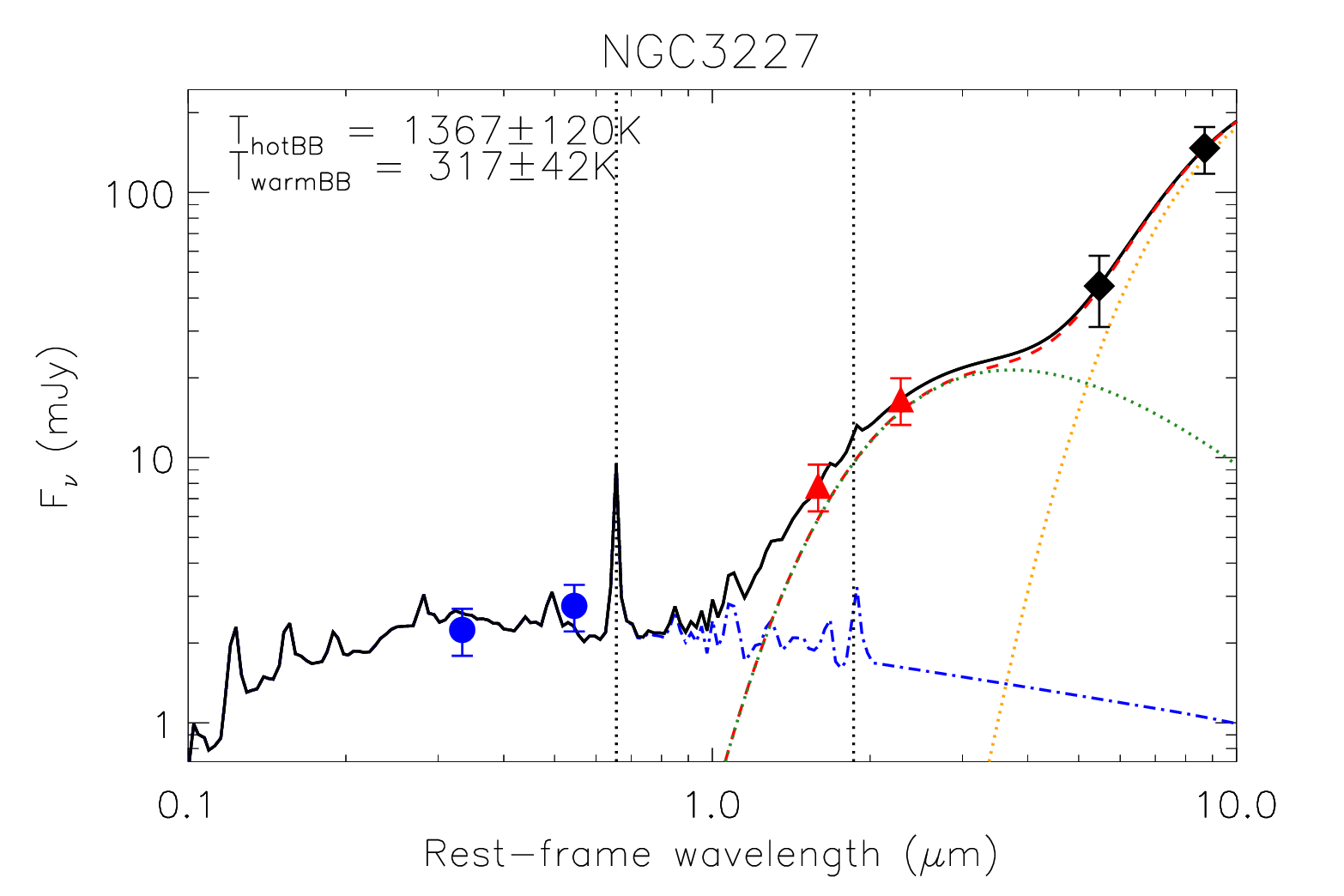}
\includegraphics[width=8.0cm]{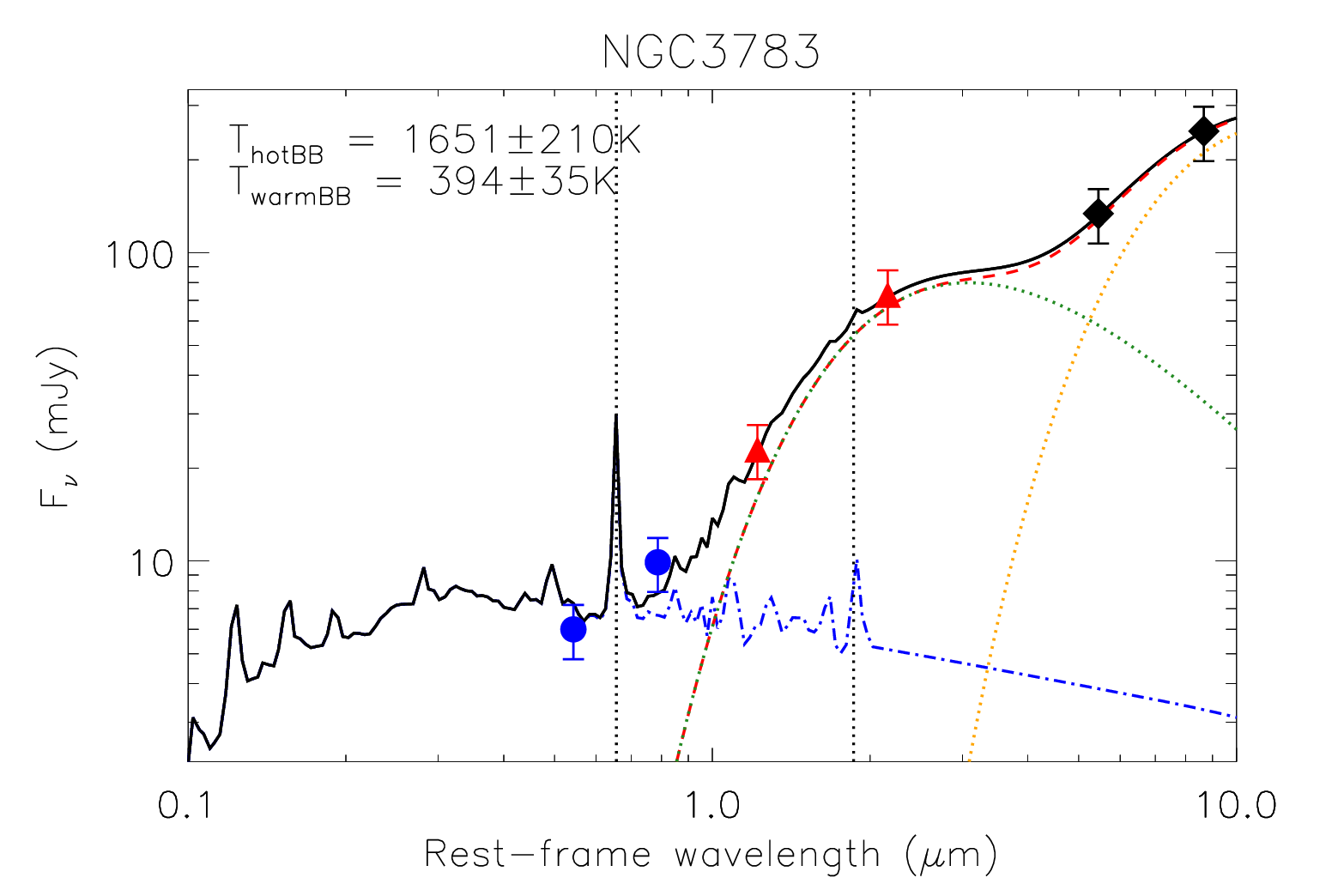}
\includegraphics[width=8.0cm]{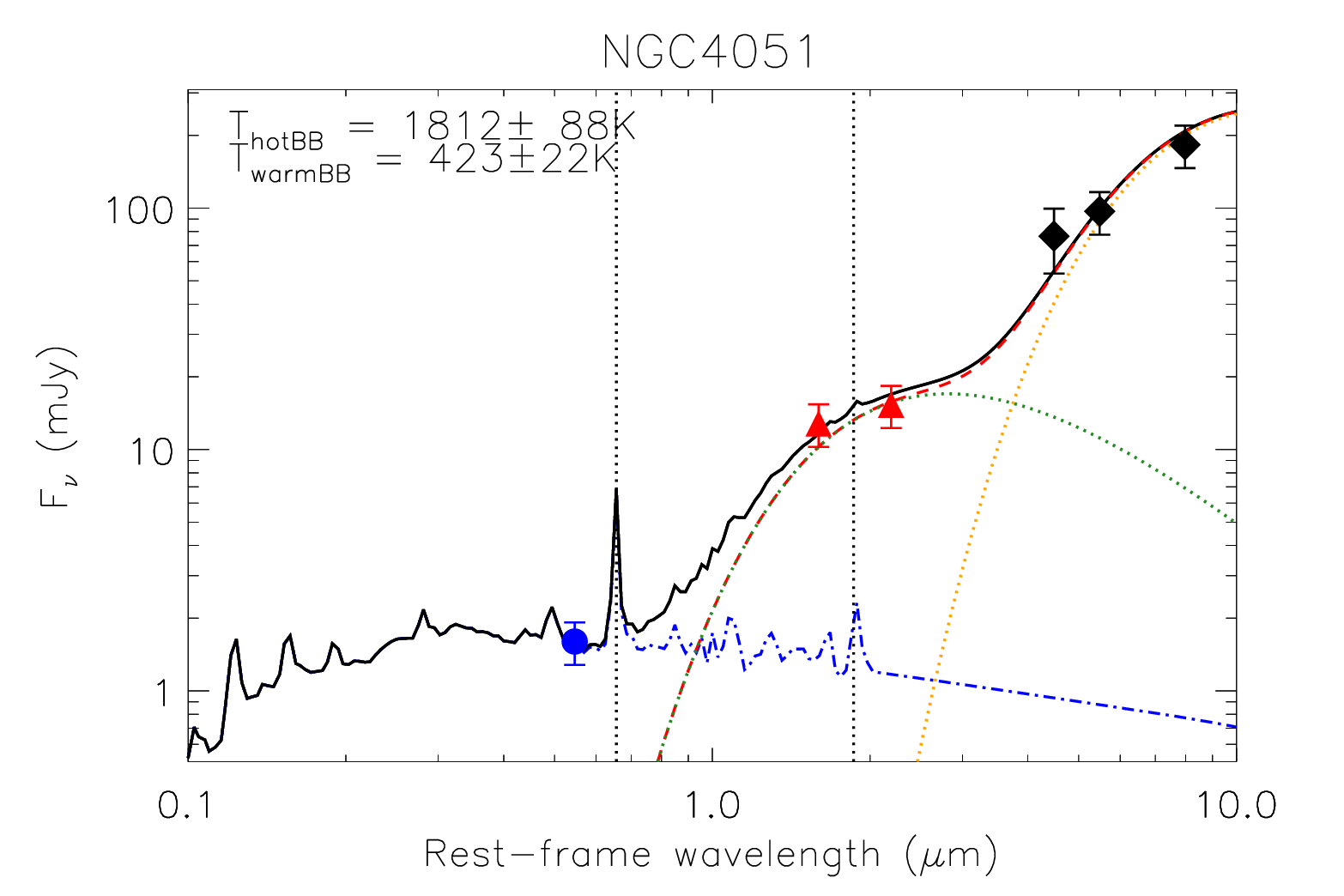}
\includegraphics[width=8.0cm]{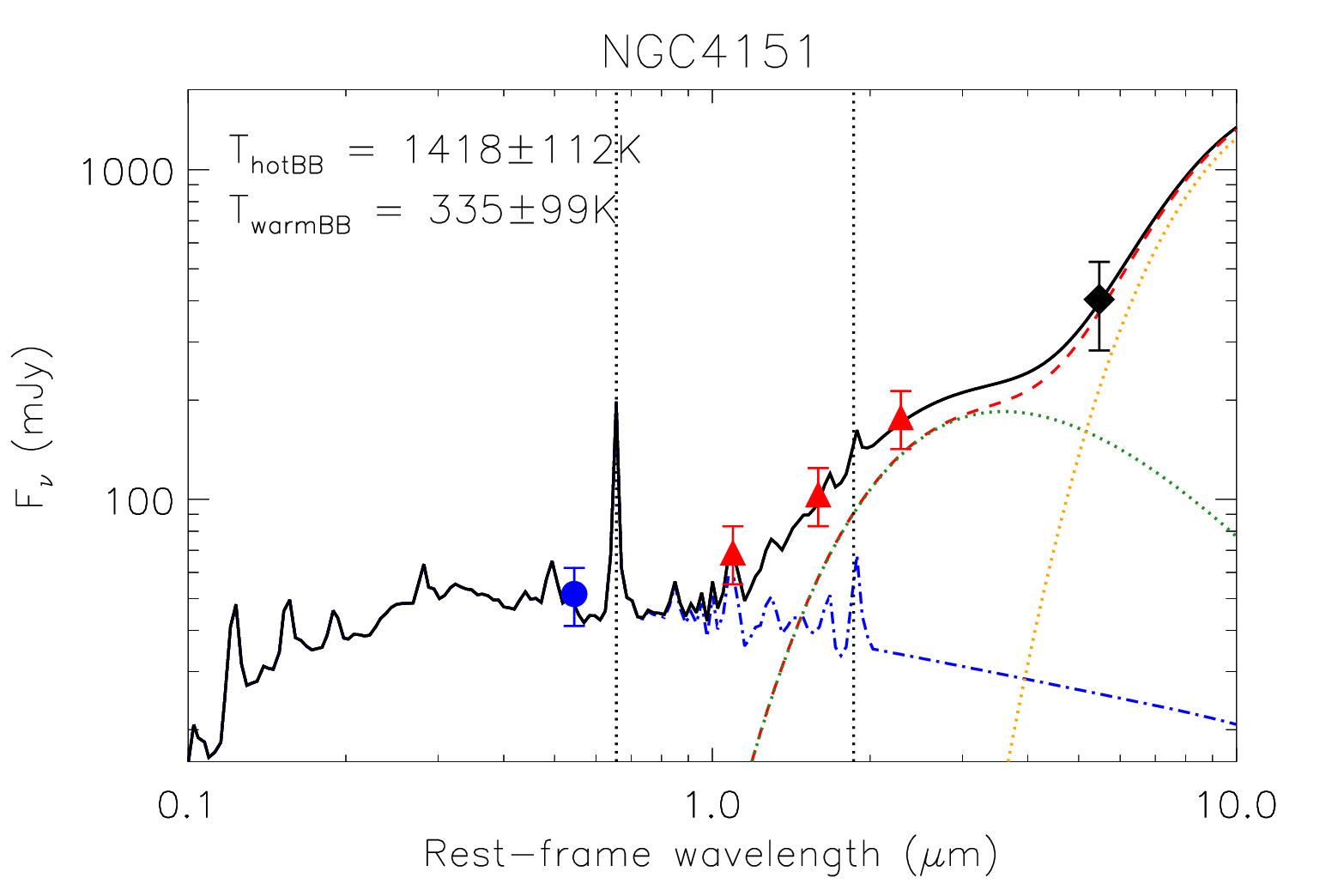}
\includegraphics[width=8.0cm]{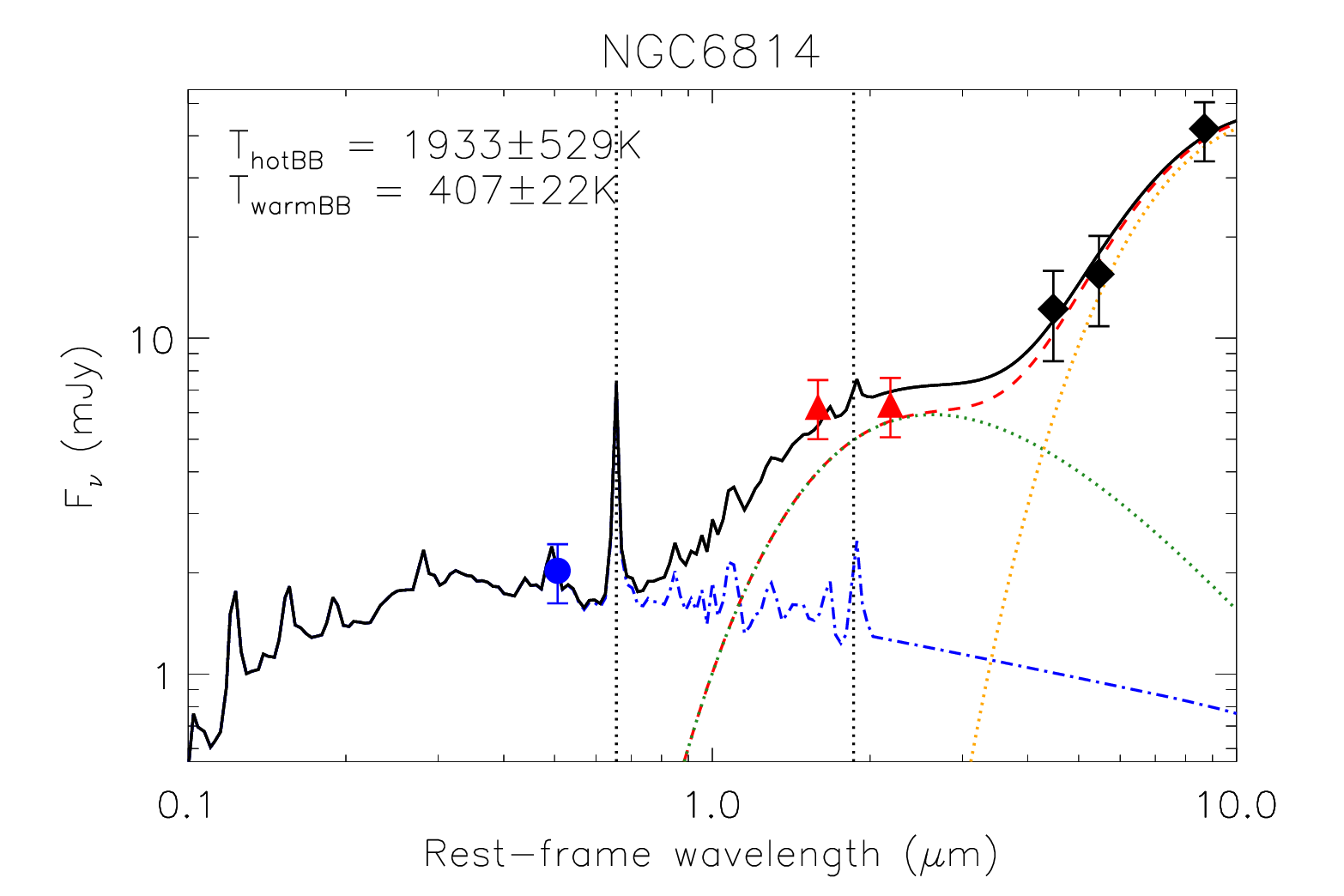}
\includegraphics[width=8.0cm]{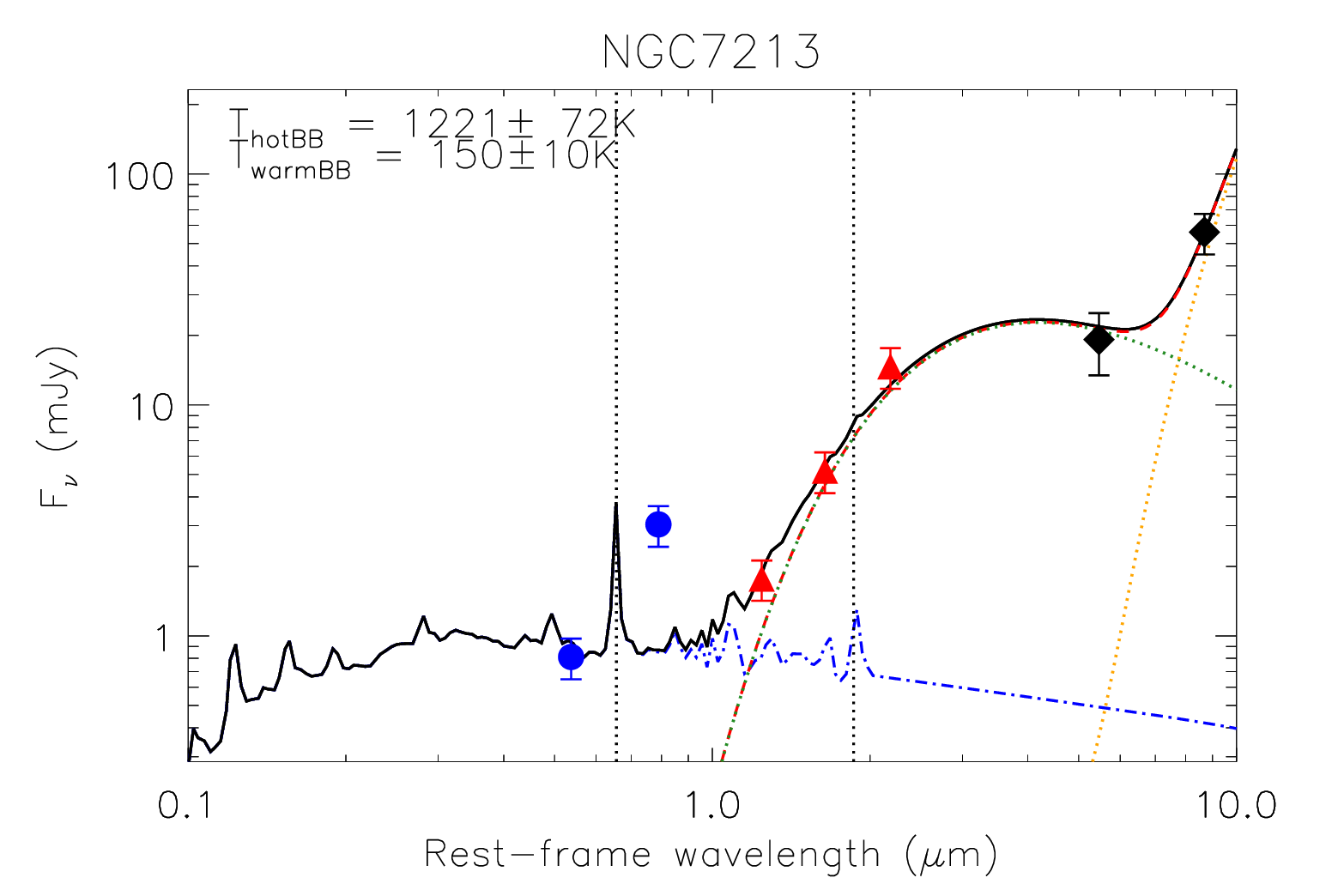}
\includegraphics[width=8.0cm]{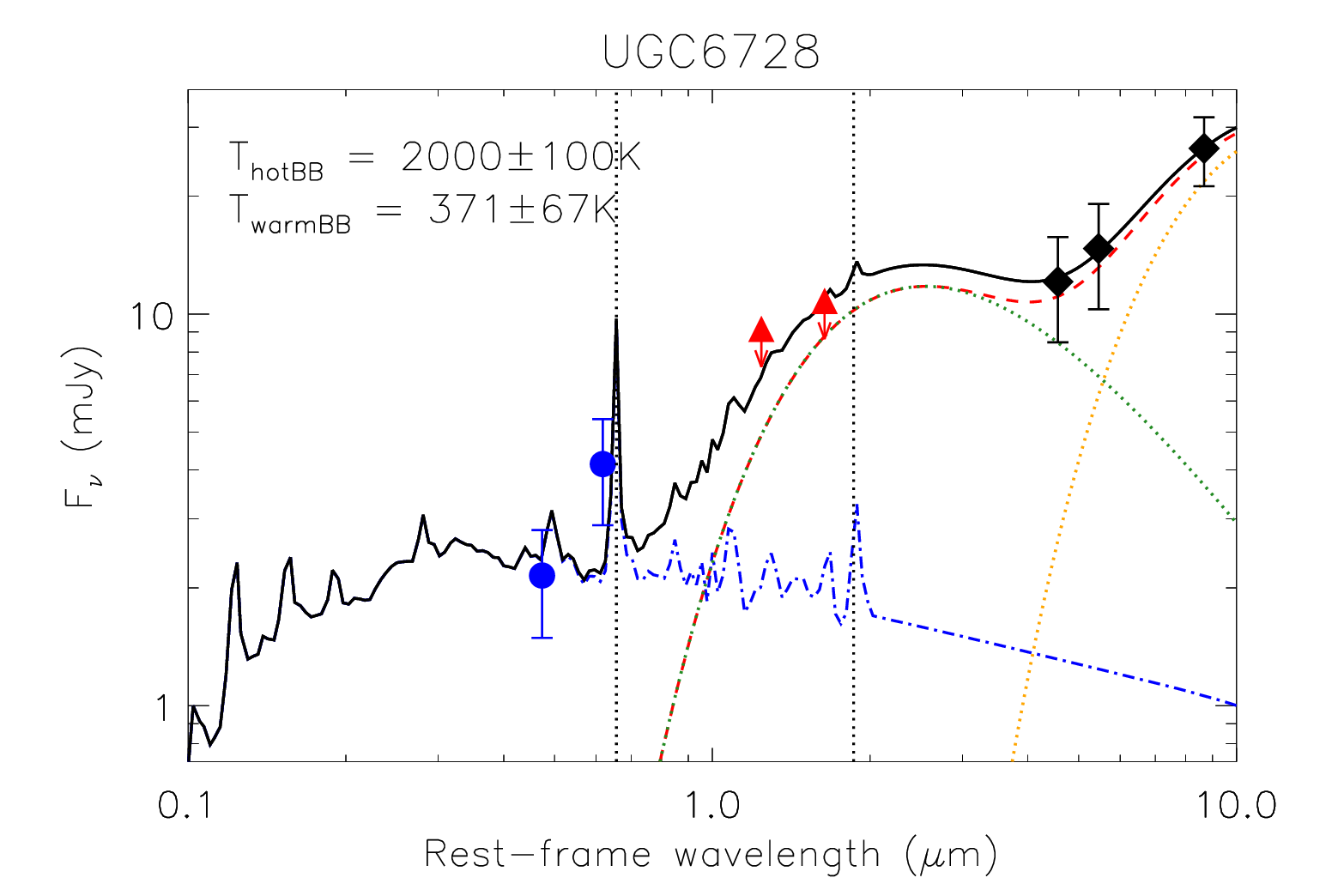}
\par}
\caption{Accretion disk emission fits of Sy1s. Blue circles, red triangles, and black diamonds represent broadband photometry in the rest-frame optical, NIR, and MIR, respectively. We show the dust components (red dashed lines), accretion disk component (blue dot-dashed line) and best fits (black solid lines). Note that the dust component is modelled as a linear combination of two black-bodies at adjustable temperatures (green and orange dotted lines). The vertical dotted lines mark the rest-frame wavelength of H$_{\alpha}$ and Pa$_{\alpha}$.}
\label{fig2}
\end{figure*}

In Fig. \ref{fig2}, we present the fitting results and in Table \ref{tab6} we list the fractional contribution of this component to the nuclear NIR emission. Using only the fits with subarcsecond resolution data, we find that the average contribution of the accretion disk to the J-, H- and K-band emission are 46$\pm$28, 23$\pm$13 and 11$\pm$5\% in $\sim$0.4\arcsec ~apertures, which are in good agreement with the values reported by \citet{Hernan-caballero2016} for the rest-frame J-, H- and K-band (48$\pm$16, 27$\pm$14 and 17$\pm$1\%) using a sample of luminous quasars. We note that the largest contribution from the accretion disk to the NIR emission is found for NGC\,4151. This is in agreement with previous works on this galaxy (e.g. \citealt{Swain03,Kishimoto07,Riffel09}).  

Since we find a significant contribution of the accretion disk emission in the NIR range of Sy1, we subtracted this component in all Sy1 galaxies prior to fitting the nuclear IR SEDs with torus models. 

\subsection{SED modelling with the CLUMPY torus models}
\label{clumpy_torus}
Using the CLUMPY models and  \textsc{BayesClumpy}, we fit all the nuclear NIR-to-MIR SEDs in our sample (See Appendix \ref{B}). A detailed description of the CLUMPY model parameters (see Table \ref{tab8}) can be found in \citet{Nenkova08a,Nenkova08b}. For approximately half of our sample (13/24 sources; see Table \ref{tab7}), we used Gaussian priors for $\sigma$ (width of clouds angular distribution) based on the opening angle of the ionization cones from published [O\,III] and/or H$_{\alpha}$ images or NLR kinematics modelling (see Table \ref{tab7} for further details). In addition, we used the IR extinction curve of \citet{Chiar06} of the local ISM to account for any possible foreground extinction from the host galaxy. This curve covers the range $\sim$1--35~$\mu$m and accounts for the two silicate features at 9.7 and 18~$\mu$m. We used different priors for the foreground extinction from the host galaxy (A$_{V}^{for}$) for the various Seyfert types, taking into account the values available in the literature (see Table \ref{tab1}). We used A$_{V}^{for}$=[0,2]~mag for Sy1 and [0,8]~mag for Sy1.8/1.9 and Sy2. Finally, we used uniform priors for the rest of the parameters. When the observed data introduce sufficient information into the fit, the resulting posteriors will clearly differ from the input uniform distributions, either showing trends or being centered at certain values within the considered intervals. 

\begin{table}
\centering
\begin{tabular}{lccc}
\hline
Name 
									& $\sigma_{literature}$ & Ref.		& Interval used 	\\
 (1)&(2)&(3)&(4)\\	

\hline
\multicolumn{4}{|c|}{Sy1 galaxies}\\
\hline
NGC\,3227		& 55$^\circ$--60$^\circ$&	a	& [45$^\circ$--65$^\circ$]\\
NGC\,3783		& 35$^\circ$--56$^\circ$&   b, c& [35$^\circ$--55$^\circ$]\\
NGC\,4051		& 67$^{\circ\dagger}$	&	d	& [50$^\circ$--70$^\circ$]\\
NGC\,4151		& 52$^\circ$--62$^\circ$&	e	& [45$^\circ$--65$^\circ$]\\
NGC\,6814		& 43$^{\circ\dagger}$	&	c	& [35$^\circ$--55$^\circ$]\\
\hline
\multicolumn{4}{|c|}{Sy1.8/1.9 galaxies}\\
\hline
NGC\,1365		& 29$^\circ$--49$^\circ$&	e	& [30$^\circ$--50$^\circ$]\\
NGC\,2992		& 25$^\circ$&		f			& [15$^\circ$--35$^\circ$]\\
\hline
\multicolumn{4}{|c|}{Sy2 galaxies}\\
\hline
NGC\,2110		& 45$^\circ$&		g			& [35$^\circ$--55$^\circ$]\\
NGC\,3081		& 75$^\circ$&		h			& [50$^\circ$--70$^\circ$]\\
NGC\,4388		& 65$^\circ$&		e			& [50$^\circ$--70$^\circ$]\\
NGC\,5128& 55$^\circ$--65$^{\circ\dagger}$&	i	& [50$^\circ$--70$^\circ$]\\
NGC\,5506		& 45$^\circ$&		j			& [35$^\circ$--55$^\circ$]\\
NGC\,7582		& 42$^\circ$--52$^\circ$& e		& [35$^\circ$--55$^\circ$]\\
\hline
\end{tabular}					 
\caption{Constraints on the torus widths derived from ionization cone opening angles. References: a) \citet{Mundell95}; b) \citet{Fischer13}; c) \citet{Muller11}; d) \citet{Christopoulou97}; e) \citet{Wilson94}; f) \citet{Garcia-Lorenzo01}; g) \citet{Rosario10}; h) \citet{Ferruit00}; i) \citet{Bryant99}; j) \citet{Wilson85}.}
$\dagger$ Derived from NLR kinematics modelling.
\label{tab7}
\end{table}

\begin{table}
\scriptsize
\centering
\begin{tabular}{lcc}
\hline
Parameter 									& Symbol 		& Interval 	\\
\hline
Radial extent of the torus 				&	Y			&	[5, 100]		\\
Width of clouds angular distribution		&	$\sigma$	&	[15$^\circ$, 70$^\circ$]	\\
Number of clouds along an equatorial ray	&	N$_0$		&	[1, 15]		\\
Index of the radial density profile		&	q			&	[0, 3]		\\
Inclination angle of the torus				&	i			&	[0$^\circ$, 90$^\circ$]		\\
Optical depth per single cloud				&	$\tau_{V}$&	[5, 150]		\\
\hline
Foreground extinction						&	A$_{V}^{for}$		& Sy1s: [0, 2] mag		\\
											&\multicolumn{2}{|c|}{  Sy1.8/1.9/2s: [0, 8] mag}		\\
\hline
\end{tabular}					 
\caption{Clumpy torus model parameters. i=0$^\circ$ is face-on and i=90$^\circ$ is edge-on. We note that the foreground extinction is unrelated to the torus.}
\label{tab8}
\end{table}

We note that for this study we used the updated version (October 2014) of the \citet{Nenkova08a,Nenkova08b} clumpy torus models\footnote{https://www.clumpy.org/}. Older versions of these models used the optical depth along the slab normal for the synthetic clouds. However, in a recent comparison with spherical clouds (3D radiative transfer), the calculations showed that the effective optical depth through a cloud was two times higher than in the former approach (Heymann, Nikutta, and Elitzur, in preparation). Although the absorption caused by clouds is not affected by this, the cloud emission does change since its source function is wavelength-dependent. As a consequence, a moderate change in the spectral shape has been reported on the CLUMPY webpage (less than 20\% at any given wavelength). 

\begin{figure}
\centering
\includegraphics[width=8.5cm]{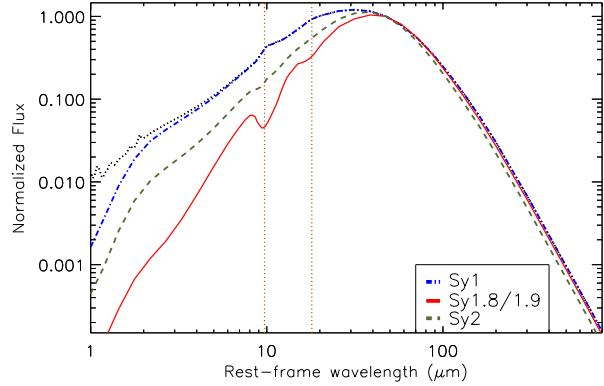}
\caption{IR torus model emission as derived from the median fitted nuclear IR SEDs. Blue dot-dashed, green dashed and red solid lines represent the average of the median SEDs of Sy1, Sy1.8/1.9, and Sy2 galaxies, respectively. We plot the average median SEDs of Sy1 including the accretion disk emission, which corresponds as the black dotted line, for comparison. The SEDs are normalized at 50~$\mu$m. The brown vertical dotted lines correspond to silicate features.}
\label{fig3}
\end{figure}

In Appendix \ref{B}, we present the results of the nuclear IR SED fitting process with the CLUMPY models (see Section \ref{clumpy_torus}), which are the marginal posterior distributions of the six parameters that define these models plus the foreground extinction and vertical shift. This shift scales with the AGN bolometric luminosity. We can also translate the posterior distributions of the parameters into a best-fitting model described by the combination of parameters that maximizes the posterior (maximum-a-posteriori; MAP) and a median model, computed with the median value of each posterior (see Appendix \ref{B}). We found different average models of each subgroup from the median fitted nuclear IR SEDs. The Sy1 average model including the accretion disk emission component (black dotted line of Fig. \ref{fig3}) shows a flat NIR slope and the shape for the Sy2 average model (red solid line of Fig. \ref{fig3}) is very steep. The Sy1.8/1.9 average model 
lies between those of the Sy1 and Sy2 models.

\begin{landscape}
\begin{table}
\scriptsize
\centering
\begin{tabular}{lccccccccccc}
\hline
Name    & log(L$_{bol}^{model}$) & log(L$_{bol}^{torus}$) & $M_{Torus}$ &r$_{Torus}$& P$_{esc}$& C$_T$ & log(N$_{\textrm{H}}^{X-rays}$) & log(L$_{\textrm{int}}^{2-10~\textrm{keV}}$) & log(L$_{\textrm{cor}}^{14-195~\textrm{keV}}$)& log($M_{BH}$/M$_\odot$)& log($\lambda_{Edd}$)\\ 
		& (erg~s$^{-1}$)& (erg~s$^{-1}$)&(M$_\odot$)&(pc)& (\%)&  & (cm$^{-2}$)& (erg~s$^{-1}$)& (erg~s$^{-1}$)&\\
 (1)&(2)&(3)&(4)&(5)&(6)&(7)&(8)&(9)&(10)&(11)&(12)\\	
		
\hline
\multicolumn{12}{|c|}{Sy1 galaxies}\\
\hline
MCG-06-30-015	& 43.33&43.02	&(0.2$\pm0.1$)$\times$10$^5$				&4.0$\pm_{2.2}^{2.3}$     &57$\pm_{23}^{21}$&0.13$\pm_{0.02}^{0.05}$    			  &20.85    & 42.74& 42.66		&7.42$^a$& -1.52\\
NGC\,3227   	& 42.91&43.08	&(1.1$\pm_{0.3}^{0.6}$)$\times$10$^5$		&1.2$\pm_{0.2}^{0.3}$     &24$\pm_{6}^{12}$ &0.86$\pm_{0.07}^{0.04}$       &20.95    & 42.10& 42.58		&6.62$^b$&-1.36\\
NGC\,3783   	& 43.48&43.72	&(8.8$\pm_{3.2}^{9.3}$)$\times$10$^5$		&3.6$\pm_{0.8}^{1.6}$     &69$\pm_{17}^{14}$&0.68$\pm_{0.10}^{0.08}$       &20.49    & 43.43& 43.28		&7.14$^a$&-0.55\\
NGC\,4051   	& 42.59&42.72	&(0.3$\pm_{0.1}^{0.9}$)$\times$10$^4$		&0.3$\pm_{0.1}^{0.5}$     &31$\pm_{13}^{17}$&0.80$\pm_{0.06}^{0.05}$       &20.00    & 41.33& 41.47		&5.60$^b$&-1.11\\
NGC\,4151   	& 43.49&43.71	&(7.5$\pm_{2.3}^{2.4}$)$\times$10$^5$		&3.3$\pm_{0.6}^{0.5}$     &71$\pm_{9}^{7}$ &0.73$\pm_{0.05}^{0.04}$       &22.71    & 42.31& 43.43		&7.43$^a$&-1.96\\
NGC\,6814   	& 42.40&42.56	&(1.9$\pm_{1.2}^{7.0}$)$\times$10$^5$		&2.1$\pm_{0.8}^{2.6}$     &80$\pm_{22}^{12}$&0.51$\pm_{0.12}^{0.10}$       &20.97    & 42.31& 42.80		&6.46$^b$&-0.99\\
NGC\,7213   	& 43.96&42.82	&(1.1$\pm_{0.5}^{1.1}$)$\times$10$^5$		&1.9$\pm_{0.4}^{0.8}$     &100				 &0.12$\pm_{0.01}^{0.03}$       &20.00    & 41.95& 42.54	&7.37$^c$&-2.26\\
UGC\,6728   	& 42.40&43.39	&(0.4$\pm_{0.3}^{0.5}$)$\times$10$^5$		&2.8$\pm_{1.5}^{1.7}$     &24$\pm_{20}^{49}$&0.58$\pm_{0.27}^{0.19}$       &20.00    & 41.80& 42.44		&5.32$^b$&-0.36\\
\hline
\multicolumn{12}{|c|}{Sy1.8/1.9 galaxies}\\
\hline
NGC\,1365		&	42.83&42.95	&(2.5$\pm_{1.2}^{5.9}$)$\times$10$^5$			&2.1$\pm_{0.6}^{1.9}$	&60$\pm_{19}^{23}$ &0.46$\pm_{0.06}^{0.13}$	    &22.21    & 42.32& 42.69			&7.92$^a$&-2.44\\
NGC\,2992		&  43.57&43.32	&(1.2$\pm_{0.5}^{1.4}$)$\times$10$^5$			&1.9$\pm_{0.5}^{1.1}	$&0.7$\pm_{0.5}^{0.8}$	   &0.46$\pm_{0.07}^{0.05}$		&21.72    & 42.00& 42.53			&5.42$^b$&-0.26\\
NGC\,4138		&  41.92&41.67	&(0.6$\pm_{0.4}^{1.2}$)$\times$10$^4$			&0.9$\pm_{0.5}^{0.8}$	&81$\pm_{32}^{17}$ &0.18$\pm_{0.06}^{0.12}$	&22.89    & 41.23& 41.59			&7.30$^b$&-2.91\\
NGC\,4395		&  39.98&39.97	&(0.2$\pm_{0.1}^{0.3}$)$\times$10$^4$			&0.2$\pm_{0.1}^{0.2}$	&0.4$\pm_{0.3}^{0.4}$	&0.74$\pm_{0.22}^{0.16}$	&21.04    & 40.50& 40.63		&4.88$^a$&-1.22\\
NGC\,5506		&  43.86&43.94	&(7.6$\pm_{1.3}^{2.6}$)$\times$10$^5$			&3.8$\pm_{0.3}^{0.6}$	&30$\pm_{10}^{9}$		 &0.91$\pm_{0.02}^{0.01}$		&22.44   & 42.99& 43.44				&8.29$^a$&-2.14\\
NGC\,7314		&  42.26&42.40	&(1.4$\pm_{0.5}^{2.4}$)$\times$10$^5$			&1.3$\pm_{0.3}^{0.9}$	&33$\pm_{13}^{16}$	&0.84$\pm_{0.31}^{0.05}$	&21.60    & 42.33& 42.44			&7.24$^b$&-1.75\\
\hline
\multicolumn{12}{|c|}{Sy2 galaxies}\\
\hline
ESO\,005-G004	& 42.43&42.22	&(0.9$\pm_{0.2}^{0.3}$)$\times$10$^5$		&1.5$\pm_{0.2}^{0.3}$		&0.002$\pm_{0.002}^{0.001}$&0.95$\pm_{0.06}^{0.03}$&24.34    & 42.78& 42.74			&6.98$^c$&-1.04\\
MCG-05-23-016	& 43.75&43.73	&(3.2$\pm_{1.6}^{8.7}$)$\times$10$^5$		&3.5$\pm_{1.2}^{4.4}$		&8$\pm_{6}^{17}$	    &0.80$\pm_{0.16}^{0.08}$			&22.18    & 43.20& 43.43			&7.98$^a$&-1.62\\
NGC\,2110		&  43.28&43.37	&(3.9$\pm_{2.3}^{11.3}$)$\times$10$^5$		&4.5$\pm_{1.9}^{5.7}$		&54$\pm26$			   &0.55$\pm_{0.07}^{0.10}$			&22.94    & 42.69& 43.49			&9.25$^b$&-3.40\\
NGC\,3081		&  43.15&43.10	&(6.6$\pm_{2.7}^{2.4}$)$\times$10$^5$		&13.5$\pm_{5.6}^{5.2}$			&0.4$\pm0.3$	   &0.95$\pm_{0.06}^{0.02}$					&23.91    & 42.72& 43.60			&8.41$^b$&-2.53\\
NGC\,4388		&  43.04&42.97	&(1.6$\pm_{0.2}^{0.3}$)$\times$10$^6$		&4.8$\pm0.4$				&0.6$\pm_{0.1}^{0.2}$  &0.99$\pm_{0.002}^{0.001}$		&23.52    & 43.05& 43.47			&6.99$^b$&-0.78\\
NGC\,4945		&  41.05&40.86	&(1.3$\pm_{0.4}^{1.4}$)$\times$10$^5$		&2.2$\pm_{0.4}^{1.0}$	&0.001$\pm_{0.001}^{0.007}$&0.97$\pm_{0.03}^{0.02}$&24.80  & 42.69  & 42.53		&7.78$^a$&-1.93\\
NGC\,5128		&  42.27&42.28	&(1.7$\pm_{0.4}^{0.6}$)$\times$10$^5$		&2.1$\pm0.3$	&10$\pm_{3}^{6}$	&0.94$\pm_{0.02}^{0.01}$		&23.02    & 42.39& 43.09			&7.94$^a$&-2.39\\
NGC\,6300		&  43.00&43.05	&(5.4$\pm_{1.7}^{2.8}$)$\times$10$^5$		&2.7$\pm_{0.5}^{0.6}$	&0.9$\pm_{0.3}^{0.7}$  &0.99$\pm_{0.002}^{0.001}$&23.31    & 41.84& 42.43			&7.01$^a$&-2.01\\
NGC\,7172		&  42.91&42.91	&(1.0$\pm_{0.3}^{0.7}$)$\times$10$^5$		&2.1$\pm_{0.4}^{0.6}$		&5$\pm_{5}^{10}$	&0.98$\pm_{0.03}^{0.02}$&22.91    & 42.76& 43.36			&8.45$^b$&-2.53\\
NGC\,7582		&  43.08&42.81	&(1.2$\pm_{0.5}^{0.7}$)$\times$10$^6$		&9.4$\pm_{2.4}^{3.0}	$&0.003$\pm_{0.003}^{0.02}$&0.83$\pm_{0.08}^{0.05}$	&24.33    & 42.86& 43.28		&7.52$^a$&-1.50\\
\hline
\end{tabular}					 
\caption{AGN and torus model properties derived from the fits and X-ray properties. The torus model properties were derived from the median values of the marginal posterior distributions. Columns 2, 3, 4, 5, 6 and 7 list the AGN bolometric luminosity (L$_{bol}^{model}$), torus bolometric luminosity (L$_{bol}^{torus}$, obtained by integrating the corresponding model torus emission), torus gas mass ($M_{Torus}$), torus outer radius ($r_{Torus}$), escape probability (P$_{esc}$) and torus covering factor (C$_T$) derived from the torus model. Note that median values are listed with their corresponding $\pm$1$\sigma$ values around the median. Columns 8, 9 and 10 correspond to the hydrogen column density, intrinsic 2-10~keV and absoption-corrected 14-195~keV X-ray luminosities taken from \citet{Ricci17}. Column 11 and 12 list the BH masses with their references and the derived Eddington Ratio following the same methodology as in \citet{Ricci17c}. References for $M_{BH}$/M$_\odot$: a) This work; b) \citet{Koss17}; c) \citet{Vasudevan10}. See Appendix \ref{D} for further information about the estimation of the BH masses.}
\label{tab9}
\end{table}
\end{landscape}

In general, the CLUMPY models provide good fits ($\chi^2/dof\textrm{(degrees~of~freedom)}<$2.0) to the majority (19/24) of the nuclear IR SEDs (see Appendix \ref{B}). While the MIR emission is well fitted for practically all the SEDs, we found that 5/24 galaxies (i.e. NGC\,3783, NGC\,4395, NGC\,5506, NGC\,7172 \& NGC\,7314) show a clear excess of emission in the NIR. This likely indicates an extra hot dust component is needed to reproduce their IR SEDs (see also \citealt{Mor09}). We note that the main goal of this work is to obtain a global statistical analysis of the clumpy torus model parameters of the various Seyfert galaxy types, rather than focussing on the individual fits (see Appendix \ref{B}). As a sanity check, we repeated the analysis using only those galaxies with the best ($\chi^2/dof<$1.0; $\sim$63\% of the sample) and good ($\chi^2/dof<$2.0; $\sim$79\% of the sample) fits and we find the same results within 1$\sigma$.

\section{Comparison of the Torus properties}
\label{Comparison_parameter}

In this section, we investigate the main differences between the clumpy torus model parameters for the BCS$_{40}$ sample. Table \ref{tab9} reports the main derived properties of the torus from the model parameters and the X-ray measurements for comparison.

\subsection{Distributions of Clumpy Torus Model Parameters}
\label{distrib_parameters}

\subsubsection{Optical Classification}
\label{Optical_Classification}

In this section we discuss the global posterior distributions and their mean values for Sy1, Sy1.8/1.9, and Sy2. To this end we apply the hierarchical Bayesian approach also used in \citet{Ichikawa15}. To be more precise, we assume that the global properties of the objects are extracted from common prior distributions and we infer the hyperparameters of these priors. Because of its flexibility, we decided to use beta distributions as these prior distributions. To take advantage of the already computed sampling of the posterior for each individual object we leverage the importance sampling trick developed by \citet{Brewer14}. Although one should ideally sample from the full hierarchical probabilistic model, we consider this approximate
technique as sufficient for our purposes.

\begin{figure*}
\centering
\par{
\includegraphics[width=7.82cm]{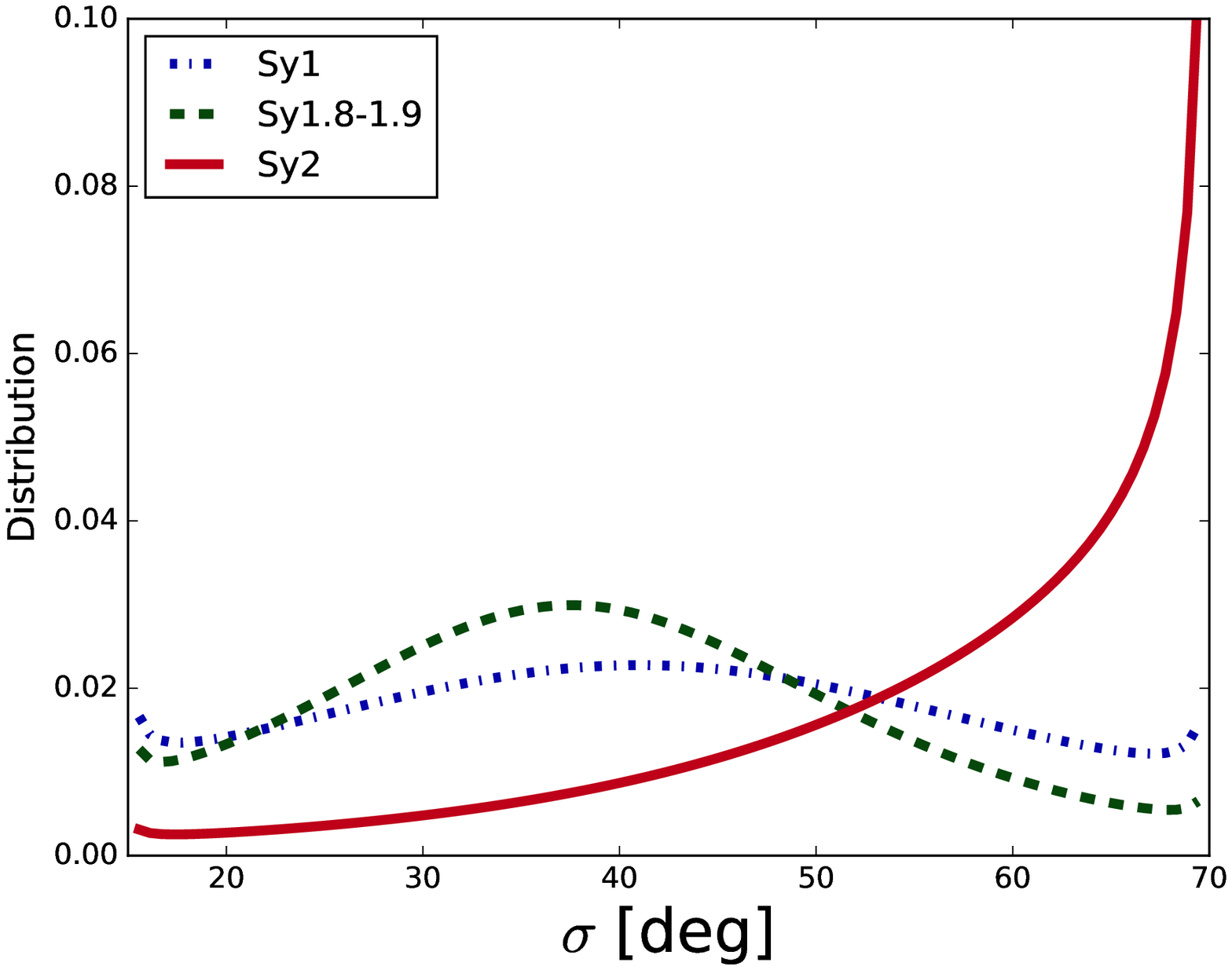}
\includegraphics[width=7.82cm]{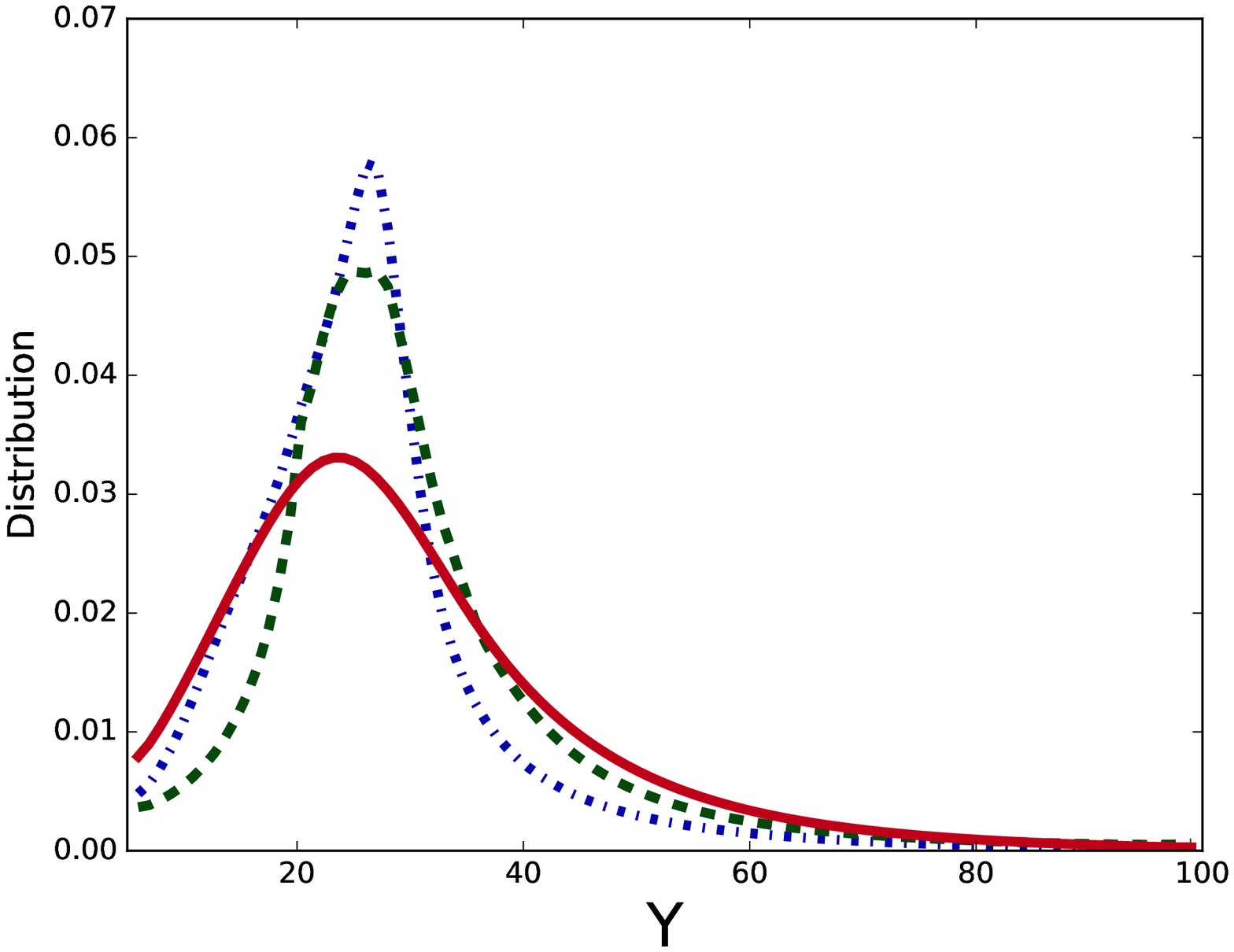}
\includegraphics[width=7.82cm]{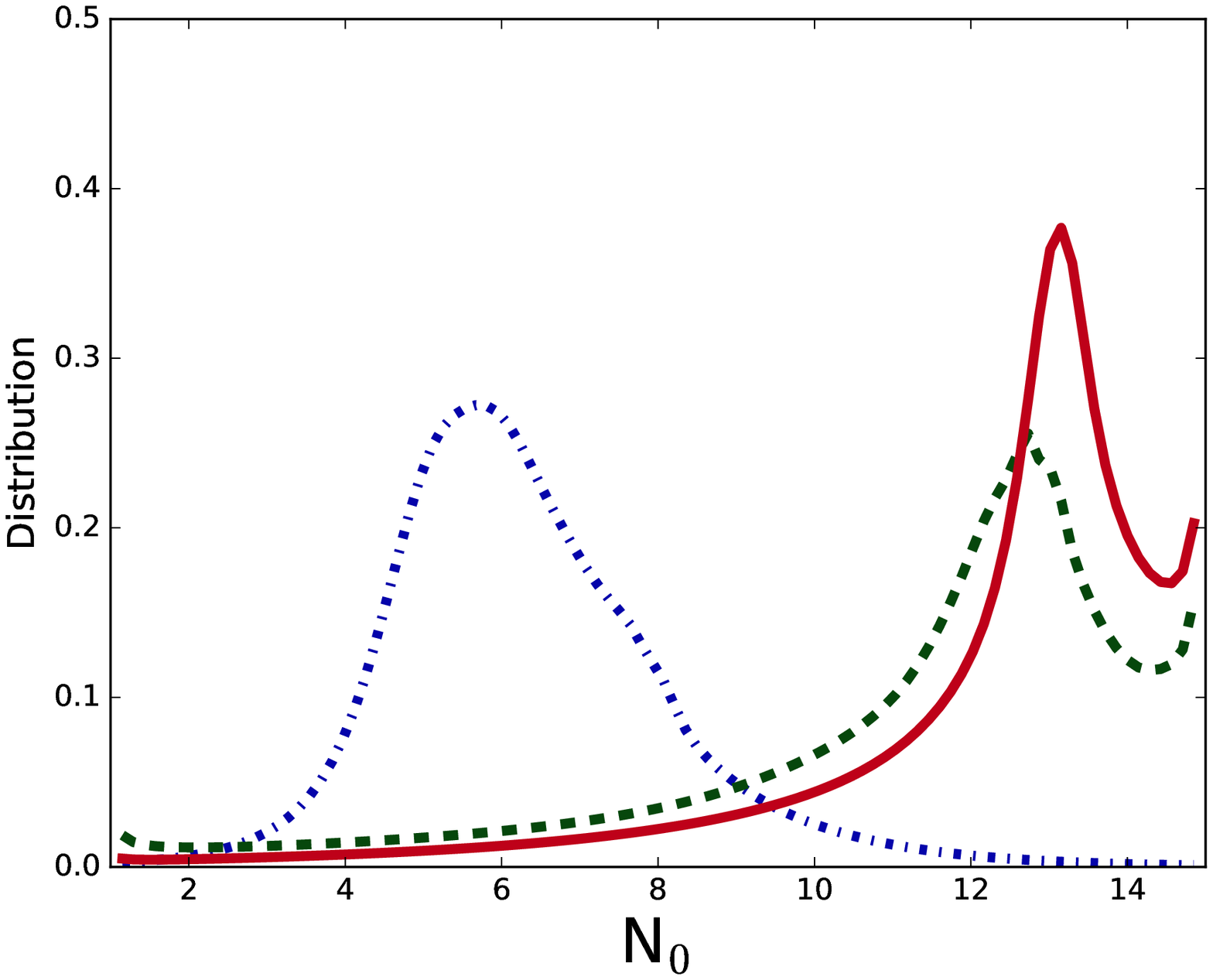}
\includegraphics[width=7.82cm]{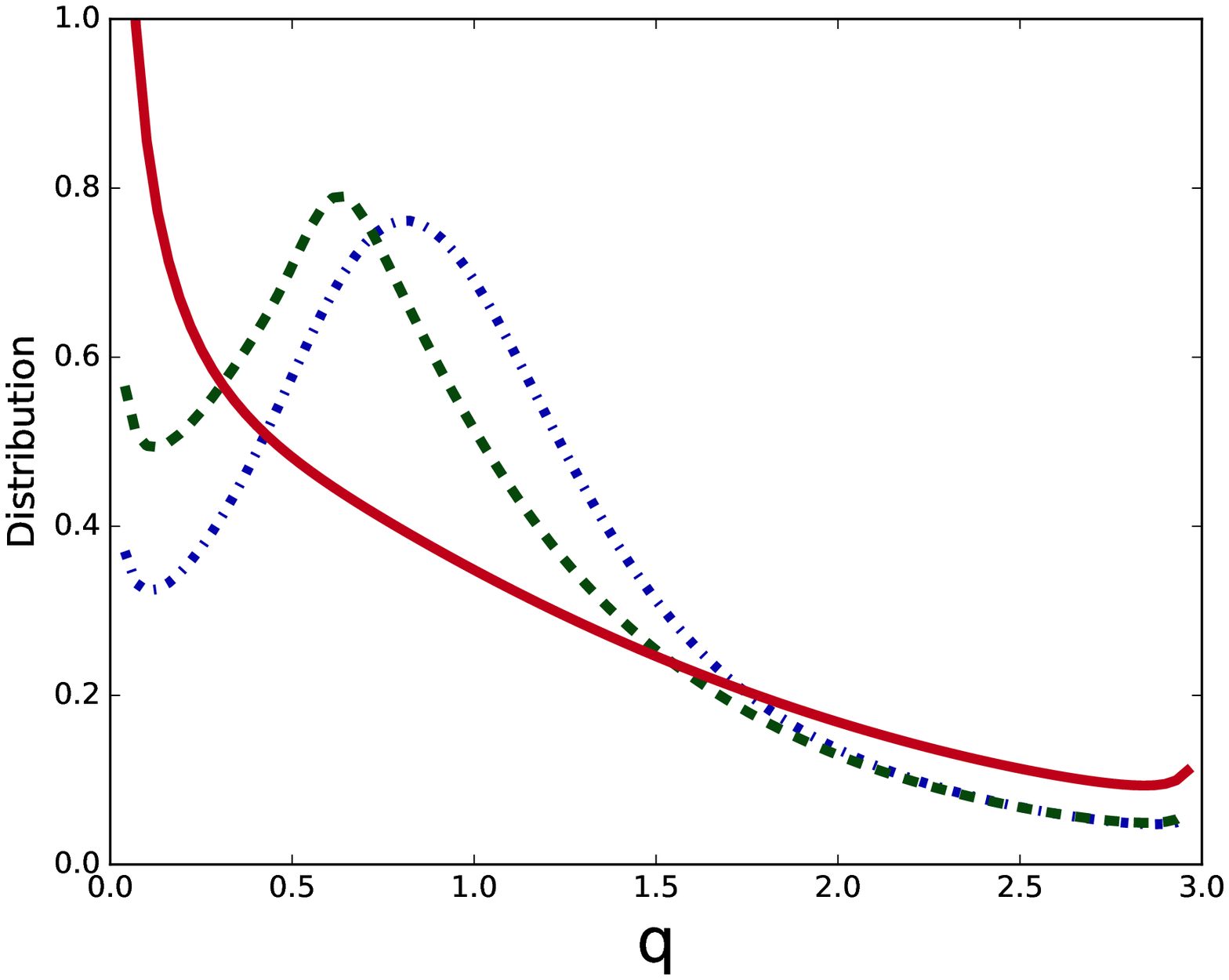}
\includegraphics[width=7.82cm]{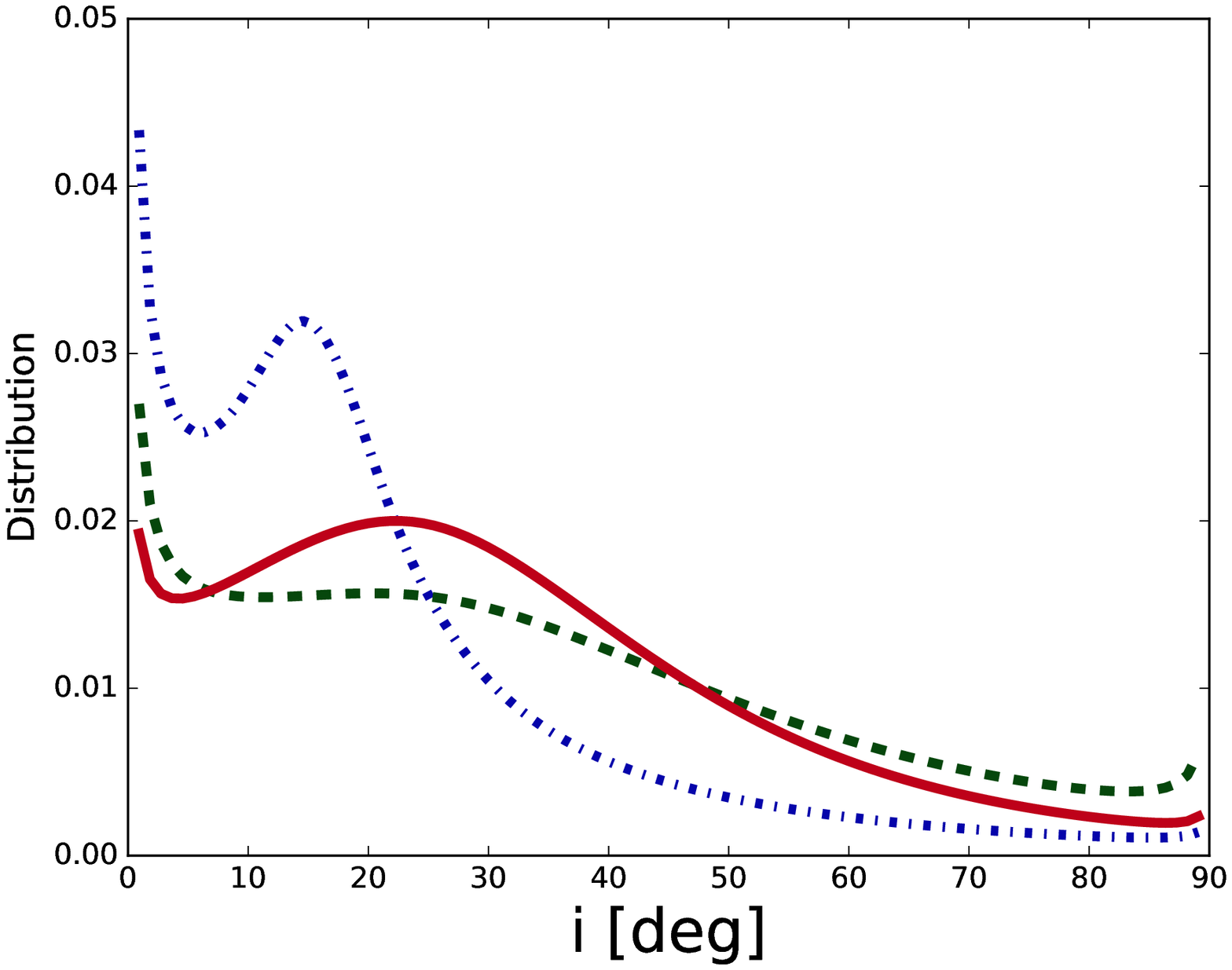}
\includegraphics[width=7.82cm]{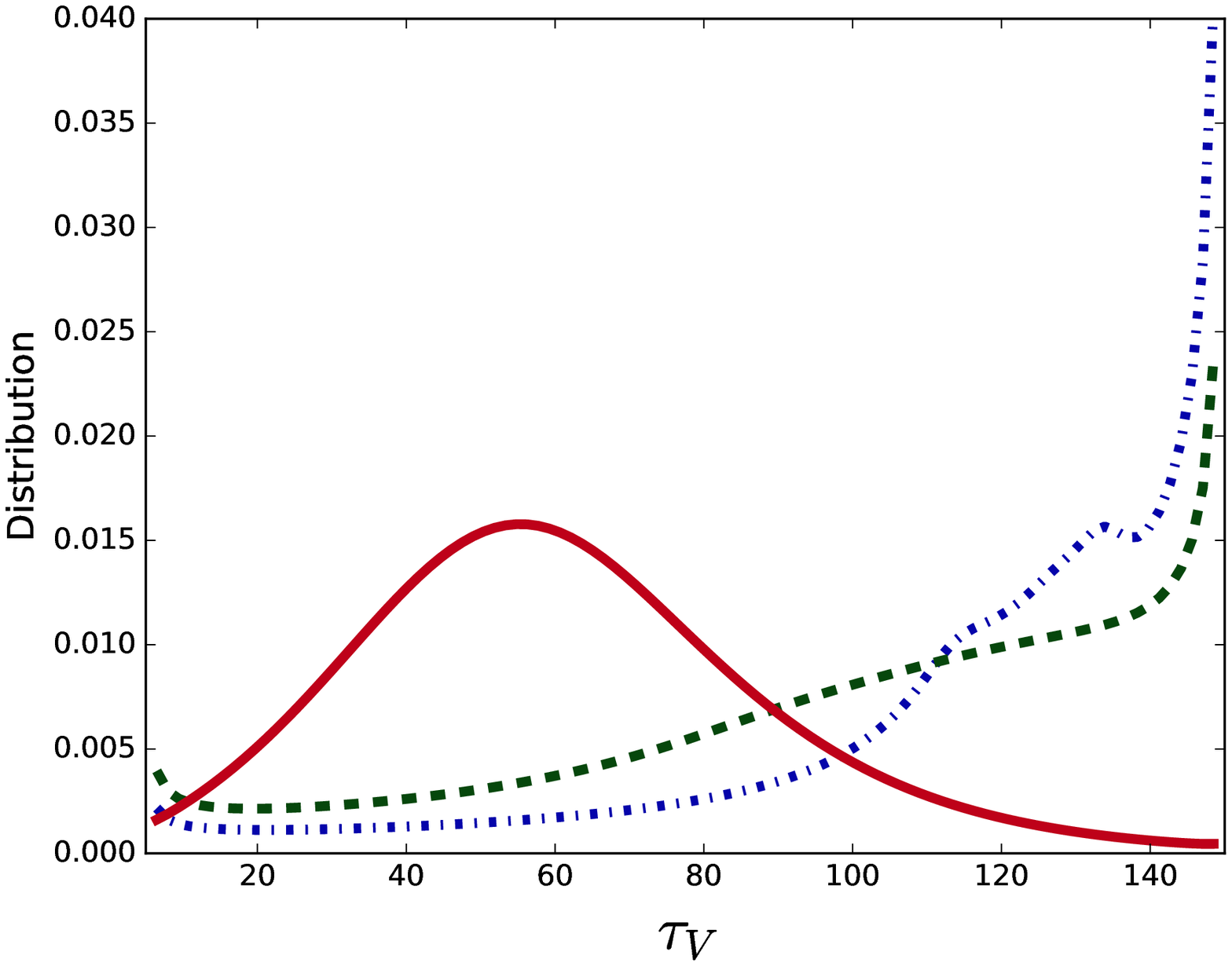}
\par}
\caption{Comparison between the clumpy torus model parameter global posterior distributions for the optical classification. Blue dot-dashed, green dashed and red solid lines represent the parameter distributions of Sy1, Sy1.8/1.9, and Sy2 galaxies, respectively.}
\label{fig4}
\end{figure*}

As can be seen from Figure \ref{fig4}, the majority of the global distributions are clearly different. To quantify these differences we use the Kullback-Leibler divergence (KLD; \citealt{Kullback51}) as in RA11. This approach takes into account the overall shape of the posterior distribution and it always has a positive value. In the case of two identical distributions it is equal to zero and the larger the values the more different the distributions. RA11 suggested that for values larger than one (bold face in Table \ref{tab10}), two posterior distributions may be considered to be significantly different. Following this, we find that the differences in $\sigma$, N$_0$ and $\tau_V$ (see Fig. \ref{fig4}) between Sy1 and Sy2 are significant according to the KLD. The same applies to $\sigma$ and $\tau_V$ between Sy1.8/1.9 and Sy2 galaxies. We note that RA11 found essentially the same differences between Sy1 and Sy2 galaxies. All these results are in good agreement with previous works (e.g. RA11, AH11 and \citealt{Ichikawa15}). However, we find smaller values of the cloud optical depth for Sy2 ($\tau_{V}\sim$56) than for Sy1 and Sy1.8/1.9 galaxies ($\tau_{V}\sim$94-114) and smaller values of i for Sy1 (i$=$19$\pm$16$^\circ$). 

\begin{table}
\tiny
\centering
\begin{tabular}{lcccccccccccc}
\hline
Subgroups	& $\sigma$	& Y	& N$_0$	& q	&$\tau_{V}$	& i & C$_T$\\
	(1) & (2) & (3) & (4) & (5)& (6)& (7)& (8)\\
\hline
Sy1s vs Sy2s		&{\bf{1.16}}	&0.19	&{\bf{5.07}}	&0.27	&{\bf{5.56}}	&0.47&{\bf{5.11}}\\
Sy1s vs Sy1.8/1.9	&0.06			&0.11	&{\bf{3.84}}	&0.04	&0.19			&0.38&0.91\\
Sy2s vs	Sy1.8/1.9	&{\bf{1.73}}	&0.10	&0.17			&0.13	&{\bf{3.63}}	&0.13&{\bf{3.21}}\\
absorbed vs unabsorbed&0.86		&0.42	&{\bf{1.74}}	&0.95	&{\bf{2.11}}	&0.10 &{\bf{1.88}}\\
\hline
\end{tabular}					 
\caption{Kullback-Leibler divergence (KLD) results for comparison of the global posterior distribution of each parameter
for the various subgroups. {\bf{In bold we indicate the statistically significant differences.}}}
\label{tab10}
\end{table}

\subsubsection{X-ray Classification}
\label{X-ray_Classification}
So far, we have compared the torus properties for Seyfert galaxies with different optical classifications. In this section we obtained the global posterior distributions of the sample divided into unabsorbed (N$_{H}<$10$^{22}$~cm$^{-2}$) and absorbed (N$_{H}>$10$^{22}$~cm$^{-2}$) Seyfert galaxies in X-rays. 

\begin{figure*}
\centering
\par{
\includegraphics[width=7.82cm]{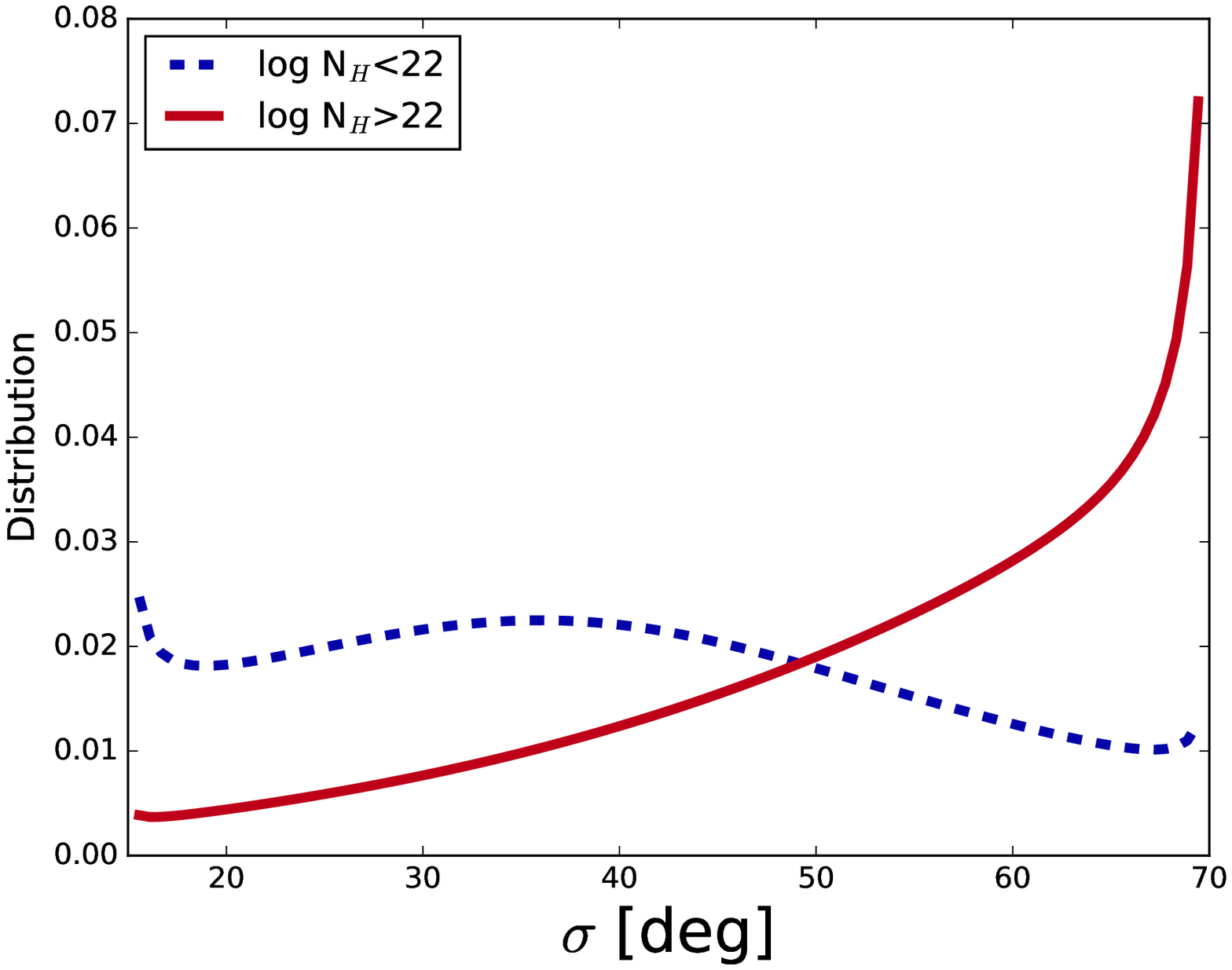}
\includegraphics[width=7.82cm]{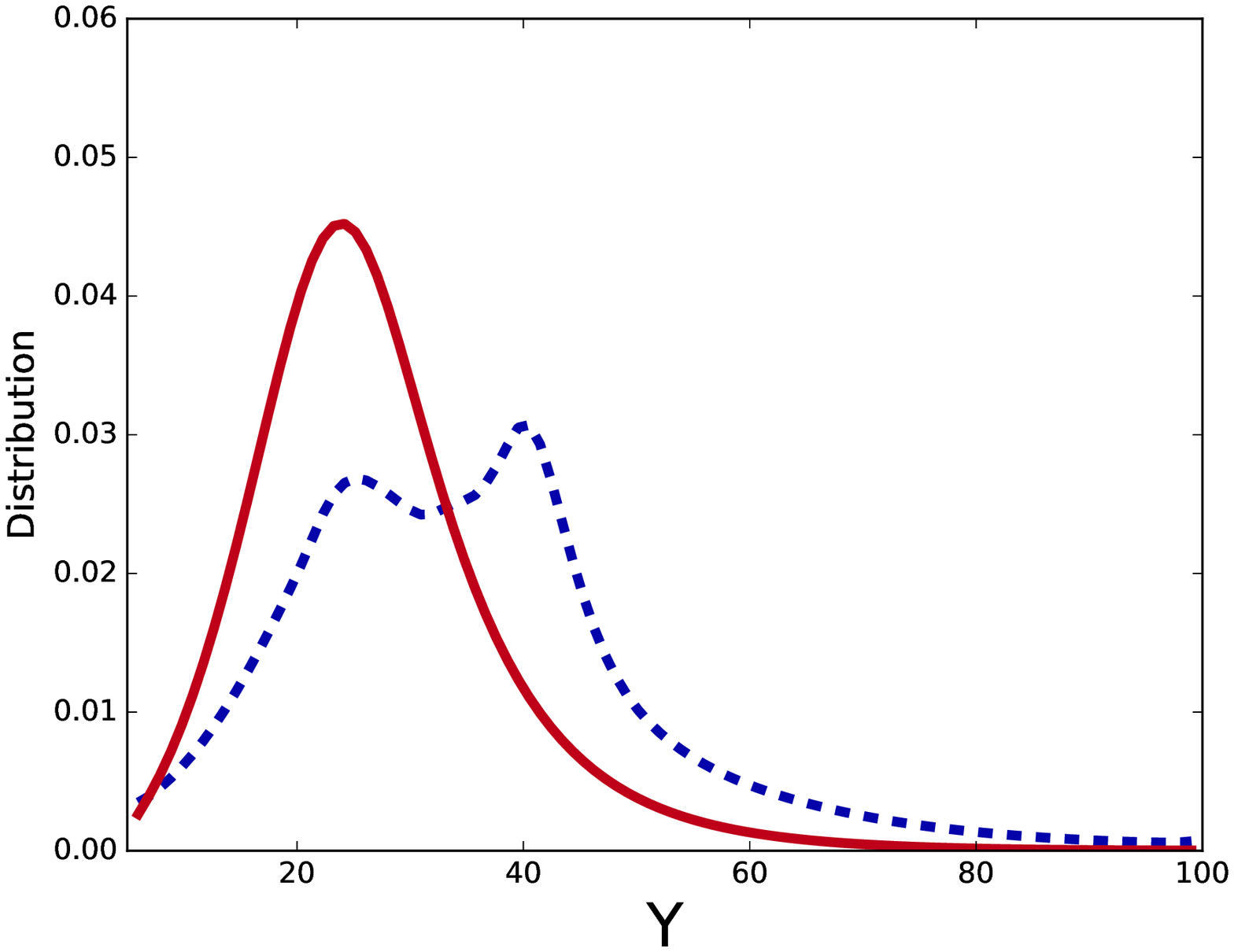}
\includegraphics[width=7.82cm]{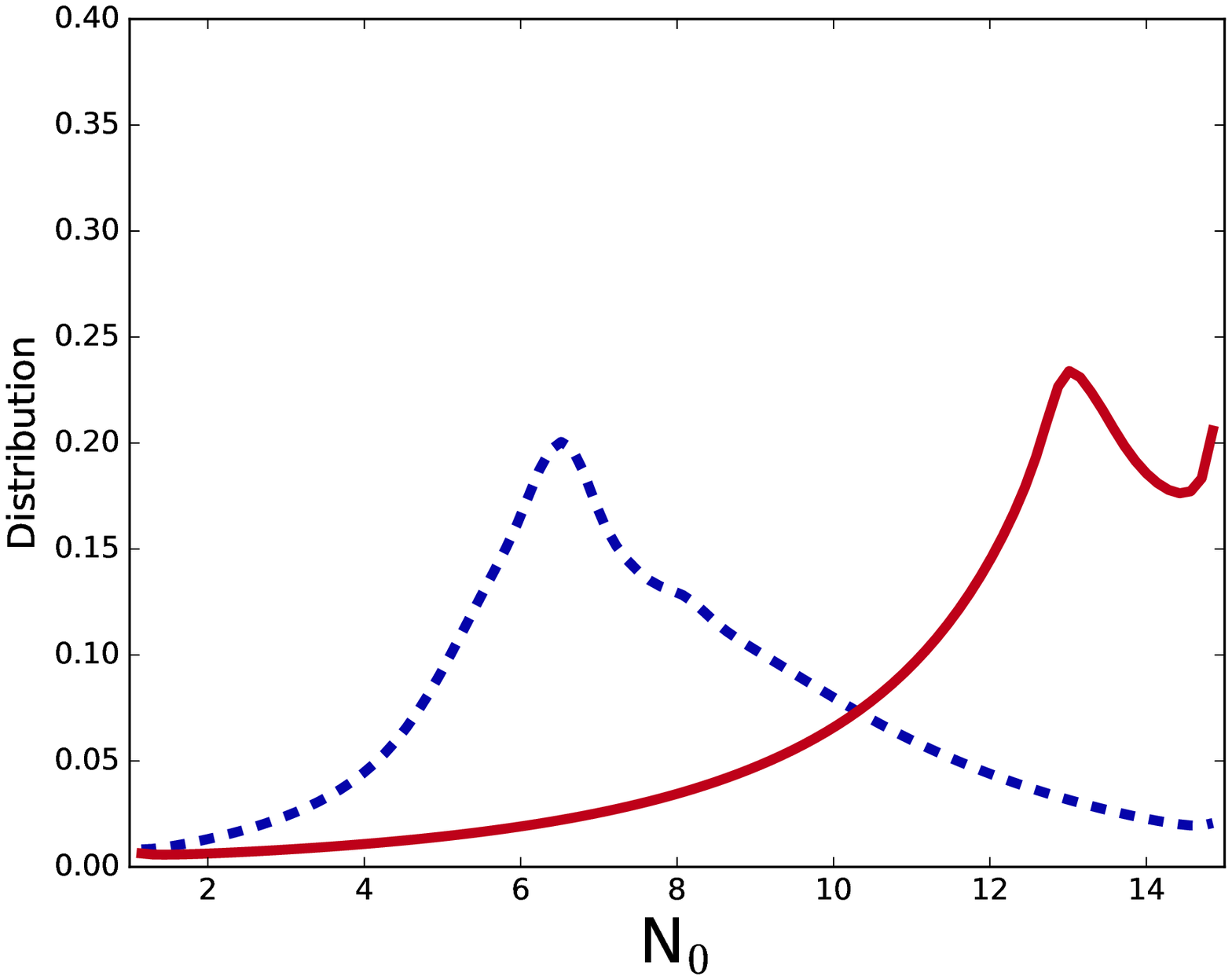}
\includegraphics[width=7.82cm]{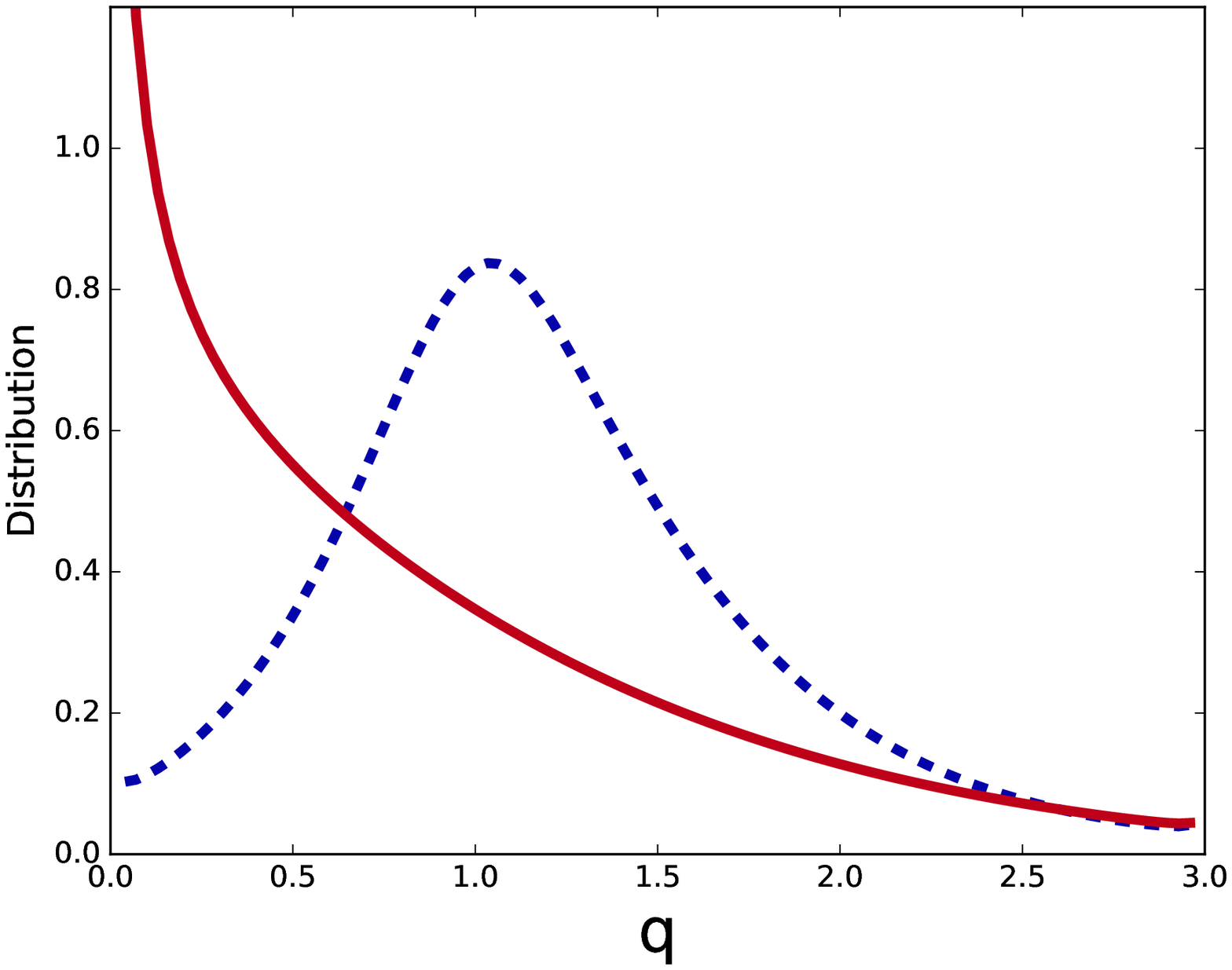}
\includegraphics[width=7.82cm]{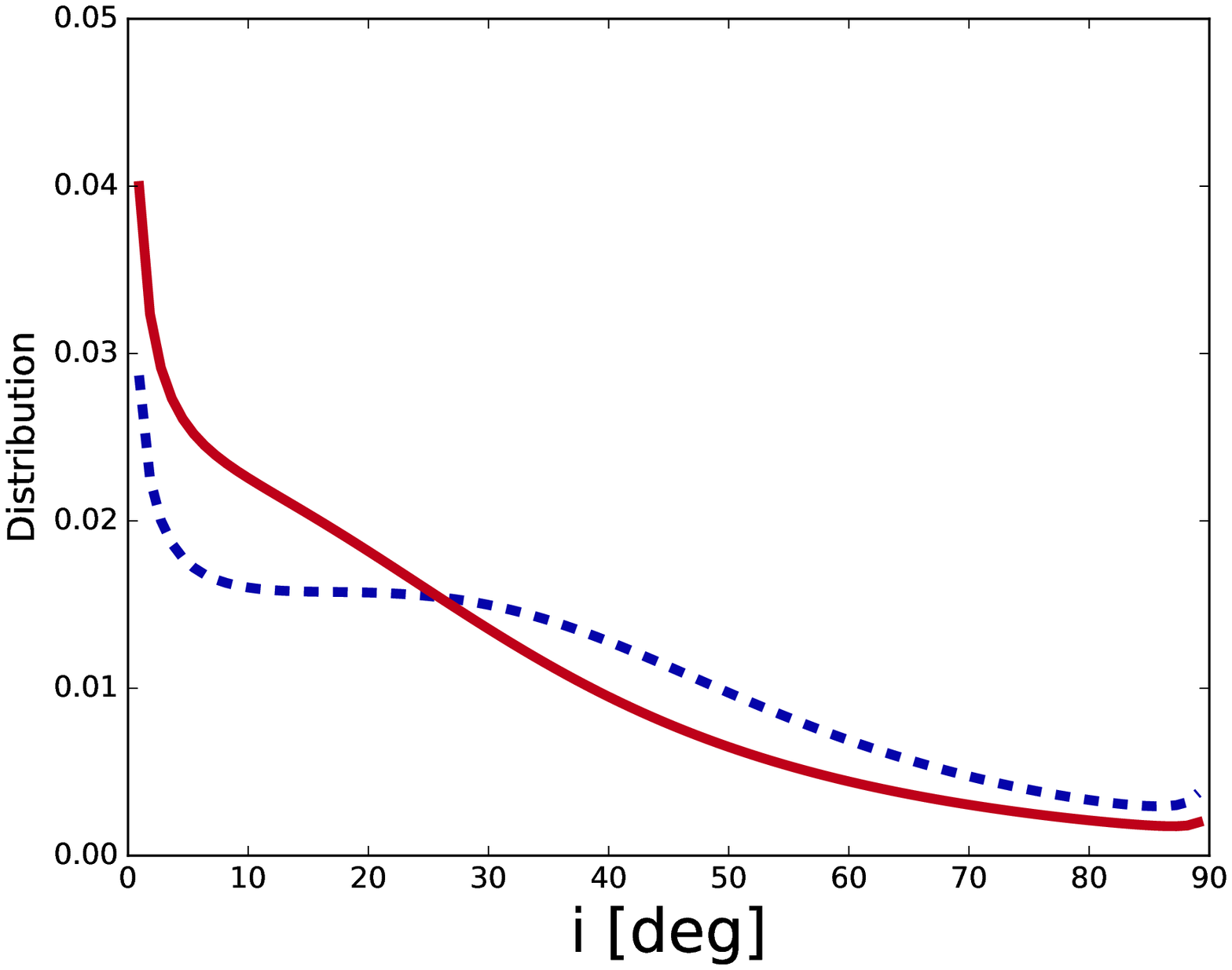}
\includegraphics[width=7.82cm]{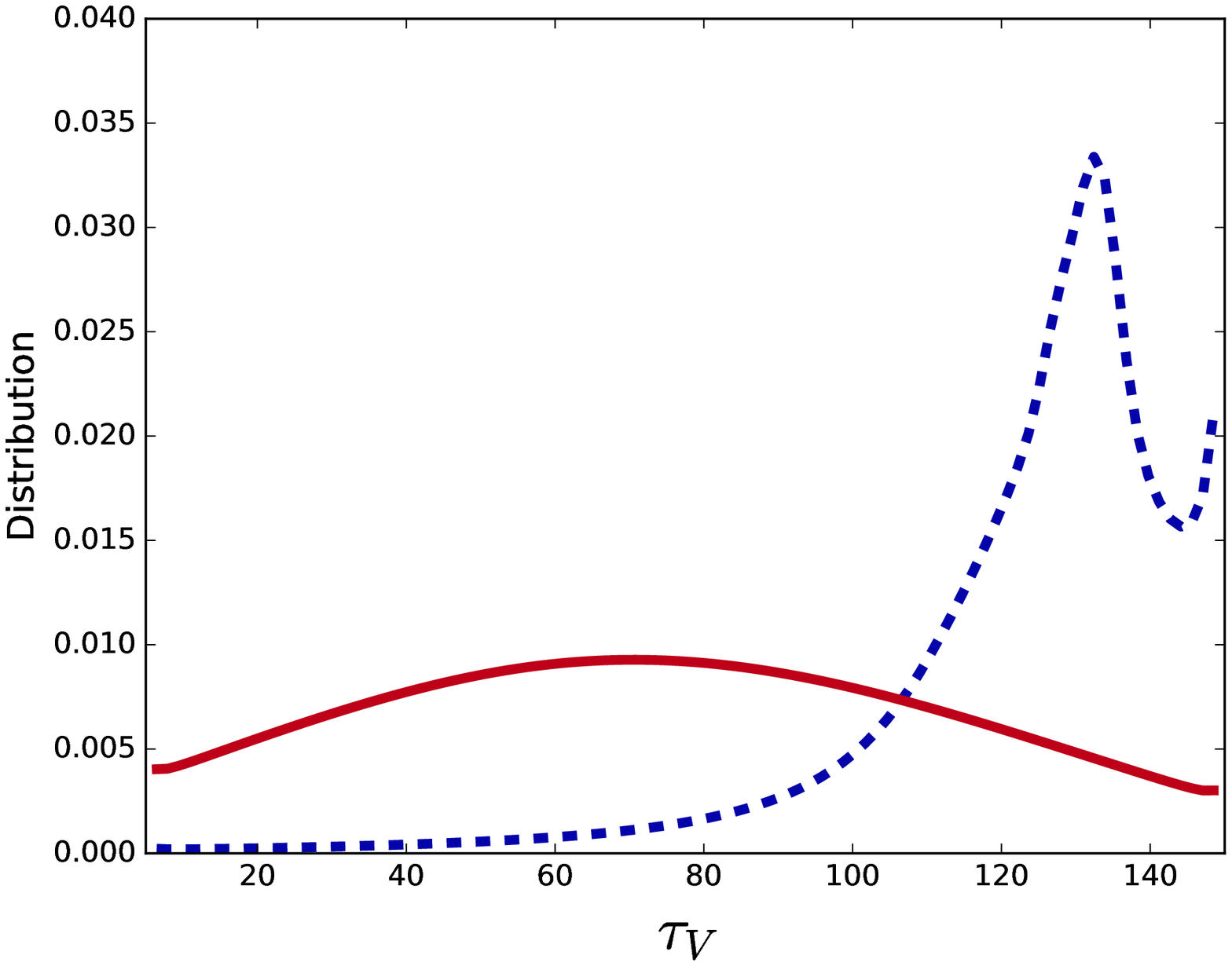}
\par}
\caption{Comparison between the clumpy torus model parameter global posterior distributions for type 1 and type 2 Seyferts according to the X-ray classification. Blue dot-dashed and red solid lines represent the parameter distributions of unabsorbed and absorbed Seyfert galaxies, respectively.}
\label{fig5}
\end{figure*}

In Fig. \ref{fig5} we show these distributions. We find essentially the same trends as when we divide the sample into Sy1 and Sy2 using an optical classification, but with less significance (see Table \ref{tab10}). This is due to the fact that the majority of optically classified Sy1 and Sy2 correspond to the unabsorbed and absorbed subgroups, respectively. Half of the Sy1.8/1.9 galaxies are classified as unabsorbed and the other half as absorbed, while only one Sy1 galaxy is absorbed (i.e. NGC\,4151). 

\subsection{Torus Size, Angular Width and Mass}
\label{AGN_proper} 
In this section we discuss the main torus model properties: torus size, angular width and mass, which can be compared to those derived from high angular resolution observations from the Atacama Large Millimeter/submillimeter Array (ALMA; e.g. \citealt{Garcia-Burillo16} and \citealt{Herrero18}). We can derive the physical radius of the clumpy torus (R$_o$) by using the radial extent of the torus (Y=R$_o$/R$_d$), the bolometric luminosity and the dust sublimation radius (R$_d$) definition (equation 1). Note that we use bolometric luminosities derived by using the 14--195 keV band and a fixed bolometric correction (L$_{bol}$/L$_{14-195~keV}$=7.42). We obtained this factor from the commonly used bolometric correction of 20 (e.g. \citealt{Vasudevan09}) at 2-10~keV and assuming a power-law slope of 1.8 (see Appendix \ref{C}). Finally, we can also estimate the torus gas mass for each AGN, using the Galactic dust-to-gas ratio from \citet{Bohlin78} and $\sigma$, N$_0$, $\tau_V$, R$_{d}$ and Y (equation 3), where I$_q=$ 1, Y/(2~ln~Y) and 1/3 for $q=$2, 1 and 0, respectively. In agreement with previous works using the CLUMPY torus models (e.g. RA09; RA11; AH11; \citealt{Lira13,Ichikawa15,Fuller16}), we find relatively compact torus sizes for all the Seyfert galaxies in our sample (R$_o<$15~pc), with median values of 2.8$\pm$1.2, 1.9$\pm$1.2 and 3.5$\pm$3.9~pc for torus radius of Sy1, Sy1.8/1.9, and Sy2 galaxies.

\begin{equation}
R_d = 0.4 ~\left( \frac{1500~K}{T_{sub}} \right)^{2.6} ~\left( \frac{L_{bol}^{AGN}}{10^{45}~erg~s^{-1}} \right)^{0.5} pc
\end{equation}

\begin{equation}
N_{H~torus}^{equatorial} = (1.9 \times 10^{21}) ~1.086 ~N_0 ~\tau_{V}~cm^{-2}
\end{equation}

\begin{equation}
{\frac{M_{torus}}{ M_\odot}} = 4 \pi~m_H  \sin (\sigma)  N_{H~torus}^{equatorial}~R_d^2~c^2~Y  I_q(Y) / 1.989 \times 10^{30}
\end{equation}

We find median values of 1.1$\pm$3.5, 1.4$\pm$2.8 and 3.9$\pm$5.1 $\times$10$^5$~M$_\odot$ for the torus gas masses of Sy1, Sy1.8/1.9, and Sy2 galaxies. \citet{Lira13} found values of the torus masses ranging from 10$^4$--10$^6$~M$_\odot$ using a sample of 48 Sy2 galaxies from the extended 12$\mu$m Galaxy Sample \citep{Rush93}. For 5/8 Sy1 galaxies, we can compare the torus gas masses with measurements corresponding to the inner 30~pc (radius) reported by \citet{Hicks09}. They derived masses of M$_{gas}^{H_2}$=3--20$\times$10$^6$~M$_\odot$ for NCG\,3227, NCG\,3783, NCG\,4051, NCG\,4151 and NCG\,6814. These masses were obtained from the H$_2$~1-0S(1) emission line at 2.12~$\mu$m. We do not have any Sy2 galaxy in common with \citet{Hicks09}, but we compare with the two Sy2 galaxies in their sample (Circinus and NGC\,1068). For Circinus using a smaller radius of 9~pc, they found a M$_{gas}^{H_2}$=1.9 $\times$10$^6$~M$_\odot$. For NGC\,1068 the latter authors reported a mass of M$_{gas}^{H_2}$=2.3 $\times$10$^7$~M$_\odot$ in the inner 30~pc. As expected, we found that the gas masses inferred from the fit of the nuclear IR SEDs in a smaller radius ($\sim$0.5-15~pc) are smaller than those measured in the inner $\sim$30~pc. Using the CO(6-5) line observed with ALMA, \citet{Garcia-Burillo16} reported a smaller gas mass (1.2$\times$10$^5$~M$_\odot$) for the inner $\sim$7-10~pc of NGC\,1068. Finally, using ALMA/CO(2-1) data, \citet{Herrero18} found that the nuclear disk of the Sy2 NGC\,5643 is a factor of $\sim$10 more massive and larger ($\sim$26~pc of diameter) than that of NGC\,1068. Therefore, we obtain comparable values of the torus mass with those derived from the highest angular resolution data.

\subsection{The Covering Factor}
\label{covering_factor}

Clumpy torus models imply that the differences between type-1 and type-2 AGN depend of whether there is a direct view of the broad line region (BLR; e.g. \citealt{Nenkova08a,Nenkova08b}). Therefore, the observed classification is the result of the probability for an AGN-produced photon to escape through the torus along a viewing angle without being absorbed. As this probability is always non-zero, it is always possible to have a direct view of the BLR, regardless of the torus orientation. Therefore, the larger the covering factor (C$_{T}$) the larger the probability of classifying an AGN as type 2. In fact, the geometrical covering factor gives the type~2/total fraction (e.g. \citealt{Mateos17}).

\begin{figure*}
\centering
\par{
\includegraphics[width=8.2cm]{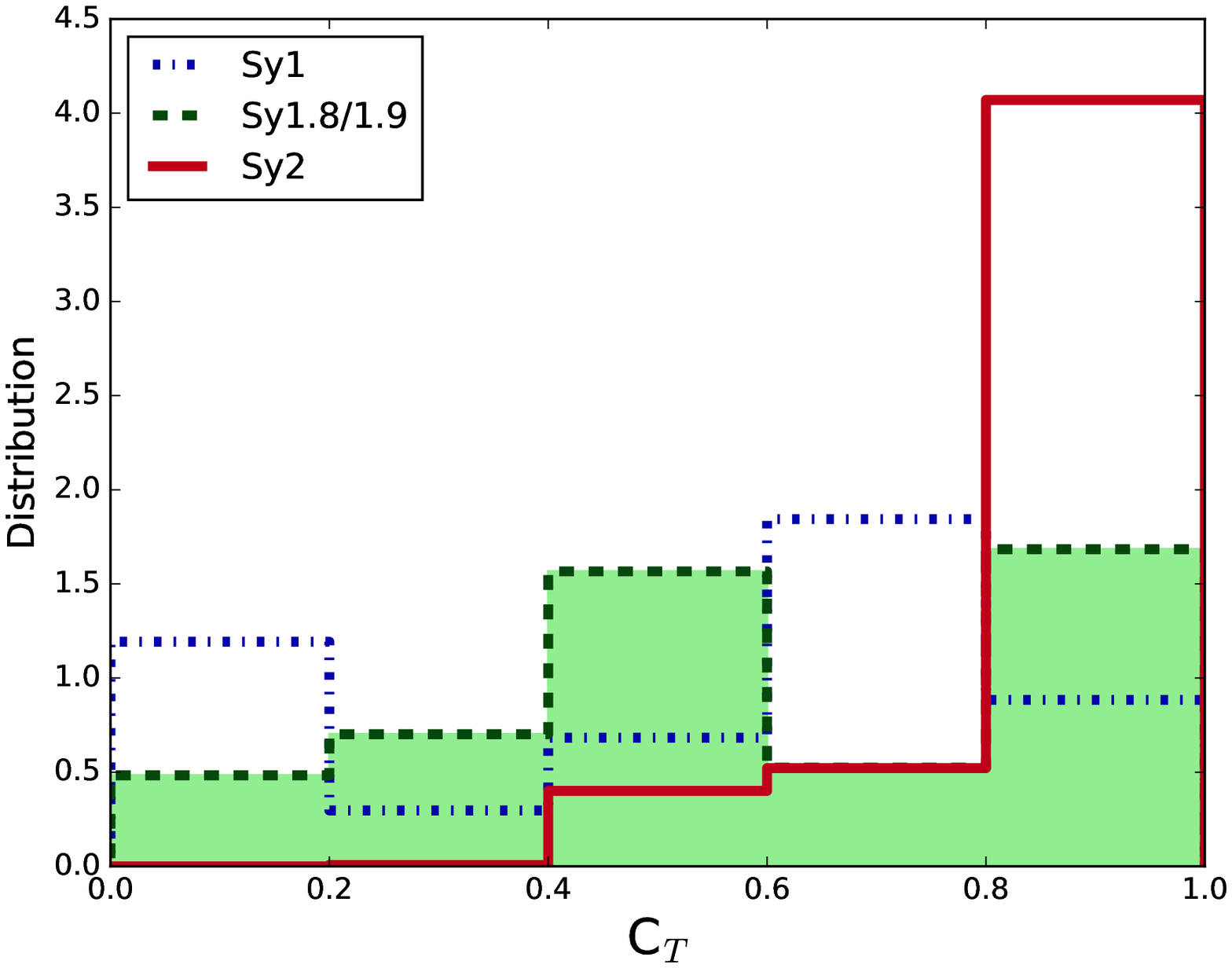}
\includegraphics[width=8.2cm]{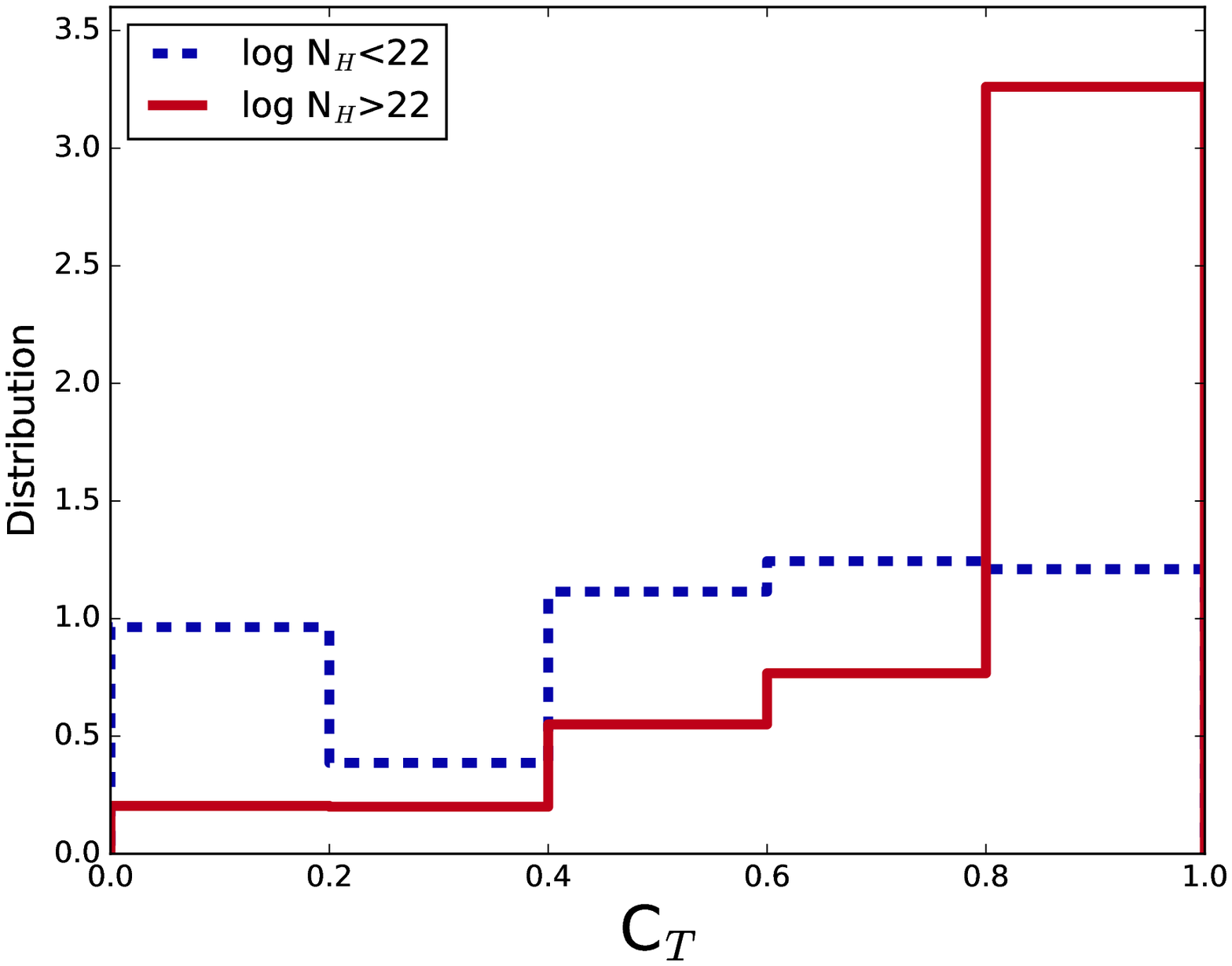}
\par}
\caption{Left panel: Comparison between the torus covering factor (C$_T$) combined probability distributions for the optical classification. Blue dot-dashed, green dashed and red solid lines represent the parameter distributions of Sy1, Sy1.8/1.9, and Sy2, respectively. Right panel: same as left panel, but for the X-ray classification, unabsorbed (log~N$_{H}<$22~cm$^{-2}$) and absorbed Seyfert galaxies (log~N$_{H}>$22~cm$^{-2}$). Blue dot-dashed and red solid lines represent the parameter distributions of unabsorbed and absorbed Seyfert galaxies, respectively.}
\label{fig6}
\end{figure*}

In order to compare the covering factors for the three subgroups we derived the combined probability distributions. To do so, we concatenated together the individual arrays of the C$_{T}$ values returned Bayesian modelling for all objects in subgroups and we computed the combined probability distributions since it is a nonlinear function of the torus model parameters ($\sigma$ and N$_0$; see left panel of Fig. \ref{fig6}). We note that we do not use the hierarchical Bayesian approach since if we use the generalized beta distribution as the prior we would introduce an extra prior in the C$_{T}$ derived quantities. Using our ultra-hard X-ray volume-limited sample of Seyfert galaxies, we confirm the results first reported by RA11 that the covering factors of Sy2 are larger than those of Sy1 galaxies. Indeed, using the optical classification, we find that Sy2 have larger median values of the covering factor combined probability distributions (C$_T=$0.95$\pm_{0.18}^{0.04}$) than Sy1 (C$_T=$0.66$\pm_{0.52}^{0.16}$) and Sy1.8/1.9 (C$_T=$0.53$\pm_{0.37}^{0.21}$) which is the same as for Sy1s, within the errors. These results are in good agreement with previous works (e.g. RA11, AH11, \citealt{Ichikawa15} and \citealt{Mateos16}). We also repeat the global posterior distribution for the covering factor using the X-ray classification (see Section \ref{X-ray_Classification}) and we find the same trend (see right panel of Fig. \ref{fig6} and Table \ref{tab10}).

\subsubsection{Dependence with AGN luminosity and Eddington ratio}
\label{receding_torus}

\begin{figure*}
\centering
\par{
\includegraphics[width=6.0cm, angle=90]{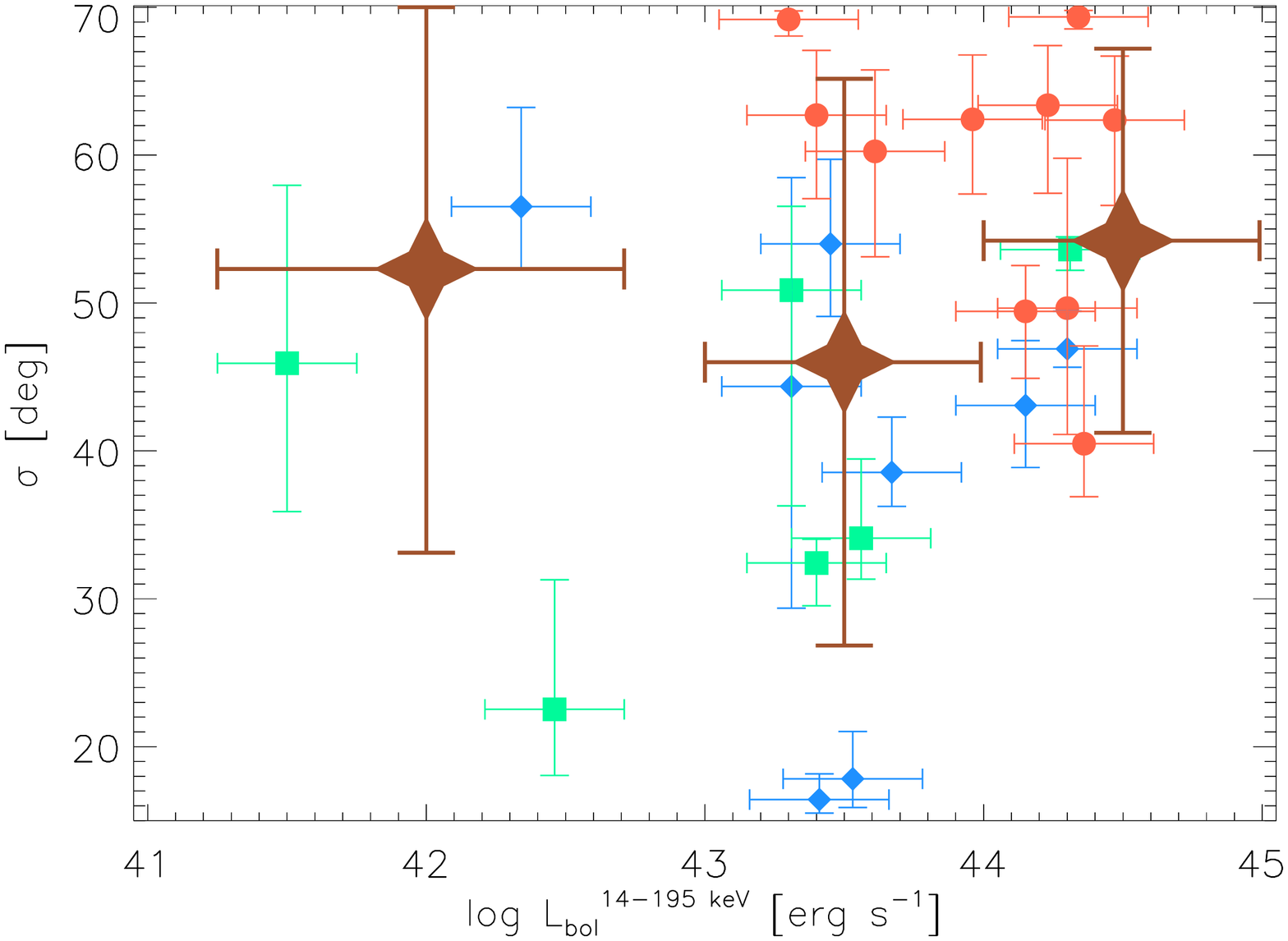}
\includegraphics[width=6.0cm, angle=90]{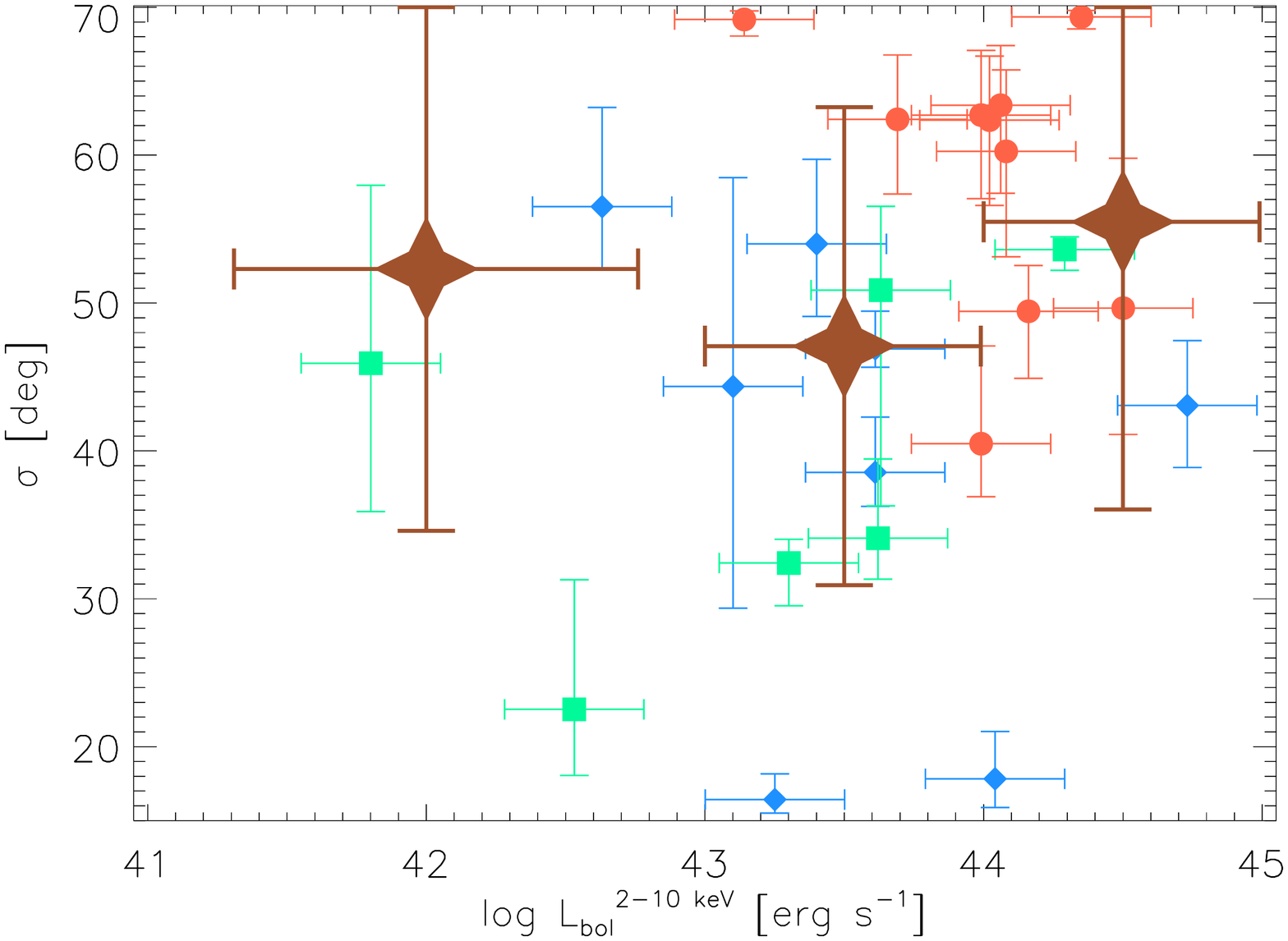}
\includegraphics[width=8.4cm]{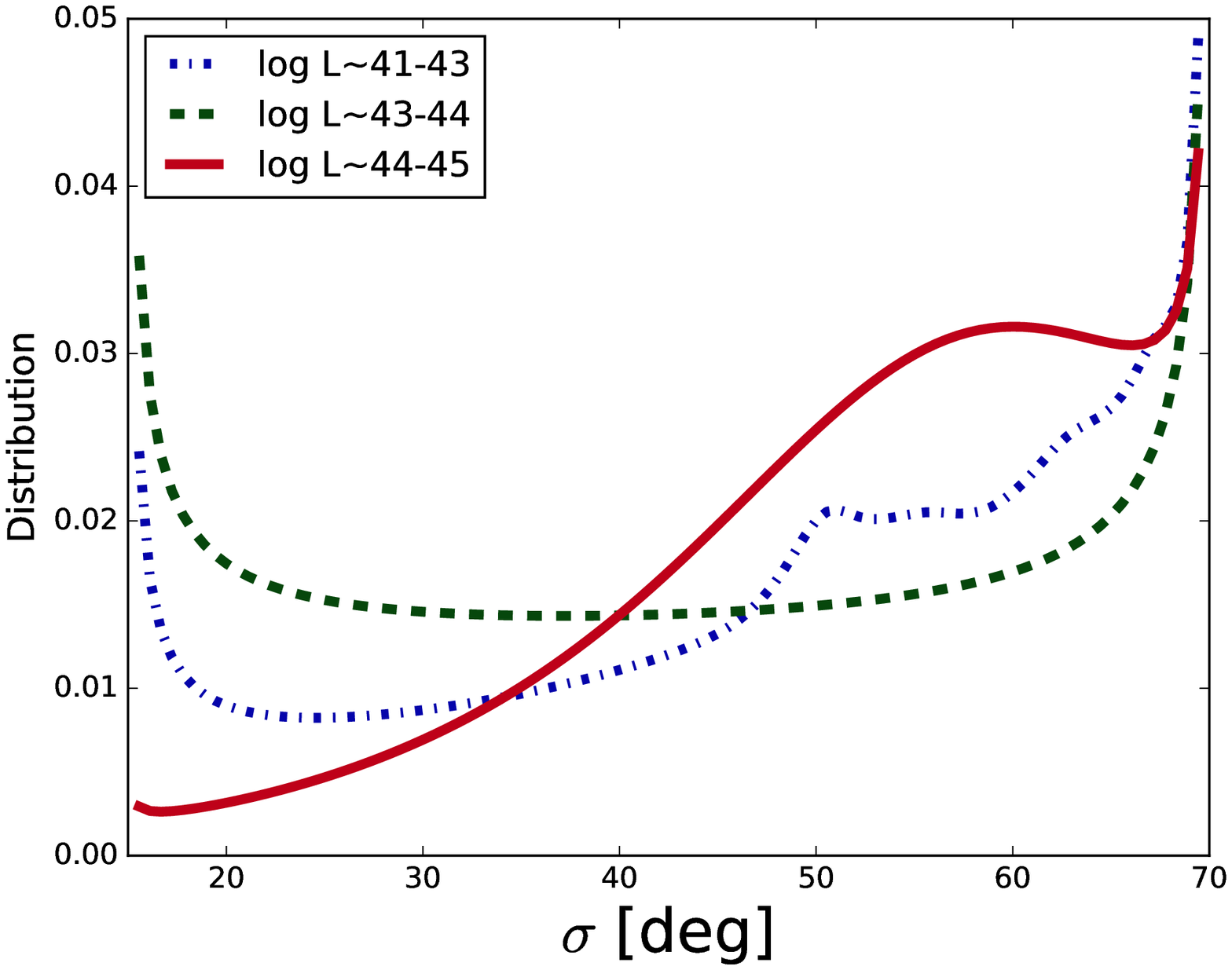}
\includegraphics[width=8.4cm]{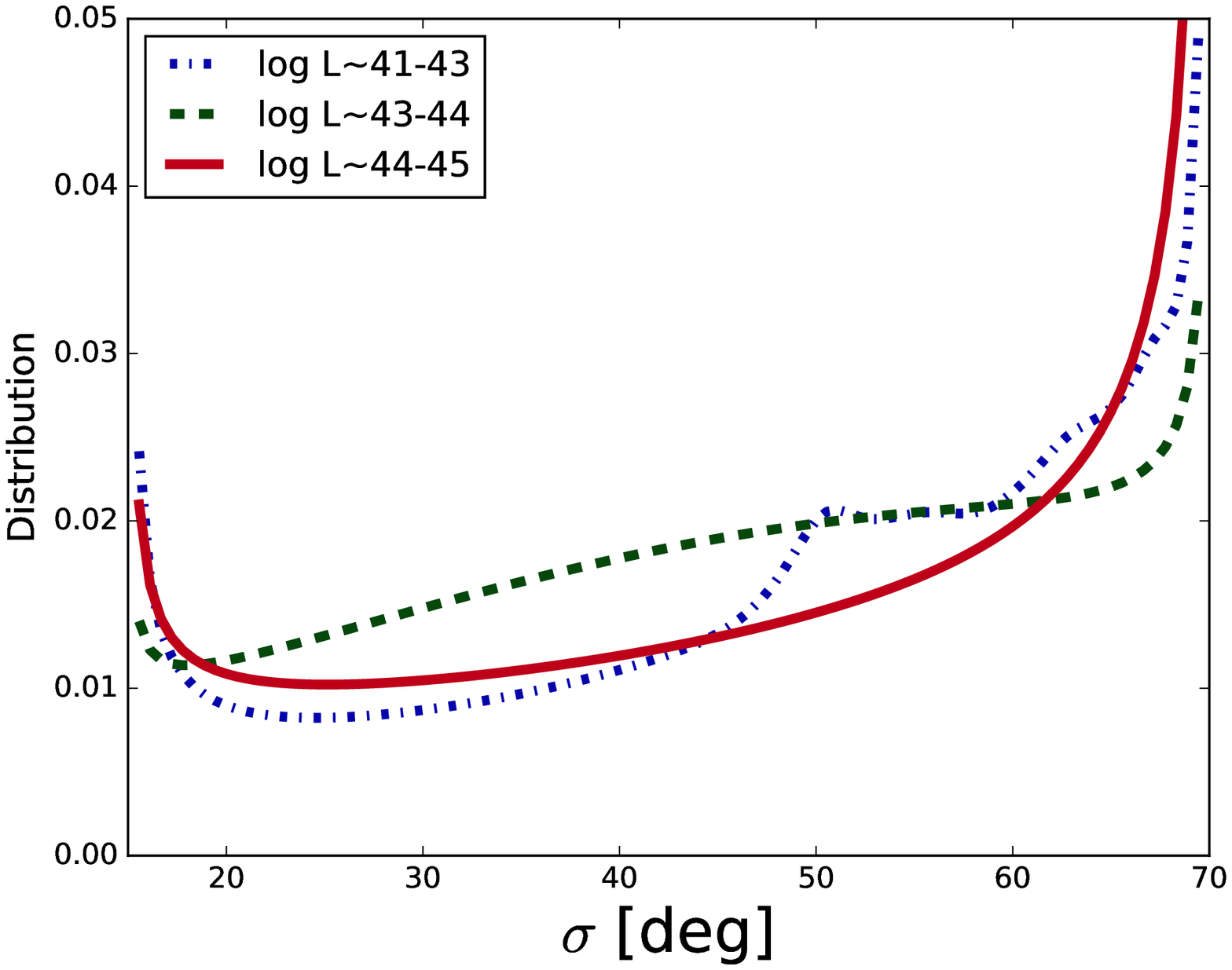}
\par}
\caption{Top panels: luminosity dependence of the torus width for the BCS$_{40}$ sample using the bolometric luminosities derived from the 14--195 keV and 2--10 keV bands. Blue diamonds, green squares and red circles represent Sy1, Sy1.8/1.9, and Sy2, respectively. Brown stars correspond to values derived from the global posterior distribution of each bin subgroup. The error bars represent the $\pm$1$\sigma$ confidence interval for the individual and average measurements. Note that for the average values the error bars in the X-axis indicate the bin width. Bottom panels: comparison between the global posterior distributions of the torus width for three ranges of bolometric luminosities derived from the 14-195 keV and 2-10 keV bands.} 
\label{fig7}
\end{figure*}

\begin{figure*}
\centering
\par{
\includegraphics[width=6.2cm, angle=90]{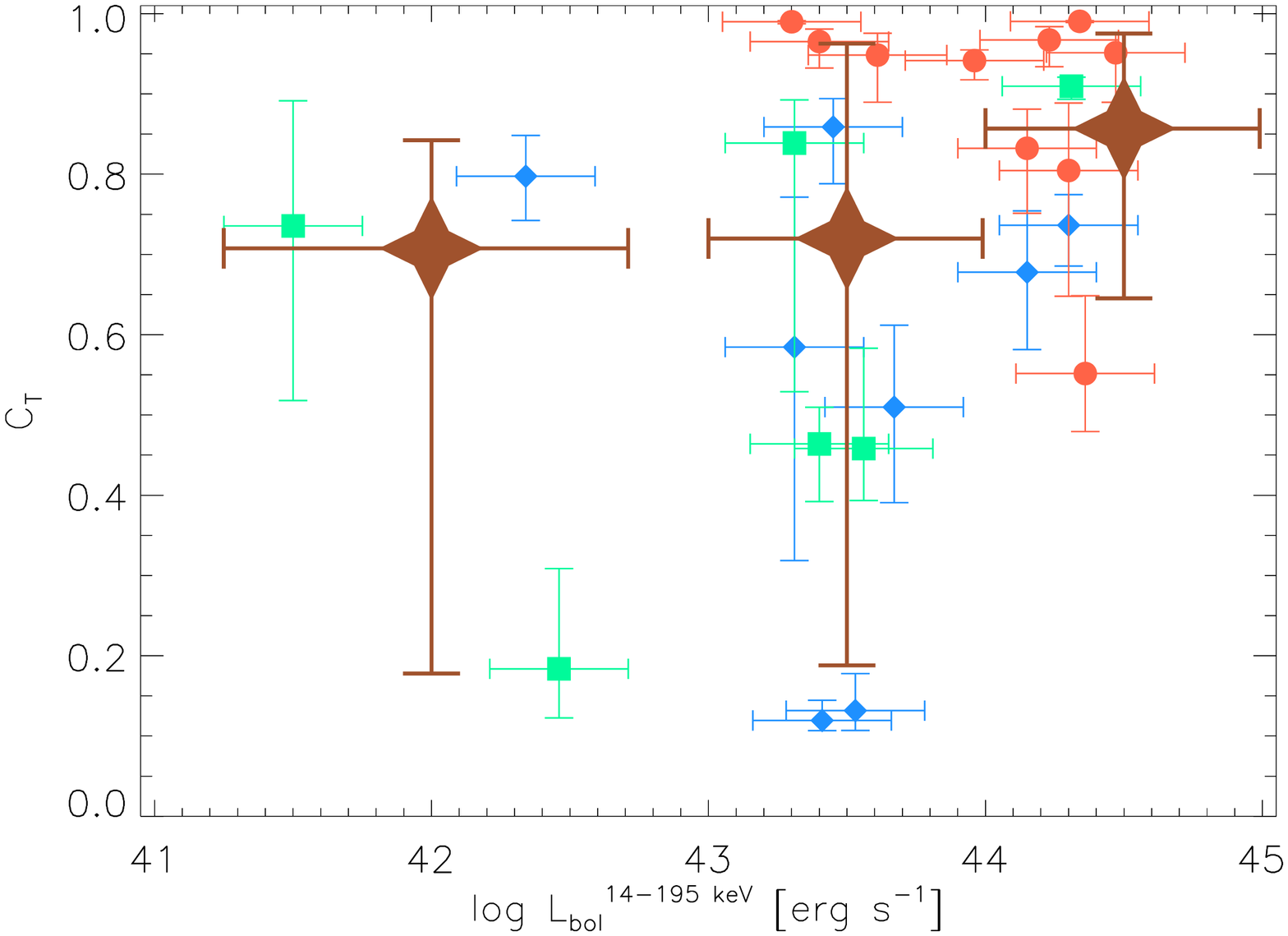}
\includegraphics[width=6.2cm, angle=90]{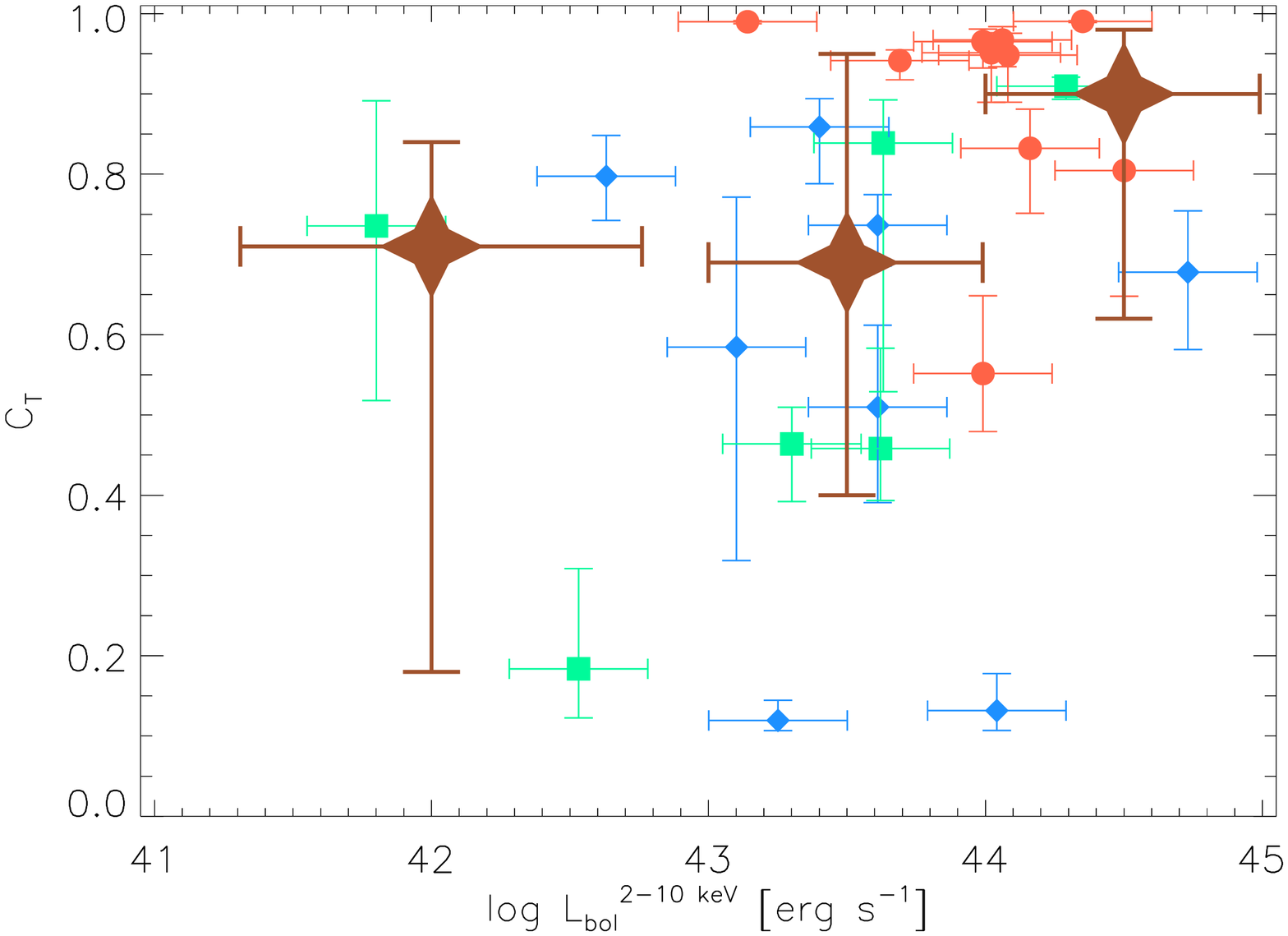}
\includegraphics[width=8.4cm]{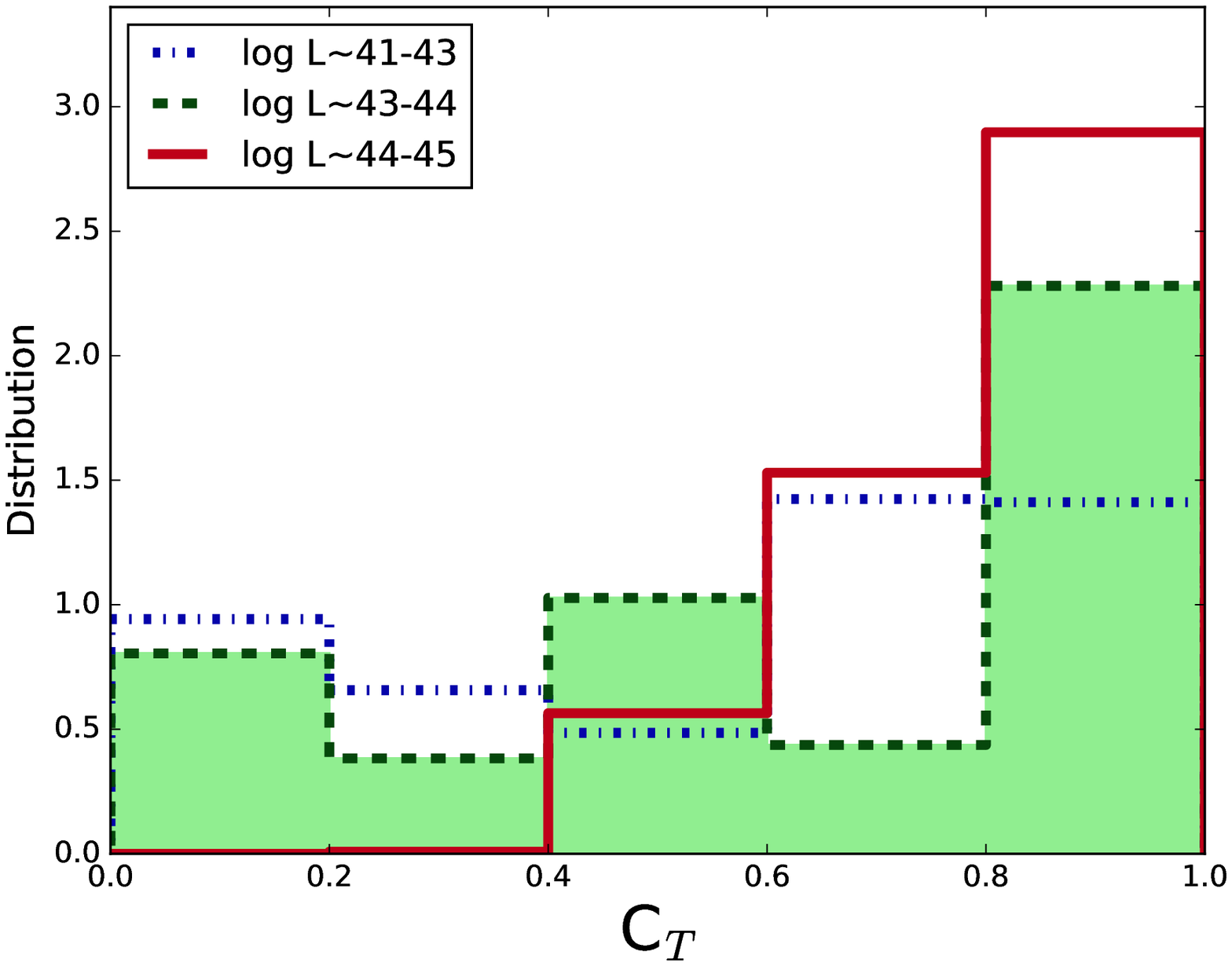}
\includegraphics[width=8.4cm]{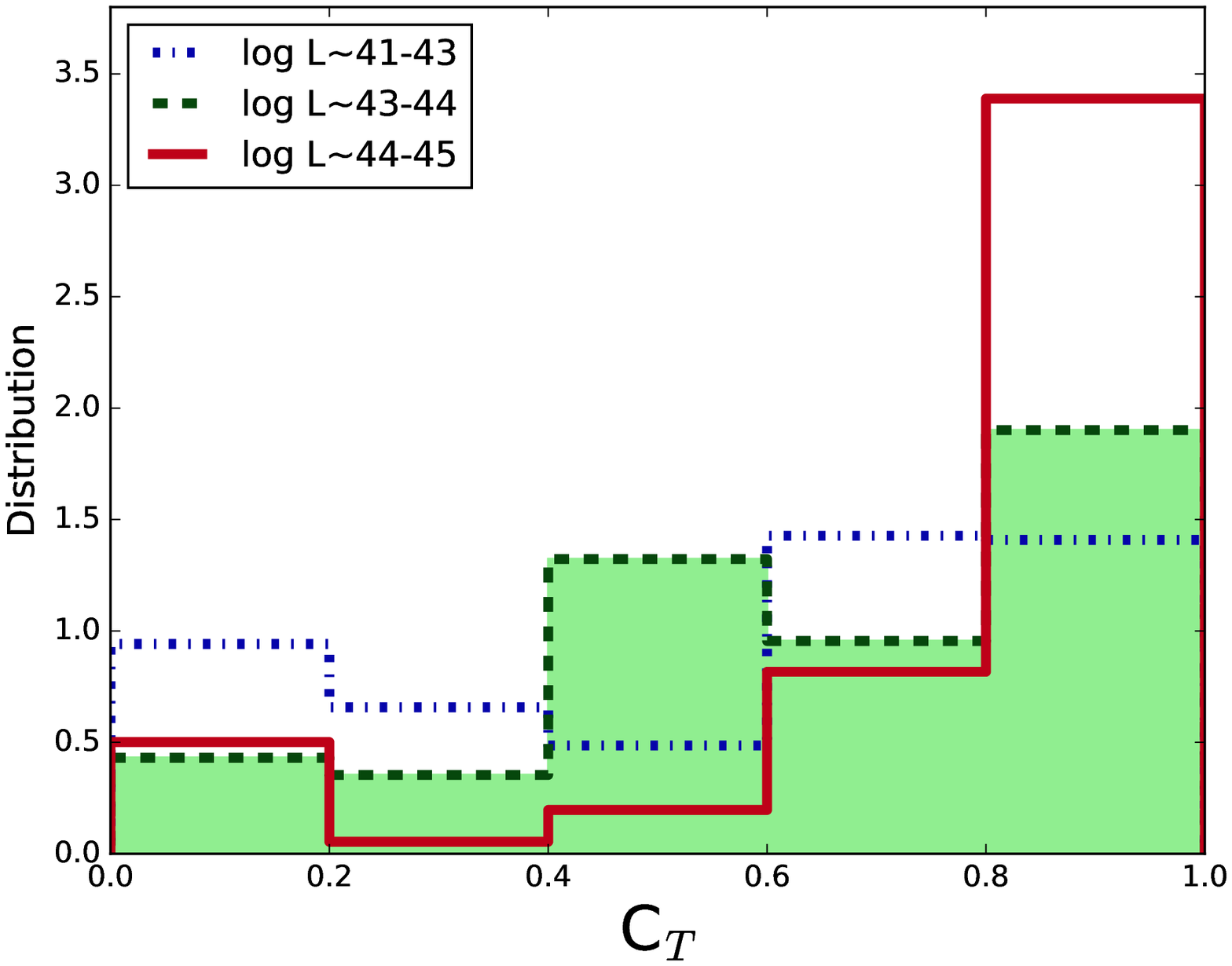}
\par}
\caption{Same as Fig. \ref{fig7} but for the covering factor (C$_T$). Bottom panels: comparison between the covering factor combined probability distributions for three ranges of bolometric luminosities. The error bars represent the $\pm$1$\sigma$ confidence interval for the individual and median measurements. Note that for the median values the error bars in the X-axis indicate the bin width.} 
\label{fig8}
\end{figure*}

To investigate the relation between the bolometric luminosity derived from the X-rays (2--10 and 14--195~keV; see Appendix \ref{C}) and the covering factor (see top panels of Fig. \ref{fig7}), we divided our sample into several luminosity bins. In the first bin we included the three sources with log(L$_{bol}^{AGN}$)$<$43, while the rest of the sample was divided in two bins of equal logarithmic width (1~dex). We find that the $\sigma$ parameter remains essentially constant within the errors, throughout our luminosity range (log(L$_{bol}^{AGN}$)$\sim$41--45~erg~s$^{-1}$; see top panels of Fig. \ref{fig7}). The same applies for the covering factor (see top panels of Fig. \ref{fig8}). We find slightly larger values of C$_T$ ($\gtrsim$0.5) in the log(L$_{bol}^{AGN}$)$\sim$44--45~erg~s$^{-1}$ luminosity range because the majority of the sources in that bin are Sy2s. Thus we do not find a statistically significant trend in the covering factor with AGN luminosity, which is in good agreement with recent studies (e.g. \citealt{Mateos16,Mateos17,Netzer16,Stalevski2016,Lani17,Ichikawa18}). 

\begin{figure*}
\centering
\par{
\includegraphics[width=6.2cm, angle=90]{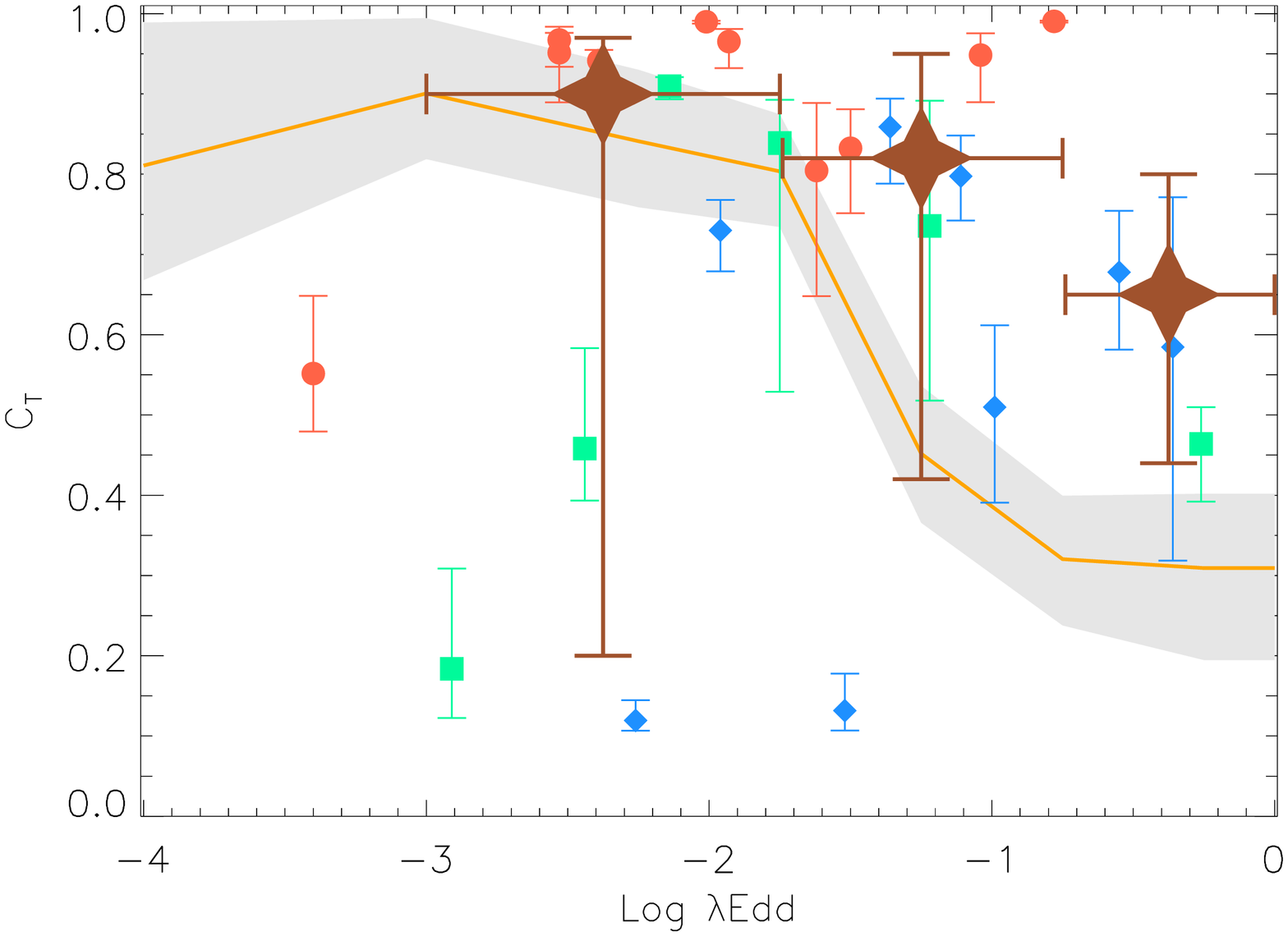}
\includegraphics[width=8.4cm]{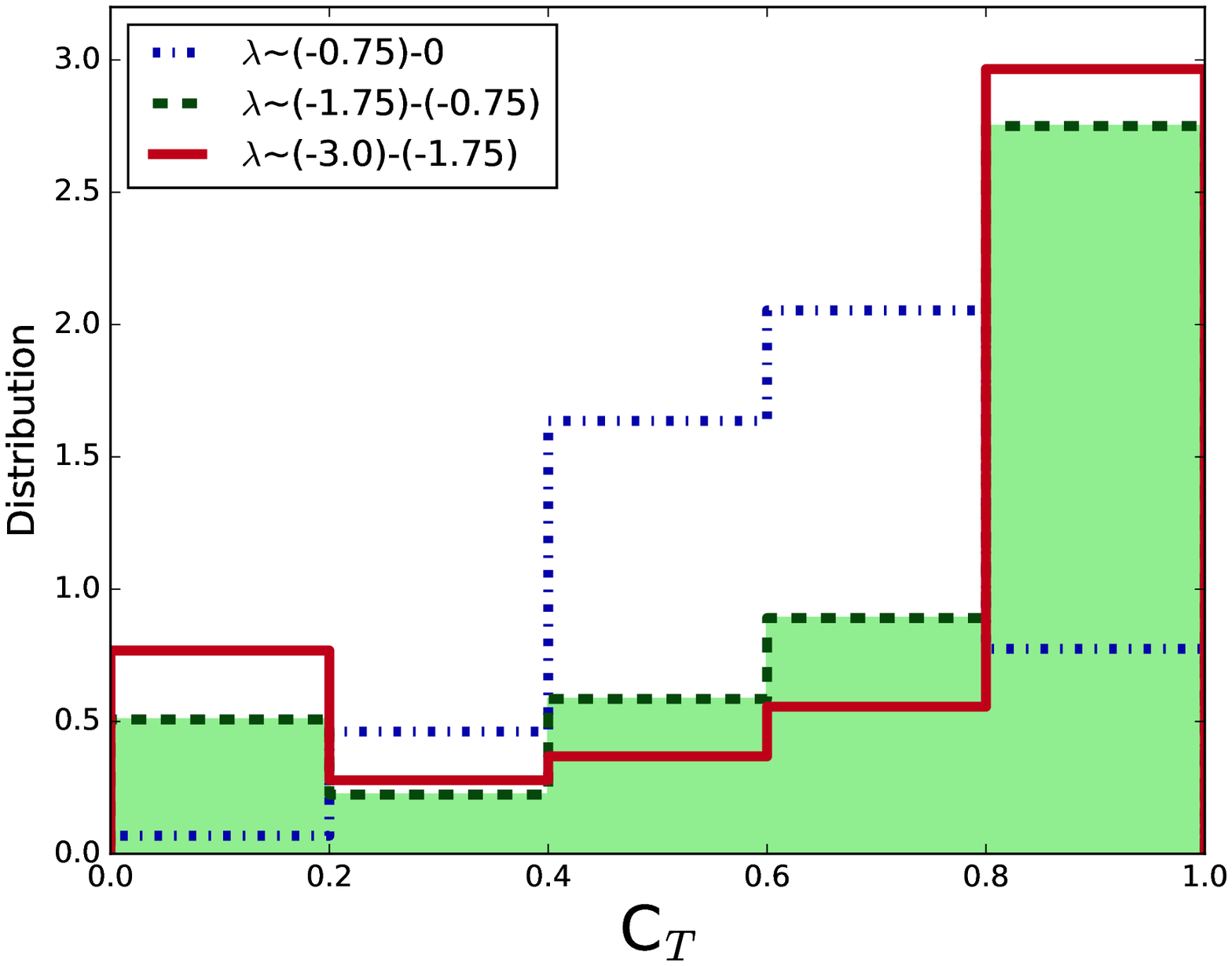}
\par}
\caption{Left panel: relation between the covering factor and the Eddington ratio for the BCS$_{40}$ sample. The solid orange line is the relation derived by \citet{Ricci17c} and the uncertainties are shown as grey shaded regions. Blue diamonds, green squares and red circles represent Sy1, Sy1.8/1.9, and Sy2, respectively. Brown stars are the median values of the combined probability distribution of each Eddington ratio bin. The error bars represent the $\pm$1$\sigma$ confidence interval for the individual and median measurements. Note that for the median values the error bars in the X-axis indicate the bin width. Right panel: comparison between the torus covering factor (C$_T$) combined probability distributions of each bin.}
\label{fig9}
\end{figure*}

More recently it has been suggested that the Eddington ratio is the key parameter determining the covering factor, instead of the bolometric luminosity (e.g. \citealt{Buchner17,Ricci17c}). \citet{Ricci17c} found that the covering factor rapidly decreases at higher Eddington ratios (see the orange solid line of Fig. \ref{fig9}). We derived Eddington ratios using the 2--10~keV bolometric X-ray luminosities and the black hole mass estimates from \citet{Ricci17} and \citet{Koss17}, respectively. For the remaining sources we estimate the black hole masses (see Appendix \ref{D}) as in \citet{Koss17}. The only exceptions are NGC\,7213 and ESO\,005-G004, for which we take their black hole masses from \citet{Vasudevan10}. The Eddington ratios of the sample are listed in Table \ref{tab9}.

Although we find higher values of the C$_T$ for lower Eddington ratios (see Fig. \ref{fig9}), we do not find a statistically significant dependence of the torus covering factor with the Eddington ratio. This result suggests, albeit for a small luminosity range and a limited number of galaxies, that the Eddington ratio would not be driving the geometrical covering factor.

\subsubsection{Missing Obscured Seyferts?}
\label{missings}

\begin{figure}
\centering
\includegraphics[width=6.2cm, angle=90]{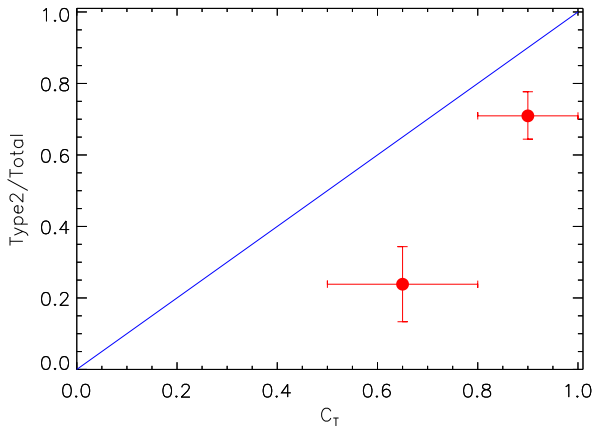}
\caption{Sy2 fraction vs. torus covering factor for the BCS$_{40}$ sample grouped in two bins. The error bars in the X-axis indicate the bin width and in the Y-axis represent the $\pm$1$\sigma$ confidence interval.} 
\label{fig10}
\end{figure}

In Fig. \ref{fig10} we show the Sy2 fraction in our sample as estimated by using two covering factor bins (0.5--0.8 \& 0.8--1.0)\footnote{Note that we use C$_T$ values larger than 0.5 due to the lack of data in the lower C$_T$ range for Sy2 galaxies (see left panel of Fig. \ref{fig6}).}. To estimate the uncertainties, we used the bootstrap error estimation generating 10$^6$ mock samples of Sy1, Sy1.8/1.9 and Sy2s by randomly selecting sources using replacements, with their corresponding covering factor distributions from the original samples. Note that the number of Sy1, Sy1.8/1.9 and Sy2s in each mock sample keep constant the observed number of Seyferts (i.e. n$_{Sy1}$+n$_{Sy1.8/1.9}$+n$_{Sy2}$). Finally, for each source, we calculate the obscured fraction in each bin by integrating its probability distribution (see e.g. \citealt{Mateos17}).
 
Our data points should follow the 1:1 blue line shown in Fig. \ref{fig10} if our sample did not miss any high covering factor source (covering factor values $\sim$1). However, the Sy2/total fraction is always below the 1:1 line. In general, we found that the most highly absorbed sources are the ones with higher torus covering factors (see also RA11, AH11 and \citealt{Mateos16}). All this suggests that even an ultra hard X-ray (14--195~keV) Swift/BAT selection is missing a significant fraction of highly absorbed type 2 sources with very high covering factor tori. This is expected since at column densities N$_H$ $>$10$^{23.5}$~cm$^{-2}$, even high energy photons (14--195keV X-ray band) are absorbed. An example of these missing sources could be NGC\,4418, which is a very highly obscured AGN at $\sim$30\,Mpc. It has a compact IR bright core with the deepest known silicate absorption but it is not detected in the Swift/BAT hard X-ray band (e.g. \citealt{Roche15} and references therein). The result presented here agrees with those reported in \citet{Ricci15} and \citet{Koss16} at energies $>$10~keV and \citet{Mateos17} at energies $>$4.5~keV. The latter authors inferred the existence of a population of X-ray undetected objects with high torus covering factor, especially at high bolometric luminosities ($>$10$^{44}$ erg~s$^{-1}$). \\

\newpage

\section{Conclusions}
\label{Conclusions}

We present for the first time a detailed modelling of the nuclear dust emission of an ultra-hard X-ray (14--195~keV) volume-limited (D$_L<$40\,Mpc) sample of 24 Seyfert galaxies. We selected our targets from the {\textit{Swift/BAT}} nine month catalog. Our sample covers an AGN luminosity range log(L$_{\textrm{int}}^{2-10~\textrm{keV}}$)$\sim$40.5--43.4~erg~s$^{-1}$. We fitted the nuclear IR SEDs obtained with high angular resolution data ($\sim$1--30~$\mu$m) with the CLUMPY models using a Bayesian approach. From these fits, we derived torus model parameters for the individual 24 galaxies. In the case of Seyfert 1s we took special care to subtract the accretion disk contribution from the observed nuclear SEDs using the type 1 QSO accretion disk template from \citet{Hernan-caballero2016}. The main goal of this work was to obtain a global statistical analysis of the clumpy torus model parameters of type 1 and 2 Seyfert galaxies. We used both optical (broad vs narrow lines) and X-ray (unabsorbed vs absorbed) classifications for our analysis. Using these classifications, we compared the global posterior distribution of the torus model parameters, rather than focusing on the individual fits.

We verified our previous results that type 2 Seyferts have tori with larger widths and more clouds than type 1/1.8/1.9s. These findings are independent of whether we use an optical or X-ray classification. We found that the covering factor is likely the main parameter driving the classification of Seyfert galaxies. We derived compact torus sizes (radius $<$15~pc), and gas masses in the 10$^4$--10$^6$~M$_\odot$ range for both types.

We derived geometrical covering factors for the individual galaxies and globally for Sy1s and Sy2s. In clumpy torus models the geometrical covering factor is a function of the angular size and the number of clouds. Using these distributions, we confirmed that Seyfert 2 galaxies have larger values of the covering factor (C$_T=$0.95$\pm_{0.18}^{0.04}$) than type 1s (C$_T=$0.66$\pm_{0.52}^{0.16}$) using, for the first time, an ultra-hard X-ray selected sample of Seyferts. We found that the torus covering factor remains constant within the errors in our luminosity range and no clear dependence with the Eddington ratio. Finally, we compared the derived covering factor with the observed type 2 fraction for our sample. From this comparison, we found tentative evidence that even an ultra hard X-ray selection is missing a significant fraction of highly absorbed type 2 sources with very high covering factor tori, as also concluded by \citet{Mateos17} at lower X-ray energies using a more distant and luminous sample of AGN.

We note that detailed studies such as this, carried out not only using larger samples of galaxies but covering wider luminosity and redshift ranges are needed to improve the statistics of the results we report here. In the future, this methodology may be applied to AGN samples using high angular resolution and sensitive MIR data, observed with the combined spectral coverage of NIRSpec and MIRI aboard the James Webb Space Telescope (JWST).

\section*{Acknowledgments}

IGB acknowledges financial support from the Instituto de Astrof\'isica de Canarias through Fundaci\'on La Caixa. IGB also acknowledges Oxford University and Durham University for their hospitality during his stays in 2017 August when this project was started and 2018 May, respectively. IGB also acknowledges Cardiff University for their hospitality from 2018 May to August. CRA and IGB acknowledge financial support from the Spanish Ministry of Science and Innovation (MICINN) through project AYA2016-76682-C3-2-P. IGB, AAH and FJC also acknowledge financial support through grant PN AYA2015-64346-C2-1-P (MINECO/FEDER). Funded by the Agencia Estatal de Investigaci\'on, Unidad de Excelencia Mar\'ia de Maeztu. CRA also acknowledges the Ram\'on y Cajal Program of the Spanish Ministry of Economy and Competitiveness. MPS acknowledges support from STFC through grant ST/N000919/1. OGM thanks to the PAPIIT UNAM project IA103118. S.M. acknowledges financial support through grant AYA2016-76730-P (MINECO/FEDER). IM and JM acknowledge financial support from the research project AYA2016-76682-C3-1-P (AEI/FEDER, UE). L.F. and C.P. acknowledge support from the NSF- grant number 1616828.

Finally, we are extremely grateful to the GTC staff for their constant and enthusiastic support, and to the anonymous referee for useful comments.

\appendix

\section{nuclear IR SED fits}
\label{B}
The main results of this work are based on the fits of the individual Sy1, Sy1.8/1.9 and Sy2, nuclear IR SEDs (see Section \ref{sed_construction}). The median and maximum-a-posteriori (MAP) values of the model parameters fitted to the individual nuclear IR SEDs are reported in Table \ref{tabA1}. In addition, the individual fits are shown in Figures (A1-A4). As previously mentioned (see Section \ref{clumpy_torus}), we can translate the results from the probability distributions into corresponding models spectra. The solid red and blue dashed lines in Figures (A1-A4) correspond to the model that maximizes their probability distributions (MAP) and the median model. Shaded regions of these Figures (A1-A4) correspond to the range of models compatible with the 68\% confidence interval for each parameter around the median. The galaxies with
subarcsecond angular resolution N-band spectra are labelled as `nuclear spectrum', while
those labelled as `AGN template' are sources that either do not have high angular resolution
nuclear spectra or are noisy/include a strong contribution from the host galaxy (see Section \ref{sed_construction}). In these cases we used the N-band ``pseudo-nuclear'' spectra.

\begin{table*}
\scriptsize
\centering
\begin{tabular}{lcccccccccccc}
\hline
		& \multicolumn{2}{|c|}{$\sigma$\,(deg)} 	& \multicolumn{2}{|c|}{Y}	& \multicolumn{2}{|c|}{N$_0$}	& \multicolumn{2}{|c|}{q}	&\multicolumn{2}{|c|}{$\tau_{V}$}	& \multicolumn{2}{|c|}{i\,(deg)} \\
Name	& \multicolumn{2}{|c|}{-----------------------}	& \multicolumn{2}{|c|}{-----------------------}& \multicolumn{2}{|c|}{-----------------------}& \multicolumn{2}{|c|}{-----------------------}& \multicolumn{2}{|c|}{-----------------------}& \multicolumn{2}{|c|}{-----------------------}\\
	& Median	& MAP	& Median	& MAP	& Median	& MAP	& Median	& MAP	& Median	& MAP	& Median	& MAP\\
\hline
\multicolumn{13}{|c|}{Sy1 galaxies}\\
\hline
MCG-06-30-015	&18$\pm^{3}_{2}$	&15	&52$\pm^{29}_{27}$	&29	&10$\pm^3$			&14 &2.4$\pm0.3$			&2.4	&120$\pm^{17}_{23}$		&131	&60$\pm^{5}_{6}$	&60	\\
NGC\,3227		&54$\pm^{6}_{5}$	&52	&18$\pm^{4}_{3}$	&15	&9$\pm^{3}_{2}$		&13 &0.3$\pm^{0.3}_{0.2}$	&0.1	&138$\pm^{7}_{10}$		&145	&15$\pm9$			&14	\\	
NGC\,3783		&43$\pm4$			&44	&24$\pm^{11}_{5}$	&20	&11$\pm^{2}_{3}$	&13 &0.4$\pm0.3$			&0.1	&132$\pm^{11}_{19}$		&149	&10$\pm^{10}_{6}$	&8		\\
NGC\,4051		&57$\pm^{7}_{4}$	&52	&19$\pm^{25}_{7}$	&10	&6$\pm^{2}_{1}$		&9 &1.1$\pm^{0.4}_{0.5}$	&0.5	&129$\pm^{13}_{15}$		&145	&16$\pm^{13}_{9}$	&13		\\
NGC\,4151		&47$\pm^{2}_{1}$	&46	&19$\pm3$			&19	&9$\pm2$		&10 &0.1$\pm^{0.3}_{0.1}$	&$<$0.1	&143$\pm^{4}_{8}$		&148	&5$\pm^{5}_{3}$		&1		\\
NGC\,6814		&39$\pm^{4}_{2}$	&37	&25$\pm^{29}_{9}$	&20	&7$\pm^{4}_{3}$		&15 &0.6$\pm^{1.0}_{0.4}$	&$<$0.1	&138$\pm^{8}_{14}$		&145	&16$\pm^{18}_{11}$	&7		\\
NGC\,7213		&16$\pm^{2}_{1}$	&15	&30$\pm^{12}_{7}$	&22	&12$\pm2$			&15 &0.3$\pm0.2$			&$<$0.1	&136$\pm^{9}_{17}$		&140	&12$\pm^{11}_{8}$	&4		\\
UGC\,6728		&44$\pm^{14}_{15}$	&62	&48$\pm^{29}_{25}$	&24	&6$\pm^{4}_{2}$		&7 &1.8$\pm^{0.4}_{0.5}$	&1.0	&46$\pm^{31}_{18}$		&21		&37$\pm^{26}_{22}$	&10		\\
\hline
\multicolumn{13}{|c|}{Sy1.8/1.9 galaxies}\\
\hline
NGC\,1365		&34$\pm^{5}_{3}$	&35	&28$\pm^{24}_{8}$		&22	&10$\pm^{3}_{2}$	&12	&0.9$\pm^{0.5}_{0.4}$	&0.5	&134$\pm^{9}_{15}$	&149	&31$\pm^{9}_{16}$		&30	\\	
NGC\,2992		&32$\pm^{2}_{3}$	&35	&31$\pm^{16}_{8}$		&27	&14$\pm^1$	&15	&1.4$\pm^{0.2}_{0.3}$	&1.0	&135$\pm^{9}_{14}$	&138	&58$\pm^{5}_{4}$		&51	\\	
NGC\,4138		&23$\pm^{9}_{5}$	&19	&40$\pm^{33}_{20}$		&72	&7$\pm^{4}_{3}$		&5	&1.5$\pm^{0.4}_{0.6}$	&1.5	&115$\pm^{21}_{32}$	&150	&48$\pm^{11}_{19}$		&57	\\	
NGC\,4395		&46$\pm^{12}_{10}$	&67	&34$\pm^{26}_{11}$		&20	&11$\pm3$			&8&1.1$\pm^{0.5}_{0.6}$	&0.1	&137$\pm^{8}_{14}$	&146	&42$\pm16$			&20	\\	
NGC\,5506		&54$\pm1$			&55	&21$\pm^{3}_{2}$		&20	&14$\pm1$		&15	&0.01$\pm0.1$			&$<$0.1	&65$\pm^{10}_{8}$	&66		&5$\pm^{5}_{3}$			&1.0 	\\	
NGC\,7314		&51$\pm^{6}_{15}$	&53	&23$\pm^{15}_{4}$		&19	&11$\pm^{2}_{3}$		&13	&0.4$\pm^{0.6}_{0.3}$	&$<$0.1	&131$\pm^{12}_{17}$	&146	&14$\pm^{23}_{9}$		&3		\\	
\hline
\multicolumn{13}{|c|}{Sy2 galaxies}\\
\hline
ESO\,005-G004	&60$\pm^{6}_{7}$	&54	&19$\pm^{3}_{2}$		&15	&13$\pm^{1}_{2}$	&14	&0.6$\pm0.4$			&0.2	&52$\pm^{13}_{10}$	&42		&70$\pm^{12}_{12}$		&79	\\
MCG-05-23-016	&50$\pm^{10}_{9}$	&45	&20$\pm^{23}_{7}$		&13	&10$\pm3$			&12&1.5$\pm^{0.4}_{0.5}$	&0.9	&93$\pm^{20}_{19}$	&81		&31$\pm17$		&40	\\	
NGC\,2110		&40$\pm^{7}_{4}$	&43	&24$\pm^{29}_{10}$		&11	&7$\pm^{3}_{2}$		&13	&1.3$\pm^{0.5}_{0.6}$	&0.2	&119$\pm^{17}_{24}$	&134	&25$\pm^{12}_{13}$		&17	\\	
NGC\,3081		&62$\pm^{4}_{6}$	&64	&61$\pm^{22}_{24}$		&94	&11$\pm^{2}_{3}$	&14	&2.3$\pm0.1$			&2.2	&111$\pm^{22}_{21}$	&98		&37$\pm^{29}_{23}$		&18	\\	
NGC\,4388		&69$\pm1$			&70	&26$\pm2$				&25	&14$\pm1$			&15	&0.1$\pm^{0.2}_{0.1}$	&$<$0.1	&72$\pm4$			&76		&19$\pm^{1}_{2}$		&16	\\	
NGC\,4945		&63$\pm^{4}_{6}$	&65	&36$\pm^{15}_{6}$		&28	&13$\pm^{1}_{2}$		&12	&0.6$\pm^{0.5}_{0.4}$	&$<$0.1	&33$\pm^{9}_{7}$	&30		&72$\pm^{11}_{14}$			&75		\\	
NGC\,5128		&62$\pm^{4}_{5}$	&67	&17$\pm^{3}_{2}$		&15	&10$\pm2$		&9	&0.2$\pm^{0.3}_{0.2}$	&$<$0.1	&63$\pm^{6}_{7}$	&66		&14$\pm^{7}_{8}$			&8		\\	
NGC\,6300		&69$\pm1$			&70	&49$\pm^{11}_{9}$		&50	&14$\pm1$		&15&0.5$\pm^{0.1}_{0.2}$	&0.5	&75$\pm^{7}_{8}$	&70		&18$\pm^{3}_{4}$		&15		\\	
NGC\,7172		&63$\pm^{4}_{6}$	&67	&13$\pm^{4}_{2}$		&13	&13$\pm^{1}_{2}$		&15	&0.5$\pm^{0.5}_{0.3}$	&0.6	&27$\pm^{11}_{7}$	&20		&13$\pm^{17}_{9}$		&2		\\	
NGC\,7582		&49$\pm^{3}_{5}$	&43	&62$\pm^{19}_{15}$		&69	&12$\pm^{1}_{2}$		&14	&1.0$\pm^{0.3}_{0.4}$	&1.1	&21$\pm^{7}_{5}$	&15		&73$\pm^{9}_{13}$		&87		\\	
\hline
\end{tabular}					 
\caption{Clumpy torus model parameters derived from the fits of Sy1, Sy1.8/1.9, and Sy2 galaxies. Median values of each posterior distribution are listed with their corresponding $\pm$1$\sigma$ values around the median.}
\label{tabA1}
\end{table*}

\begin{figure*}
\begin{center}
    \subfigure[MGC-06-30-015]{\includegraphics[width=7.5cm]{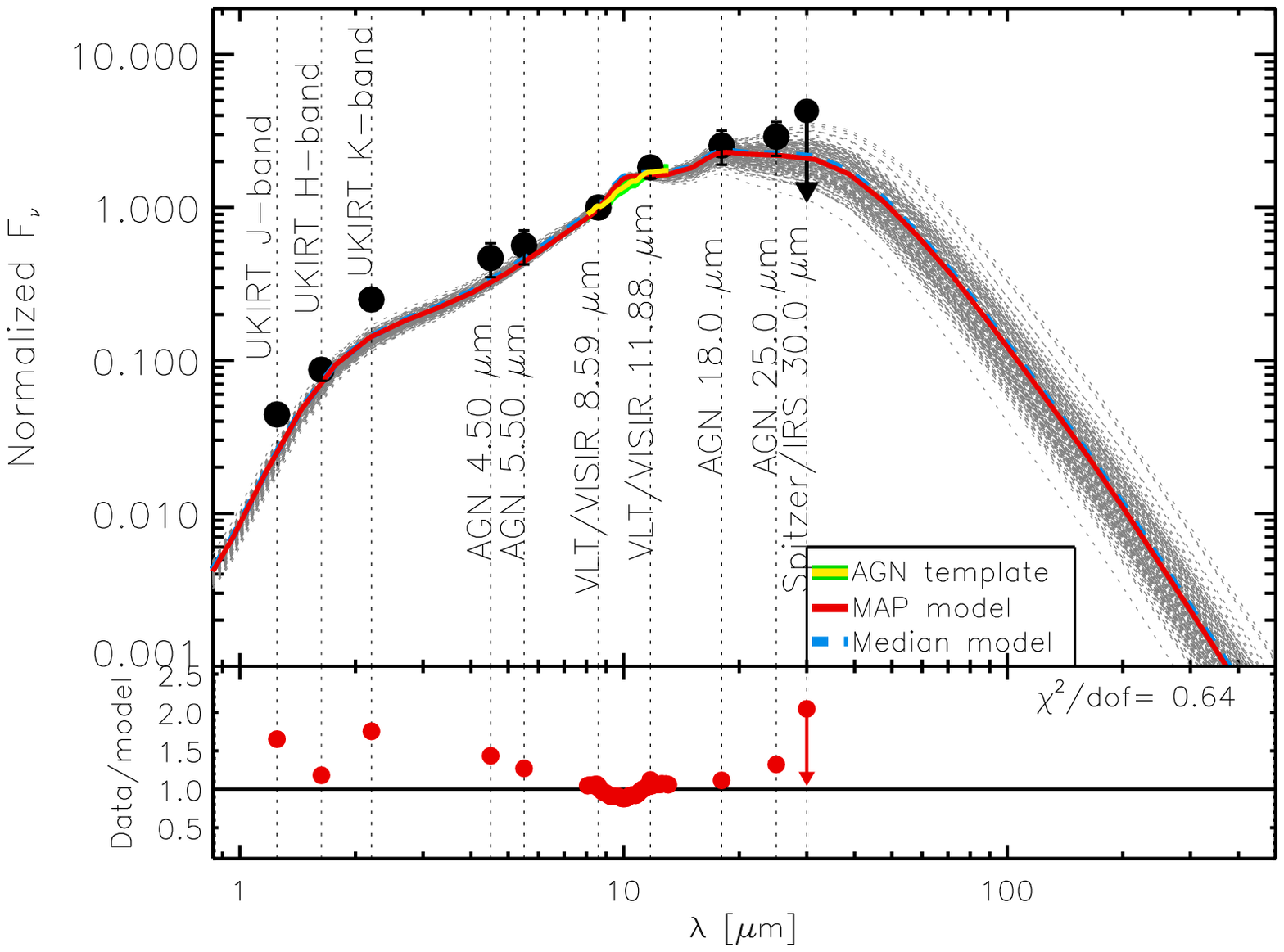}}    
    \subfigure[NGC\,3227]{\includegraphics[width=7.5cm]{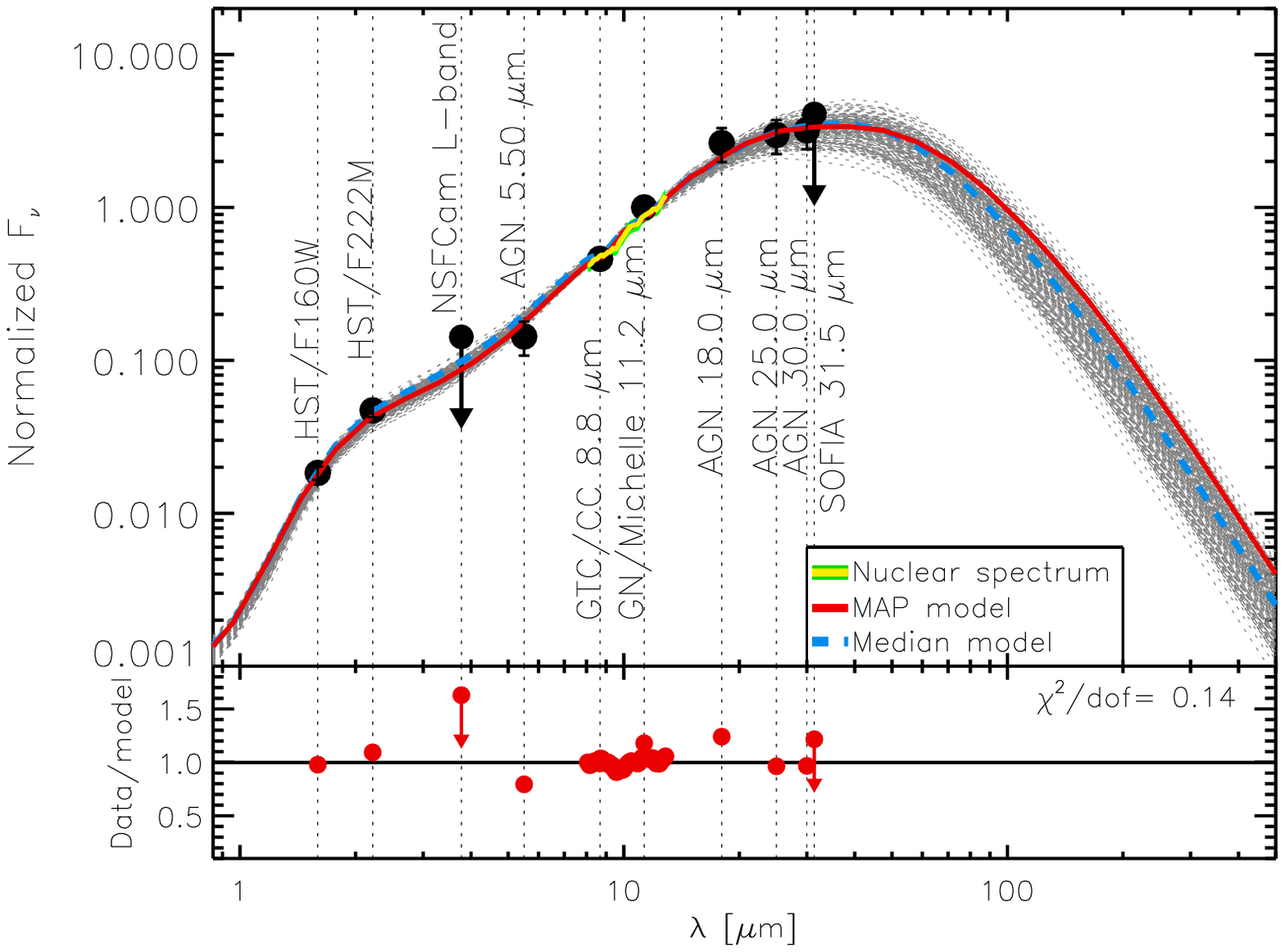}}        
    \subfigure[NGC\,3783]{\includegraphics[width=7.5cm]{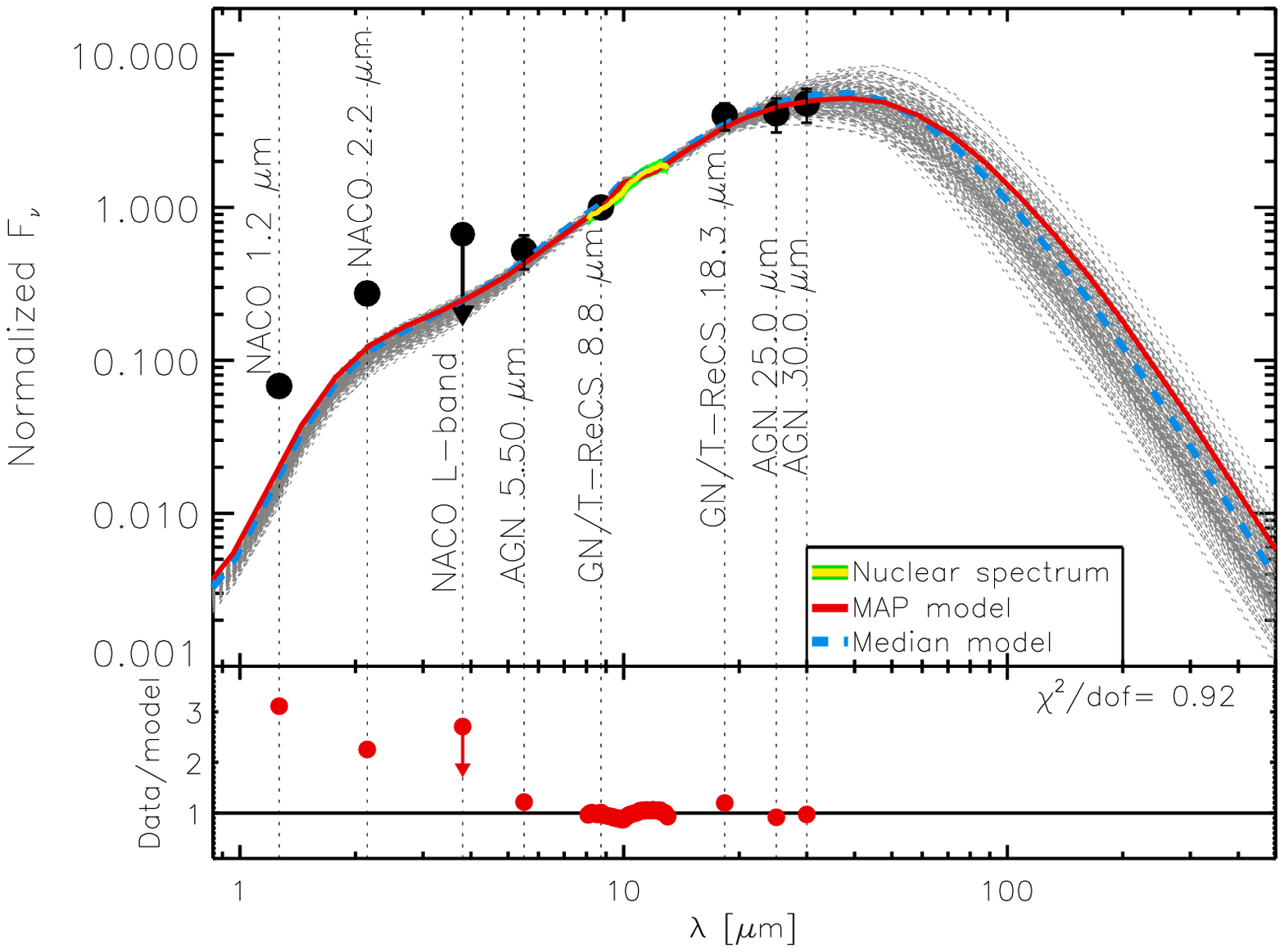}}    
    \subfigure[NGC\,4051]{\includegraphics[width=7.5cm]{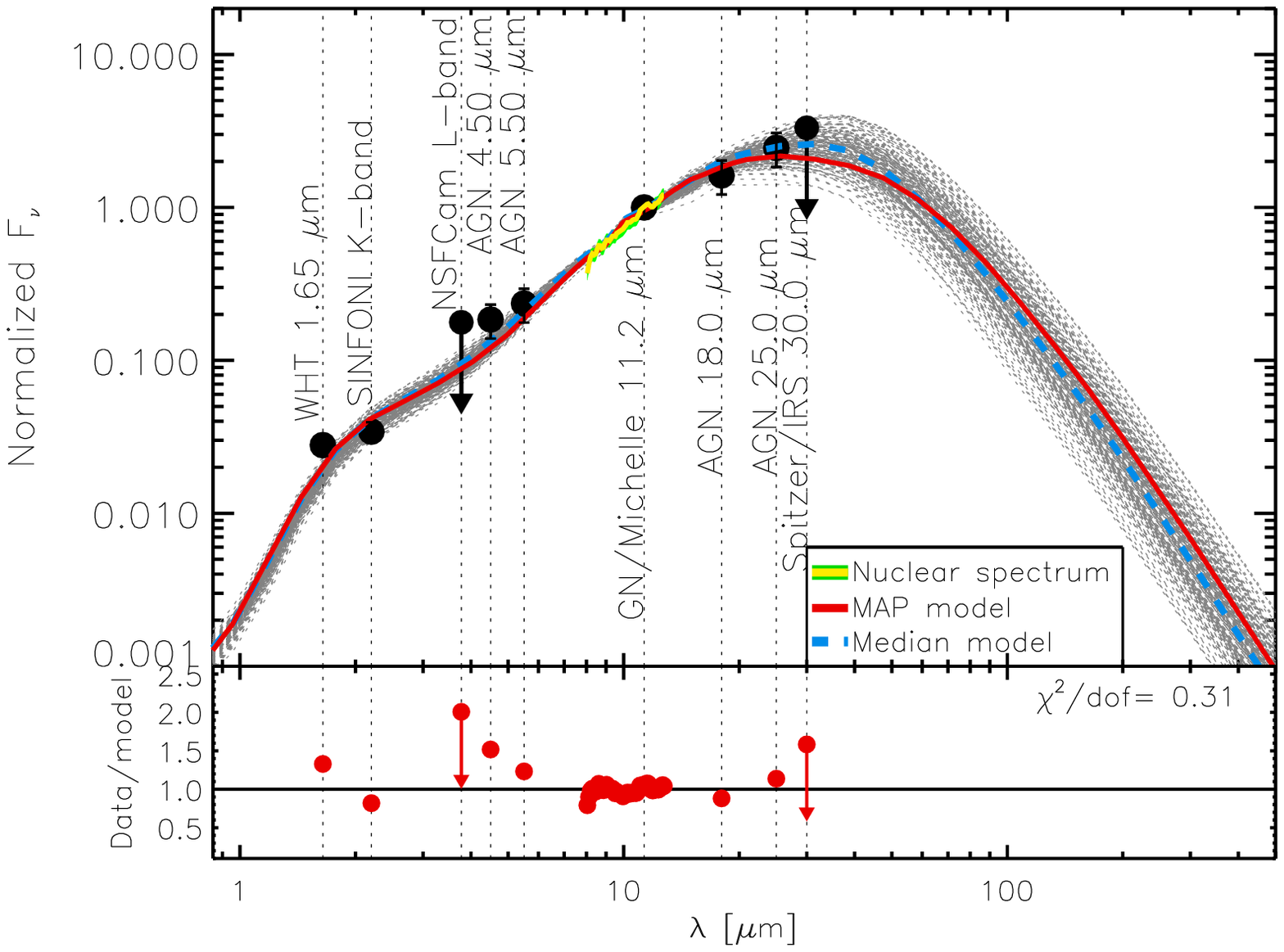}}
    \subfigure[NGC\,4151]{\includegraphics[width=7.5cm]{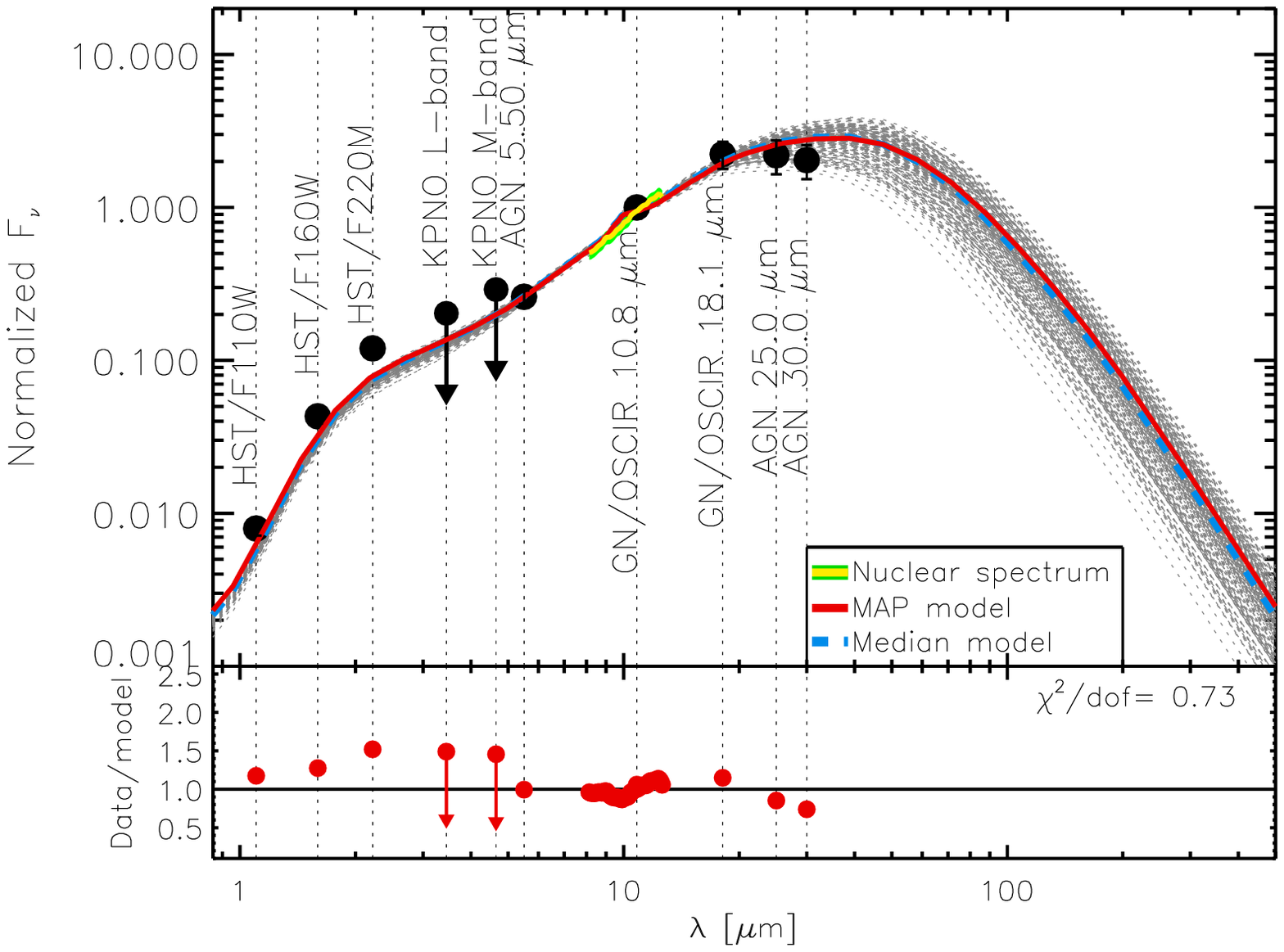}}
    \subfigure[NGC\,6814]{\includegraphics[width=7.5cm]{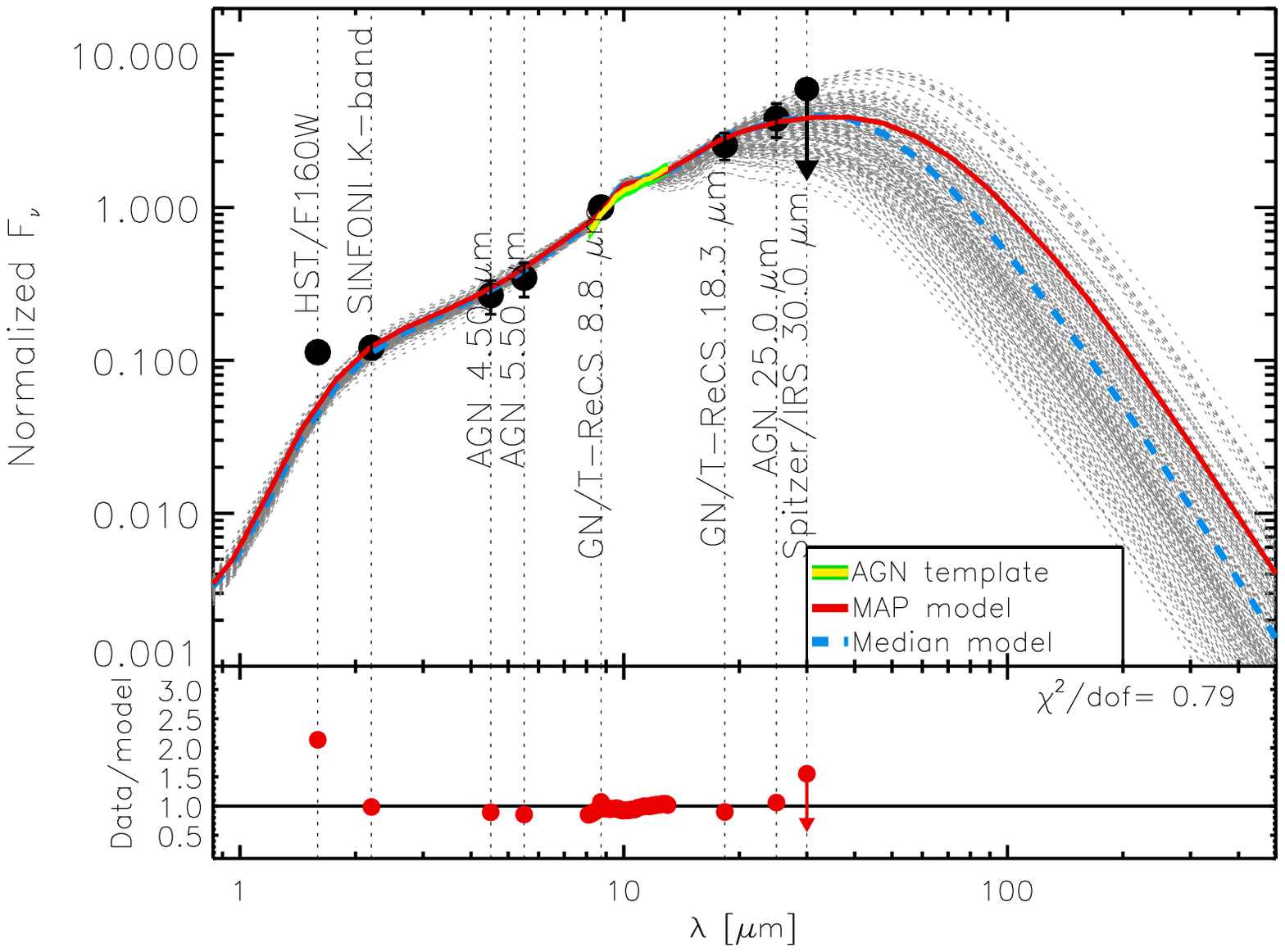}}
    \subfigure[NGC\,7213]{\includegraphics[width=7.5cm]{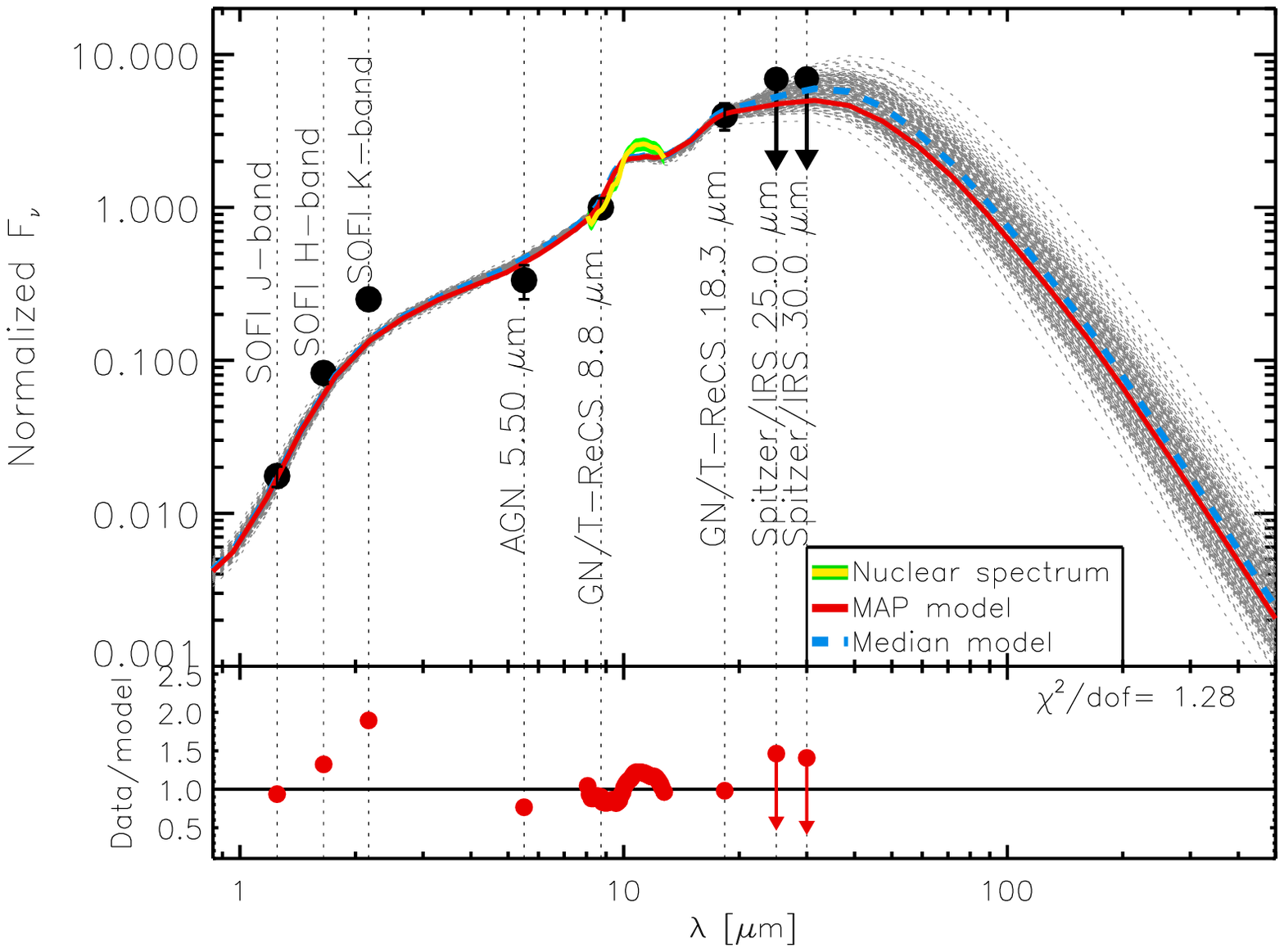}}
    \subfigure[UGC\,6728]{\includegraphics[width=7.5cm]{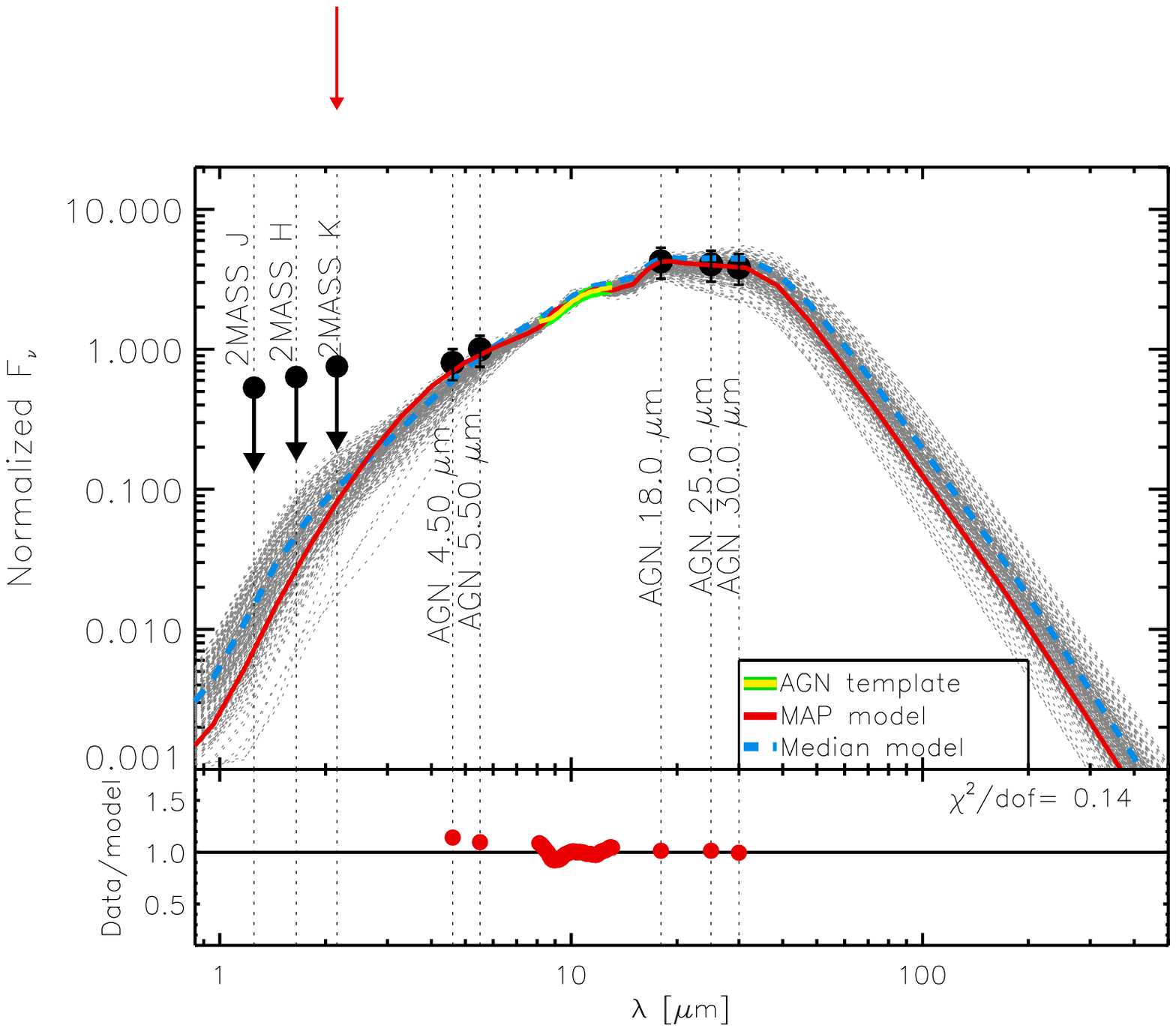}}
\end{center}
\caption[Short caption.] {\label{fig:figure_label} Nuclear IR SEDs of the Sy1 galaxies in the sample normalized at 11.2~$\mu$m. Solid red and dashed blue lines correspond to the MAP and median models
respectively. Grey curves are the clumpy models sampled from the posterior and compatible with the data at 1$\sigma$ level.}
\end{figure*}

 \begin{figure*}
\begin{center}
	 \subfigure[NGC\,1365]{\includegraphics[width=7.5cm]{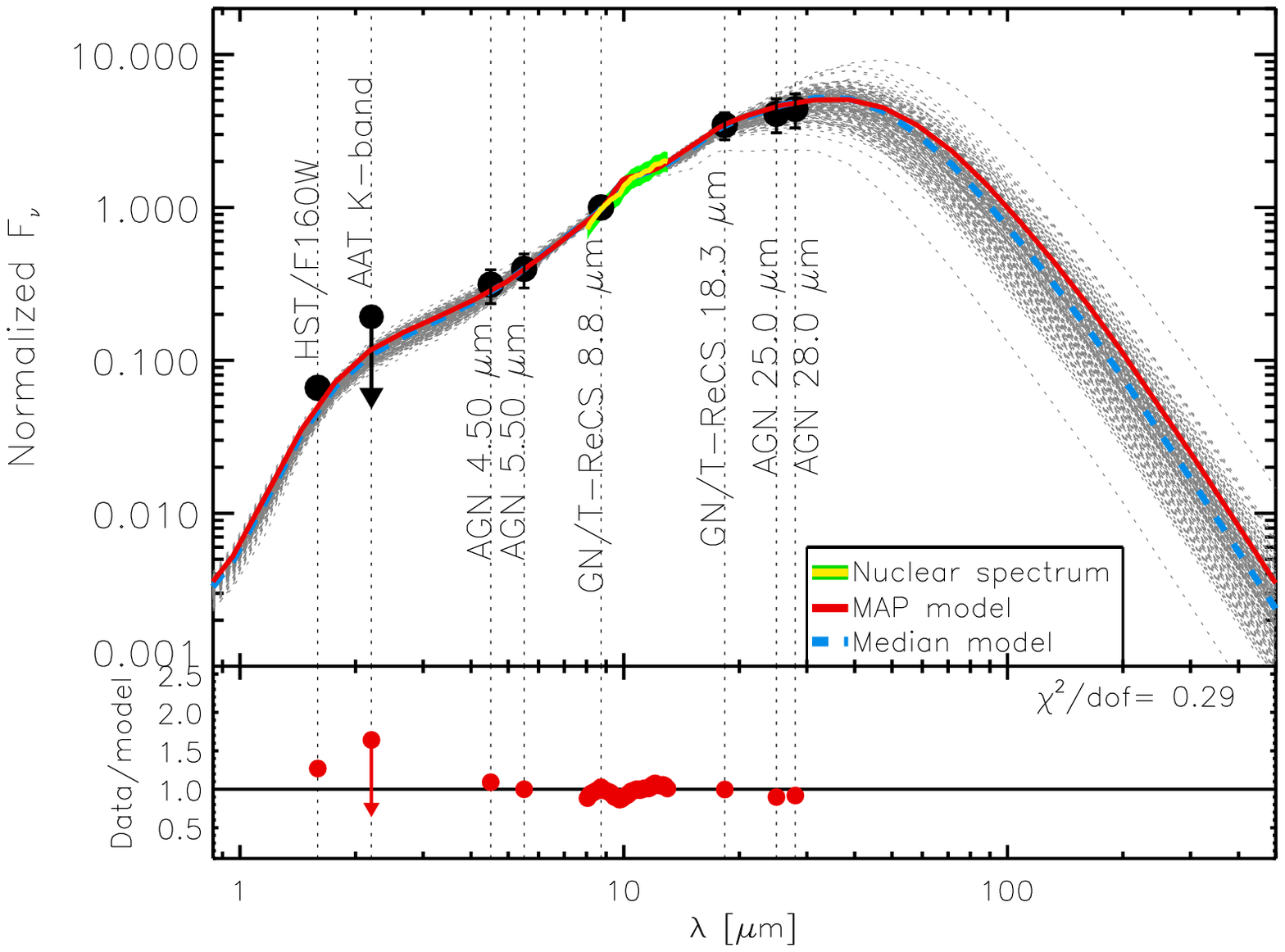}}
    \subfigure[NGC\,2992]{\includegraphics[width=7.5cm]{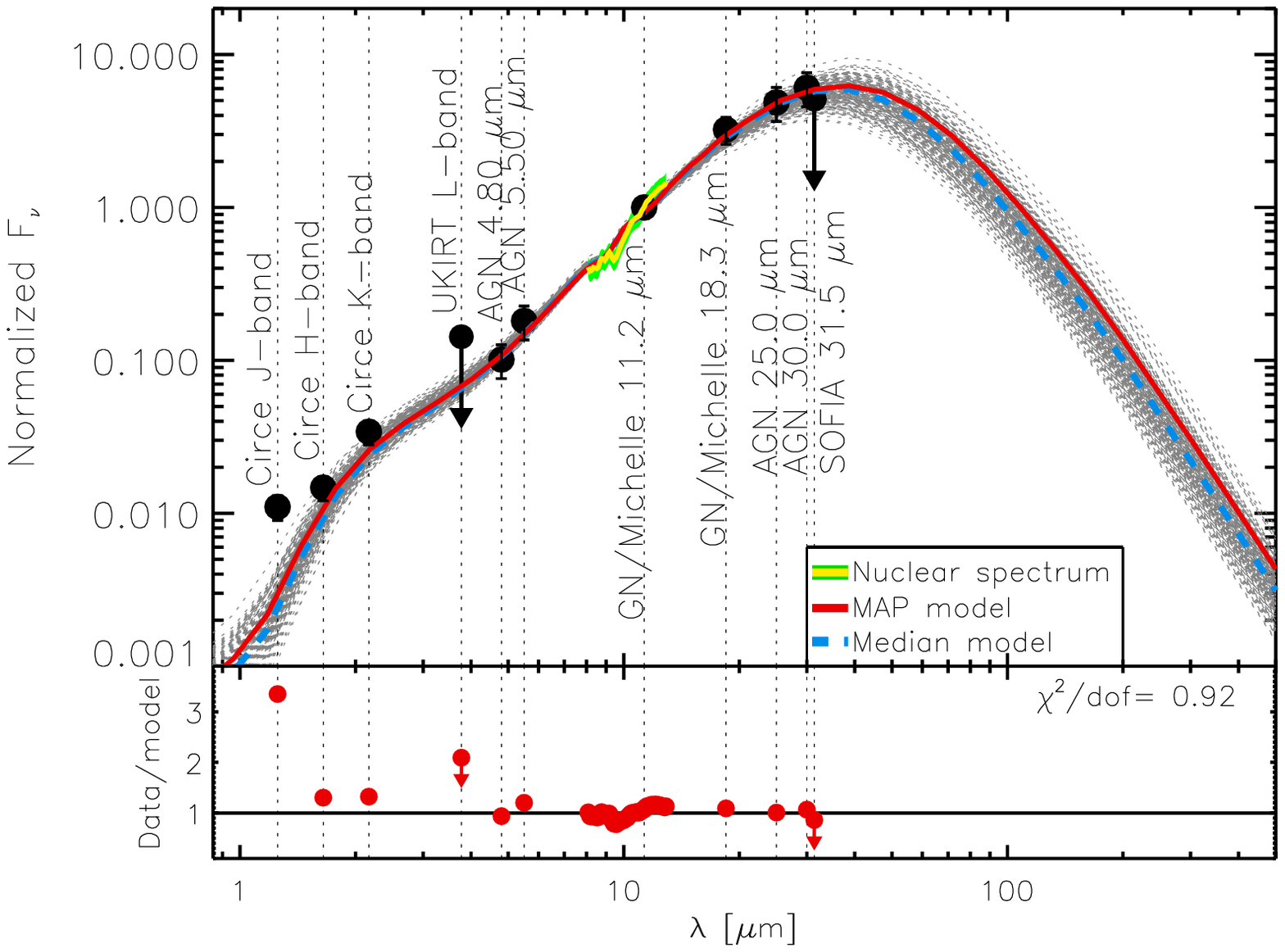}}
    \subfigure[NGC\,4138]{\includegraphics[width=7.5cm]{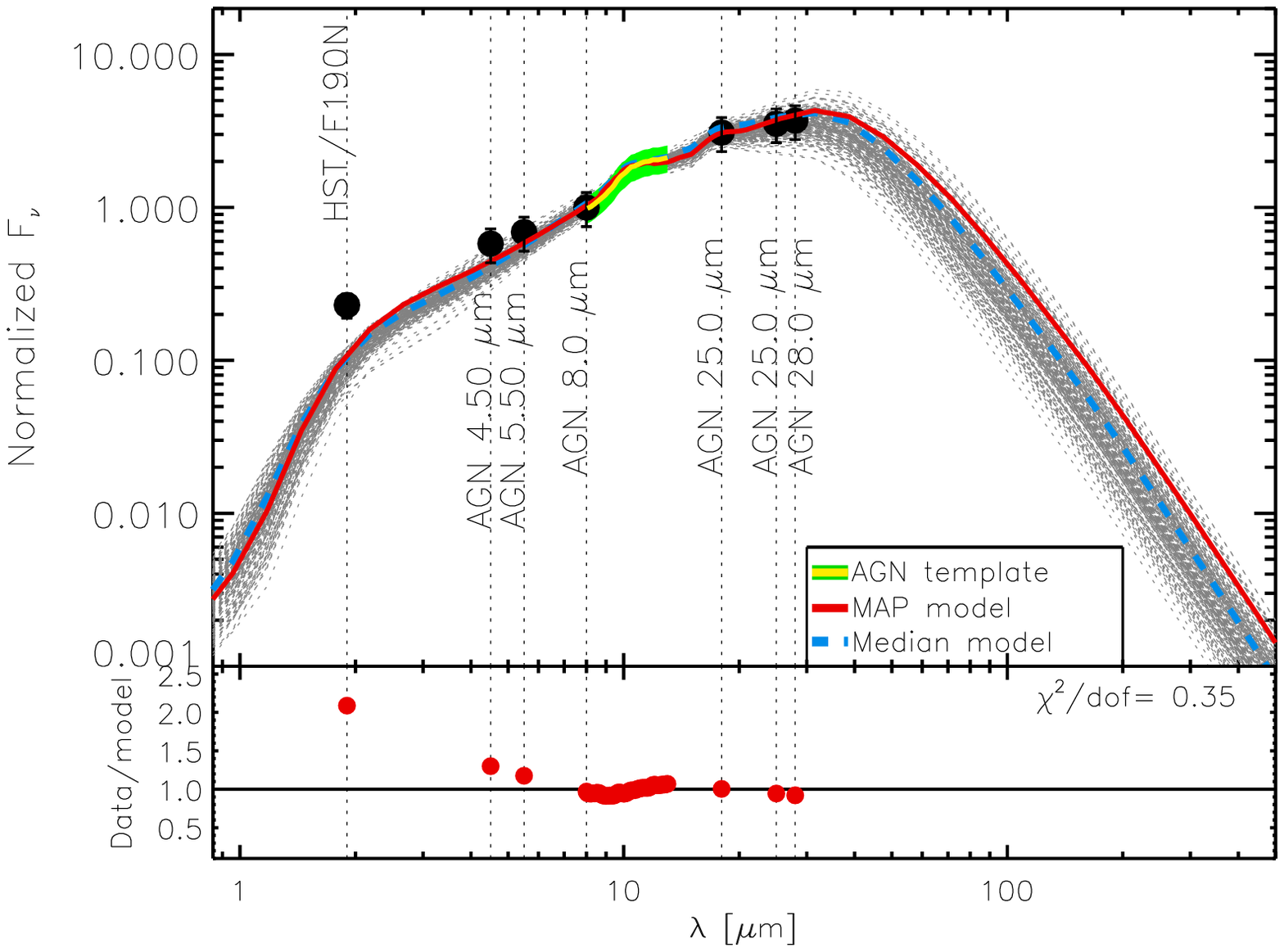}}
    \subfigure[NGC\,4395]{\includegraphics[width=7.5cm]{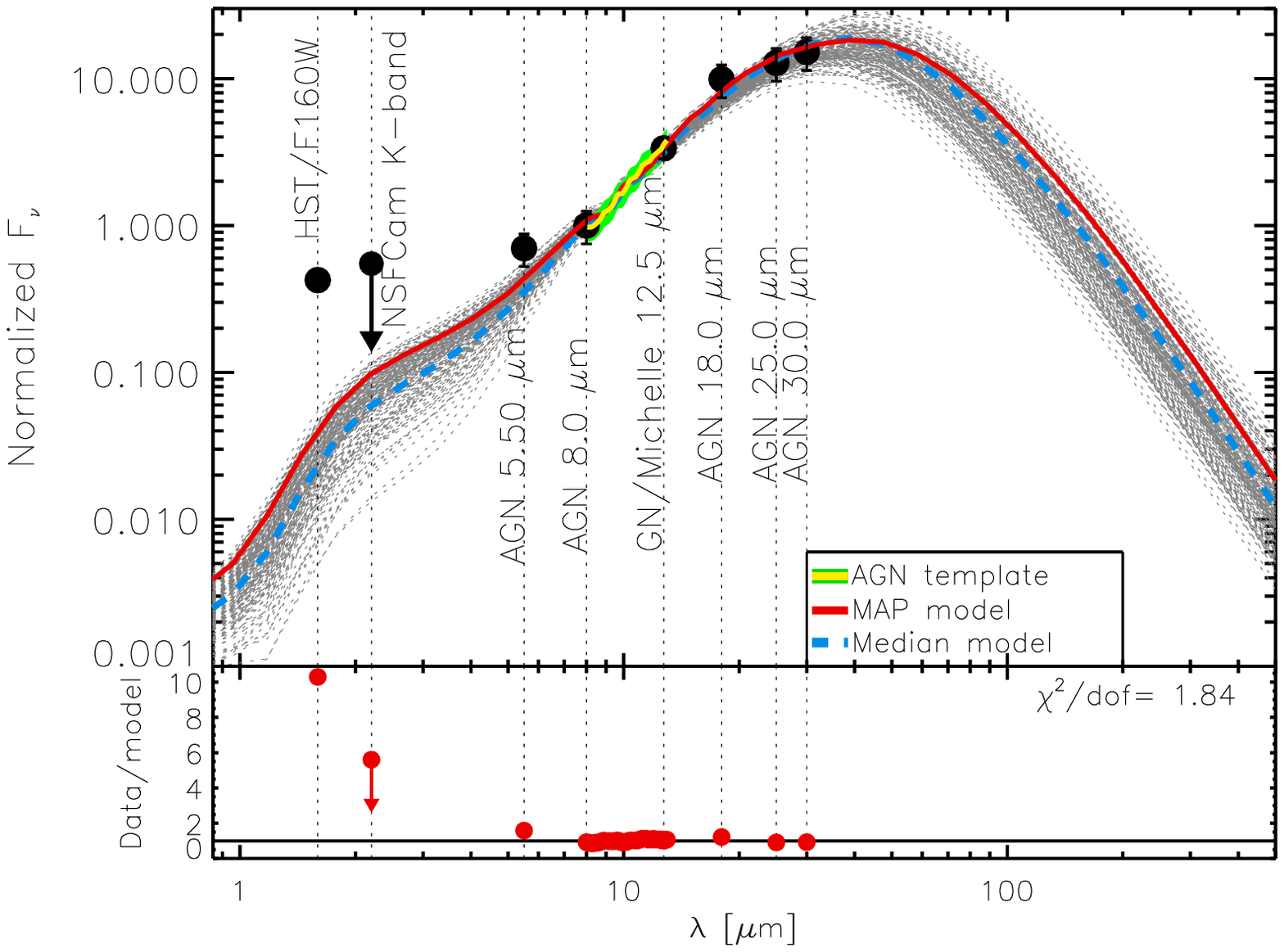}}
    \subfigure[NGC\,5506]{\includegraphics[width=7.5cm]{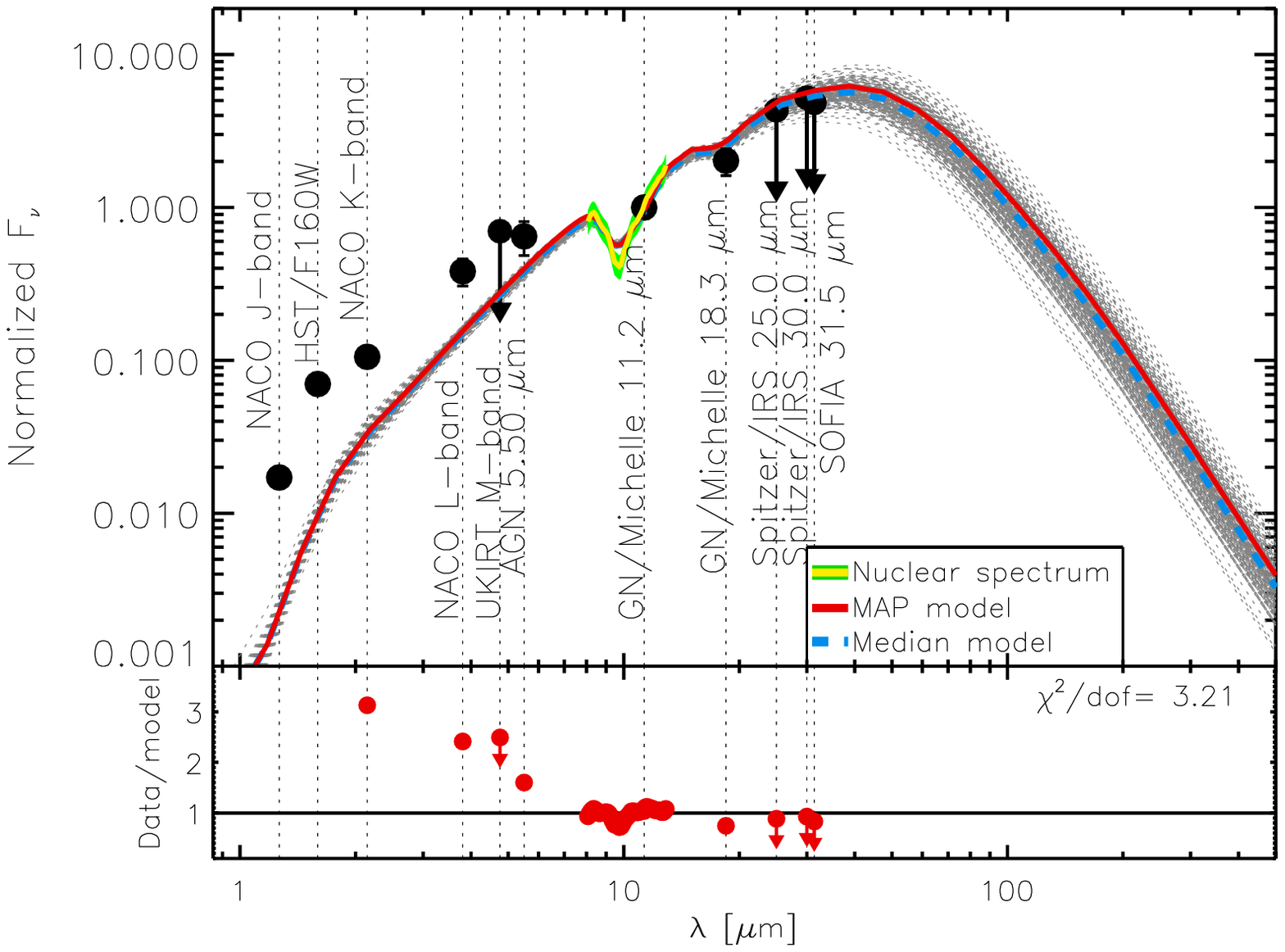}}
    \subfigure[NGC\,7314]{\includegraphics[width=7.5cm]{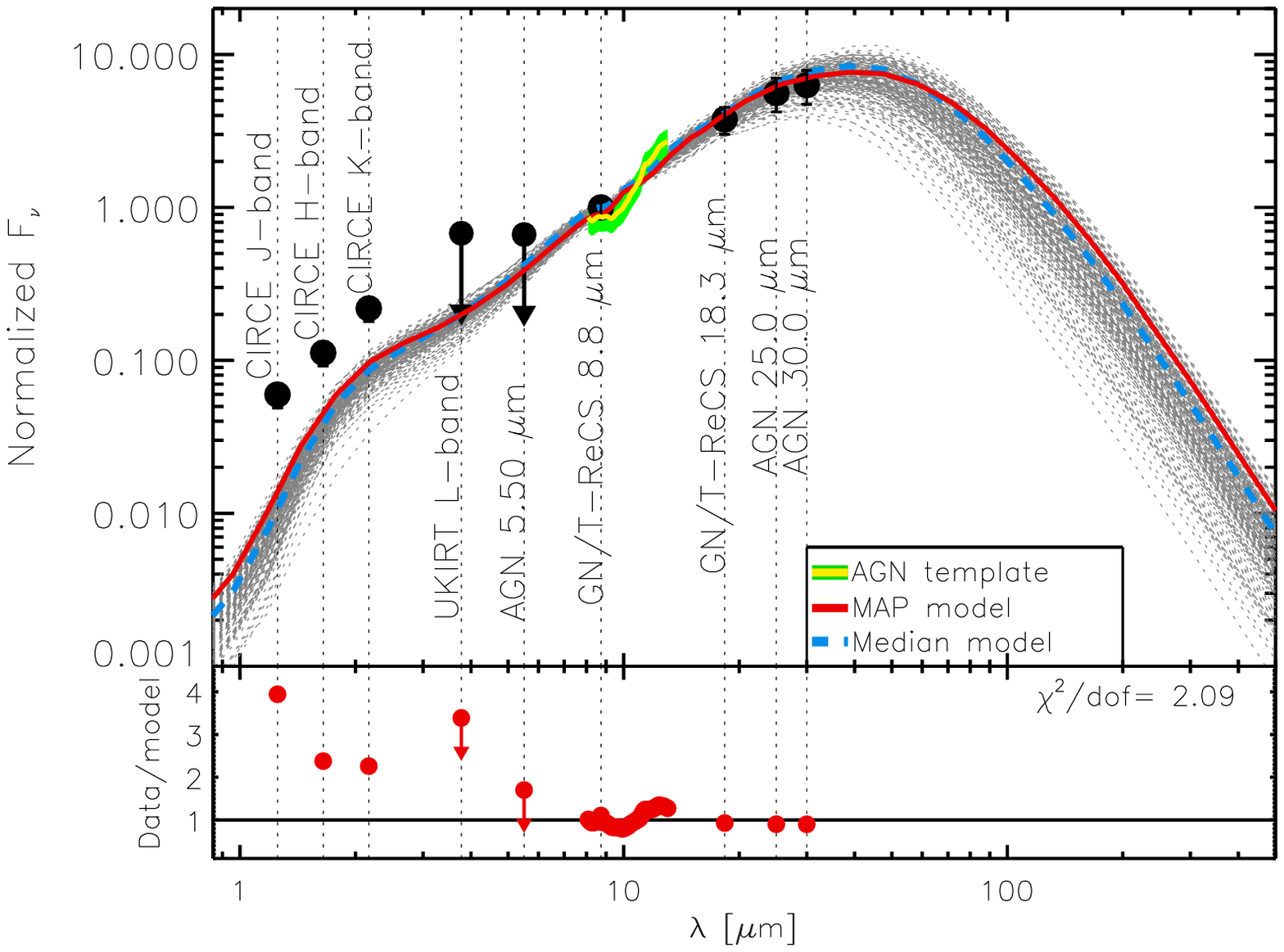}}
\end{center}
\caption[Short caption.] {\label{fig:figure_label} Nuclear IR SED of the Sy1.8/1.9 galaxies in the sample normalized at 11.2~$\mu$m. Solid red and dashed blue lines correspond to the MAP and median models
respectively. Grey curves are the clumpy models sampled from the posterior and compatible with the data at 1$\sigma$ level. A}
\end{figure*}

 \begin{figure*}
\begin{center}
    \subfigure[ESO\,005-G004]{\includegraphics[width=7.5cm]{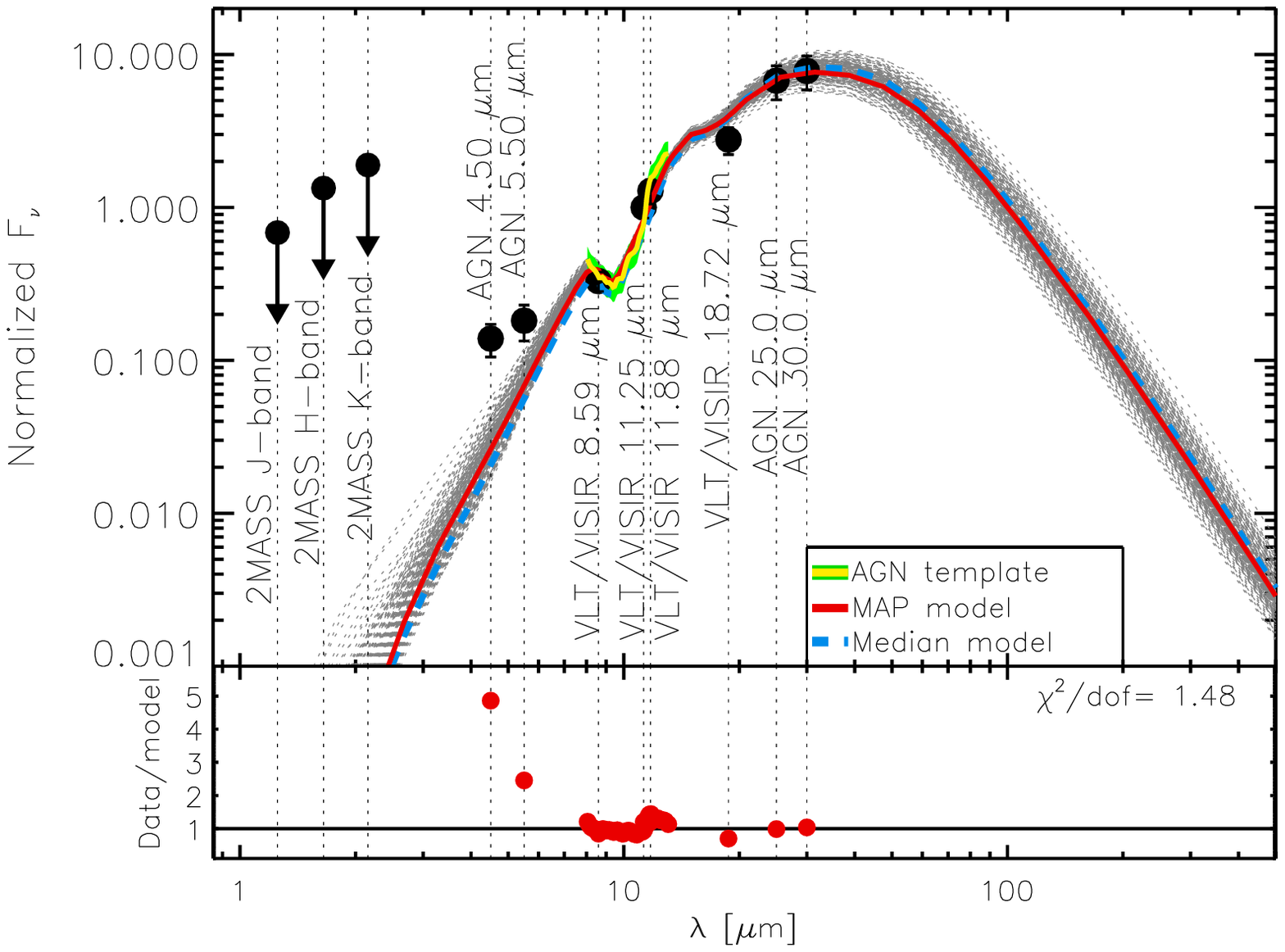}}    
    \subfigure[MGC-05-23-016]{\includegraphics[width=7.5cm]{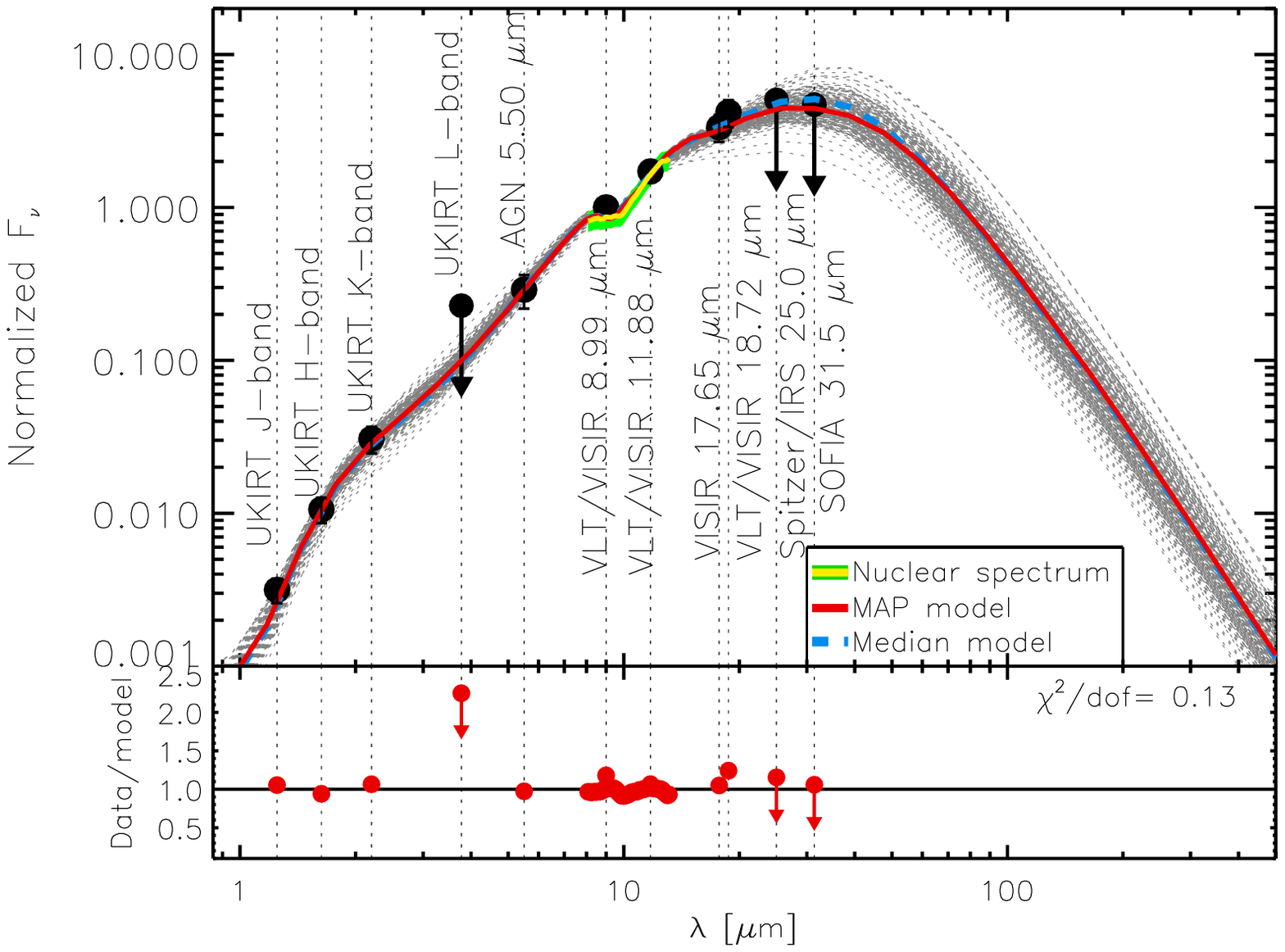}}      
    \subfigure[NGC\,2110]{\includegraphics[width=7.5cm]{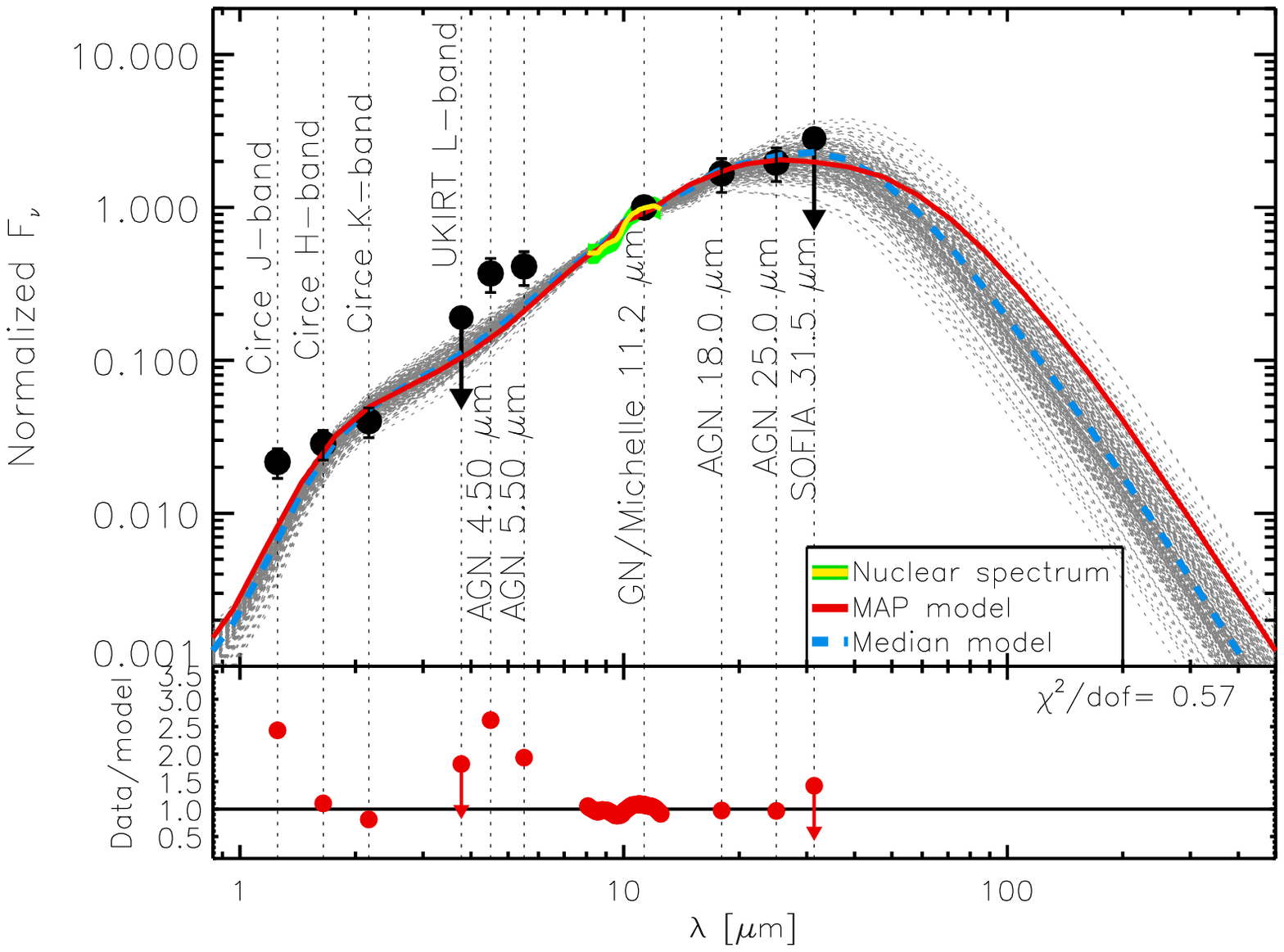}}
    \subfigure[NGC\,3081]{\includegraphics[width=7.5cm]{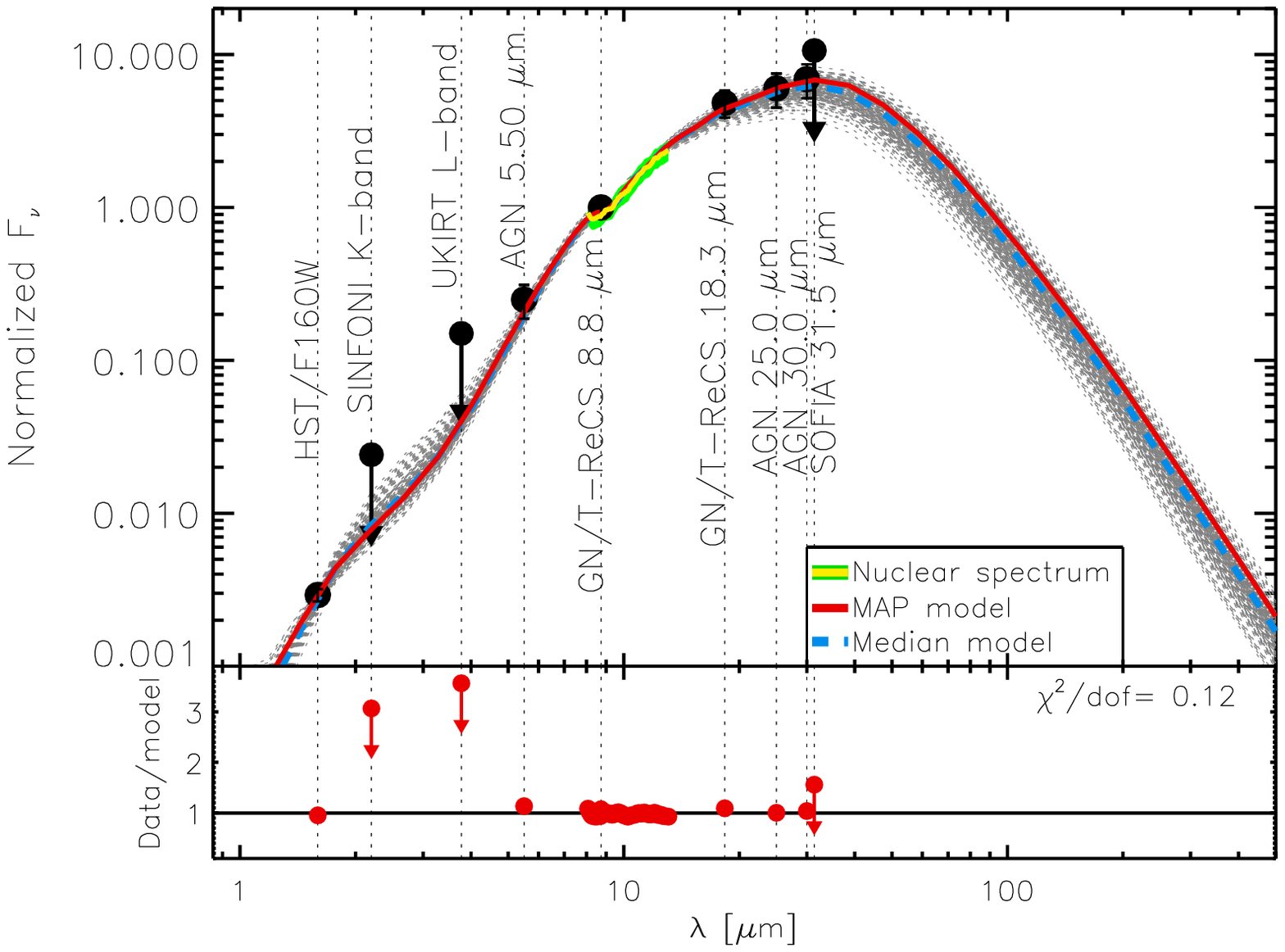}}
    \subfigure[NGC\,4388]{\includegraphics[width=7.5cm]{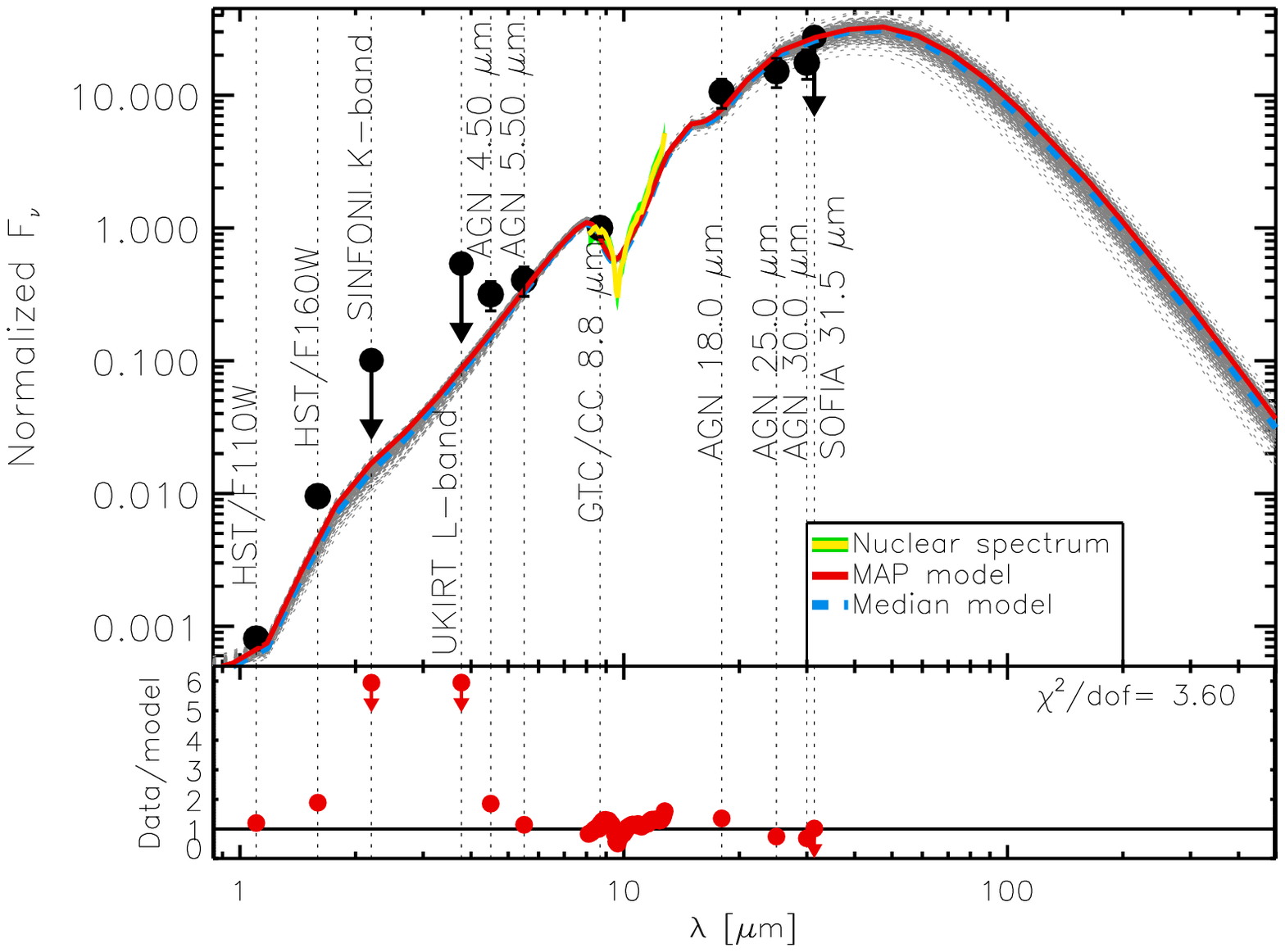}}
    \subfigure[NGC\,4945]{\includegraphics[width=7.5cm]{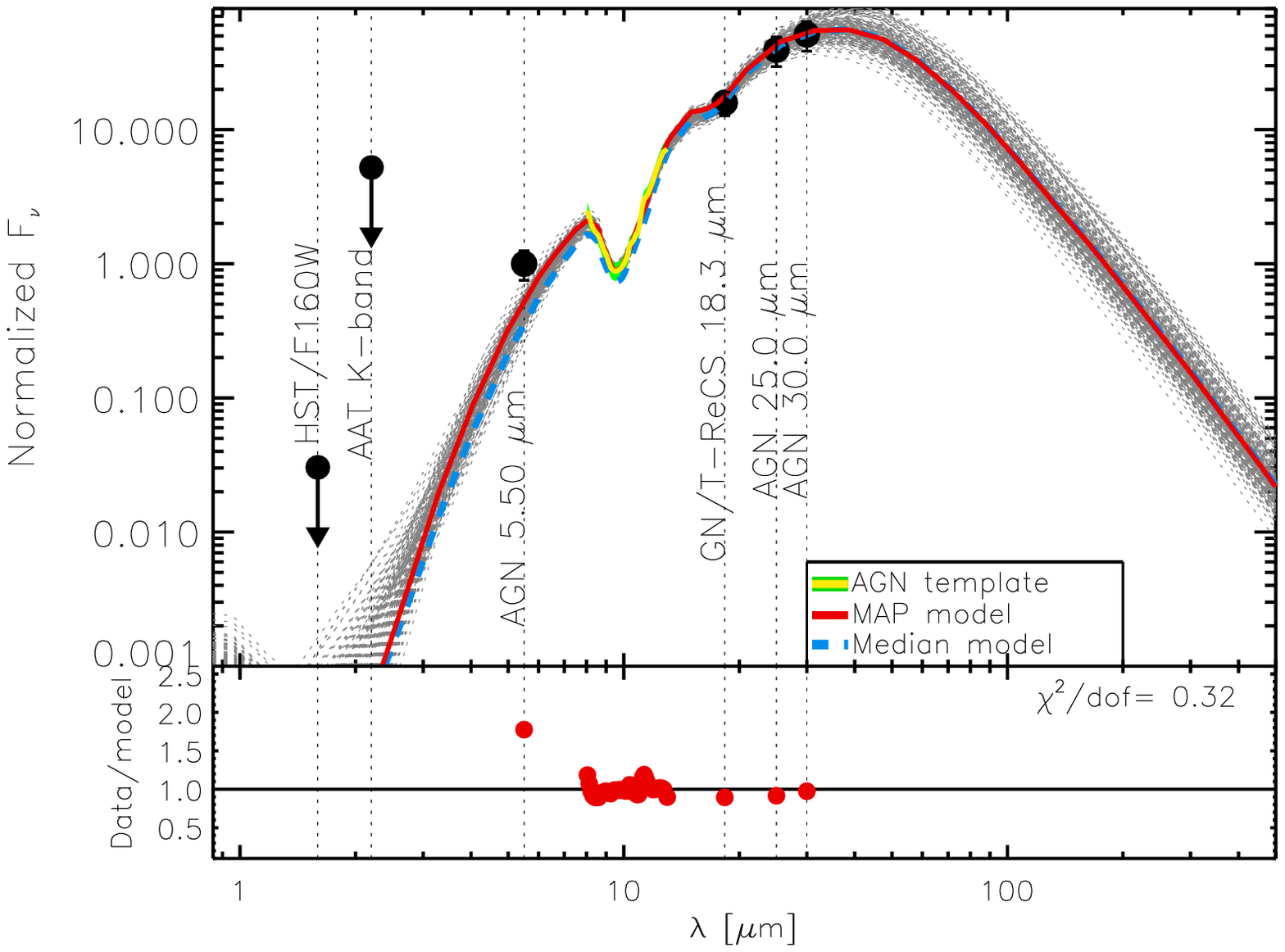}}
    \subfigure[NGC\,5128]{\includegraphics[width=7.5cm]{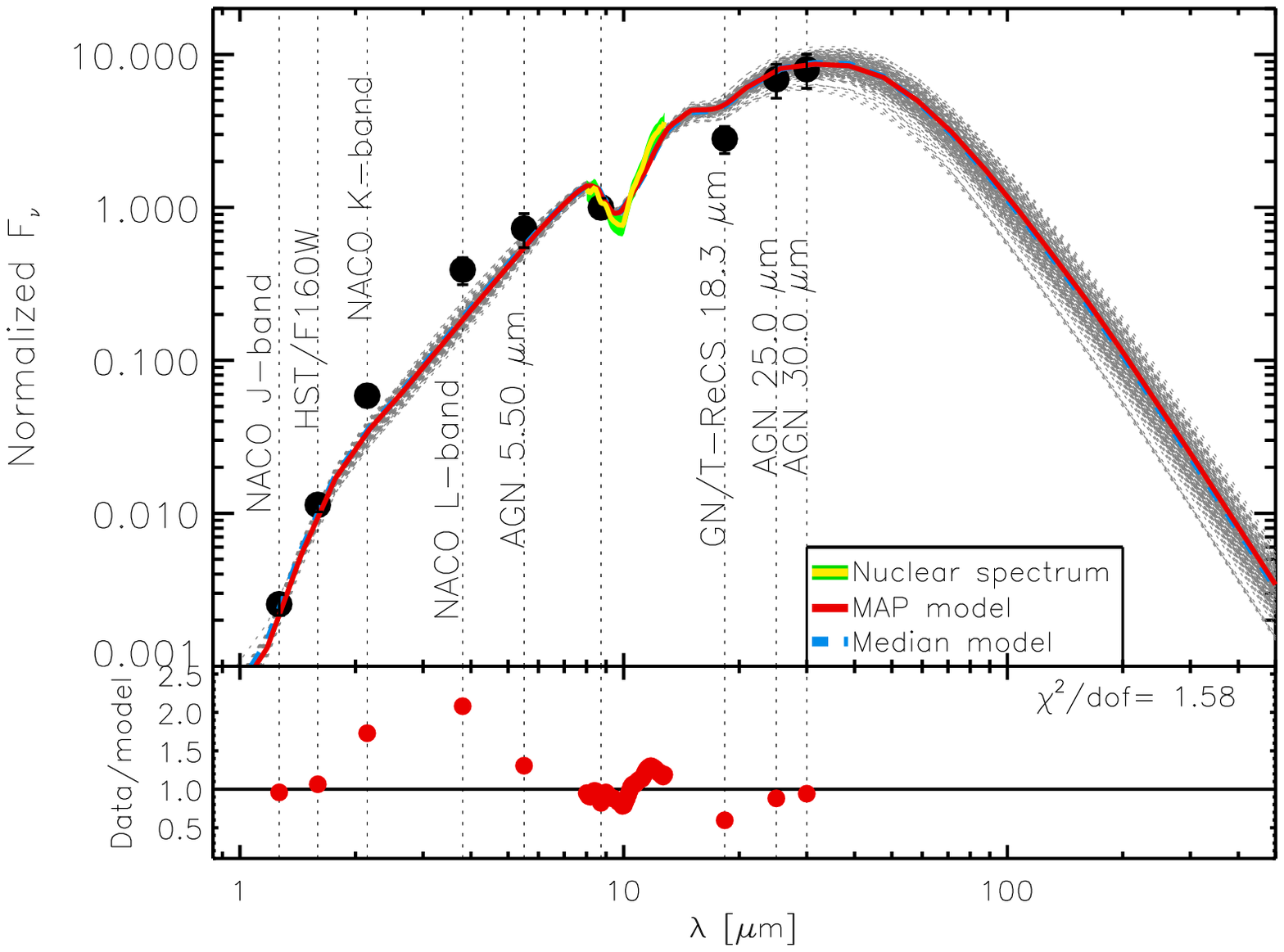}}
    \subfigure[NGC\,6300]{\includegraphics[width=7.5cm]{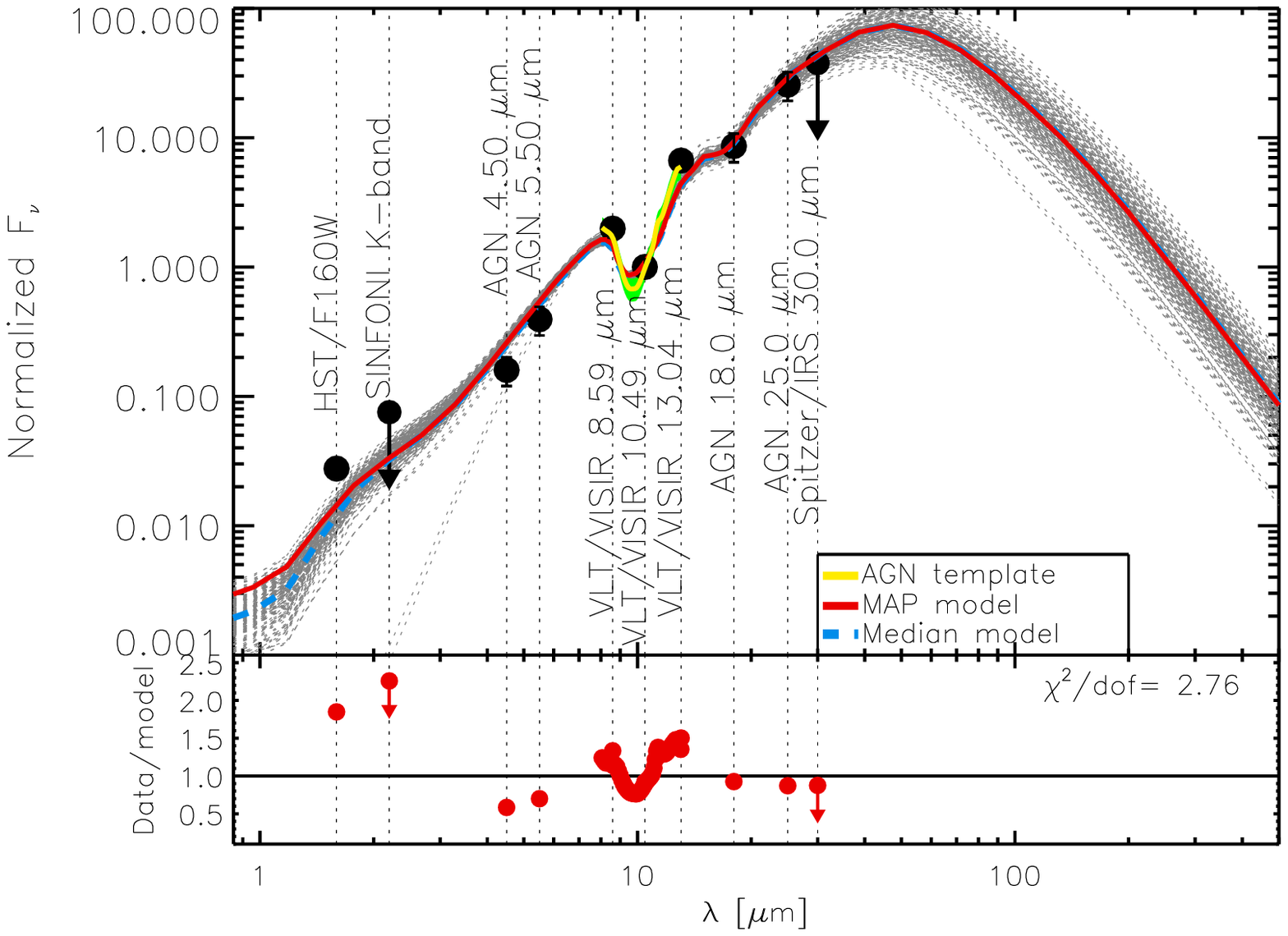}}

\end{center}
\caption[Short caption.] {\label{fig:figure_label} Nuclear IR SED of the Sy2 galaxies in the sample normalized at 11.2~$\mu$m. Solid red and dashed blue lines correspond to the MAP and median models
respectively. Grey curves are the clumpy models sampled from the posterior and compatible with the data at 1$\sigma$ level. A}
\end{figure*}   

 \begin{figure*}
\begin{center} 
    \subfigure[NGC\,7172]{\includegraphics[width=7.5cm]{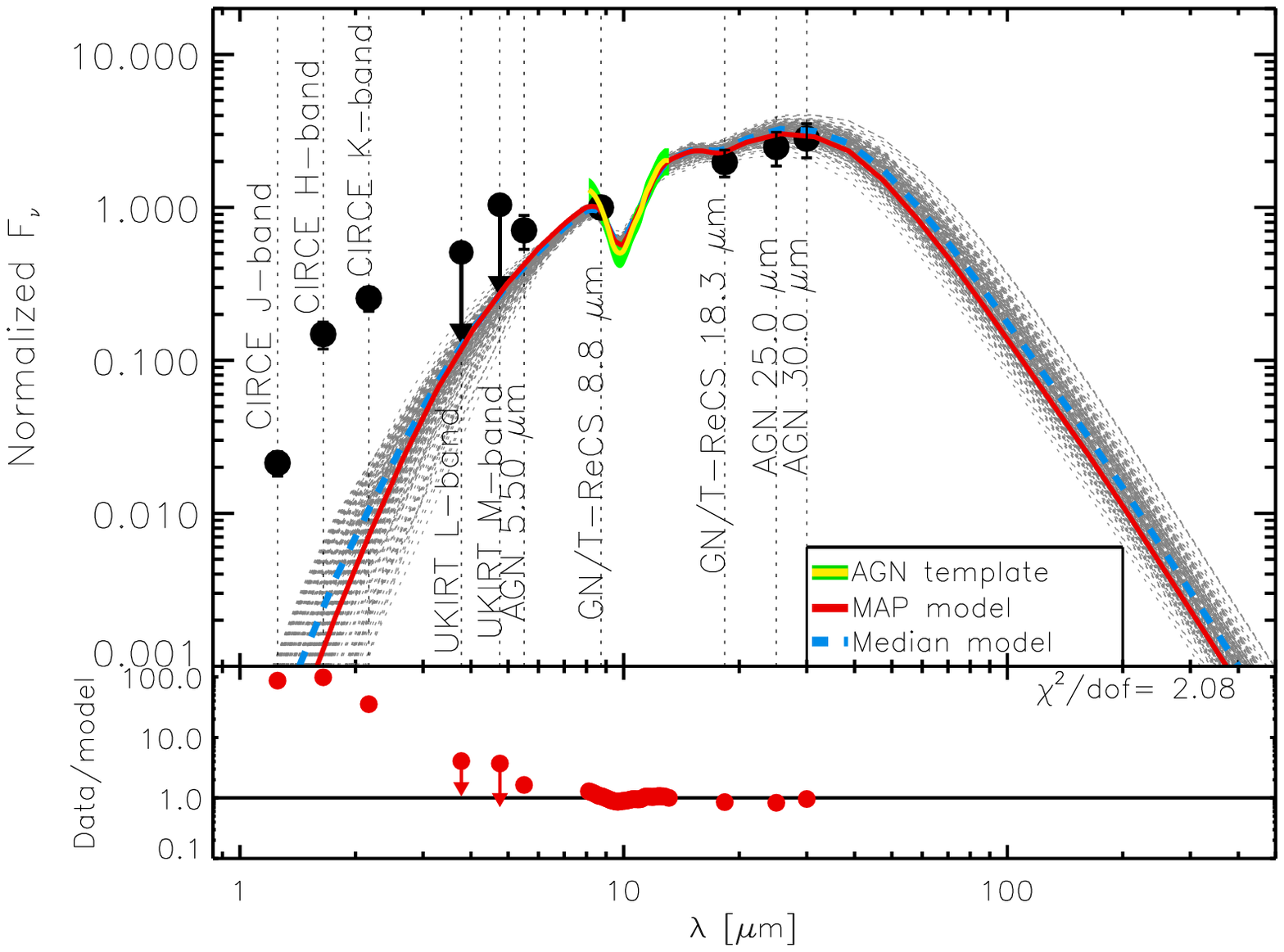}}
    \subfigure[NGC\,7582]{\includegraphics[width=7.5cm]{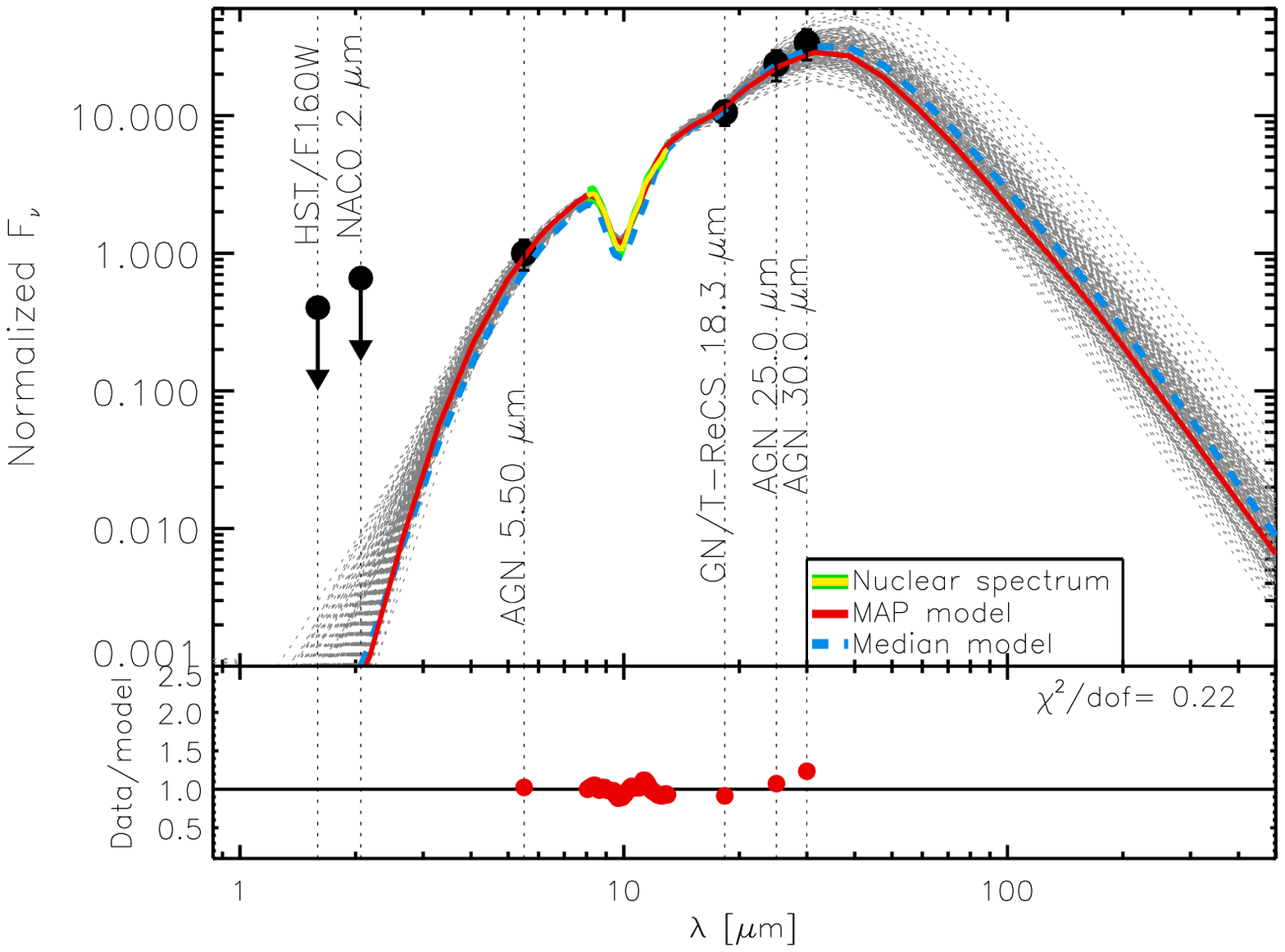}}
\end{center}
\caption[Short caption.] {\label{fig:figure_label} Nuclear IR SED of the Sy2 galaxies in the sample normalized at 11.2~$\mu$m. Solid red and dashed blue lines correspond to the MAP and median models
respectively. Grey curves are the clumpy models sampled from the posterior and compatible with the data at 1$\sigma$ level. A}
\end{figure*}  

\begin{figure*}
\centering
\par{
\includegraphics[width=7.82cm]{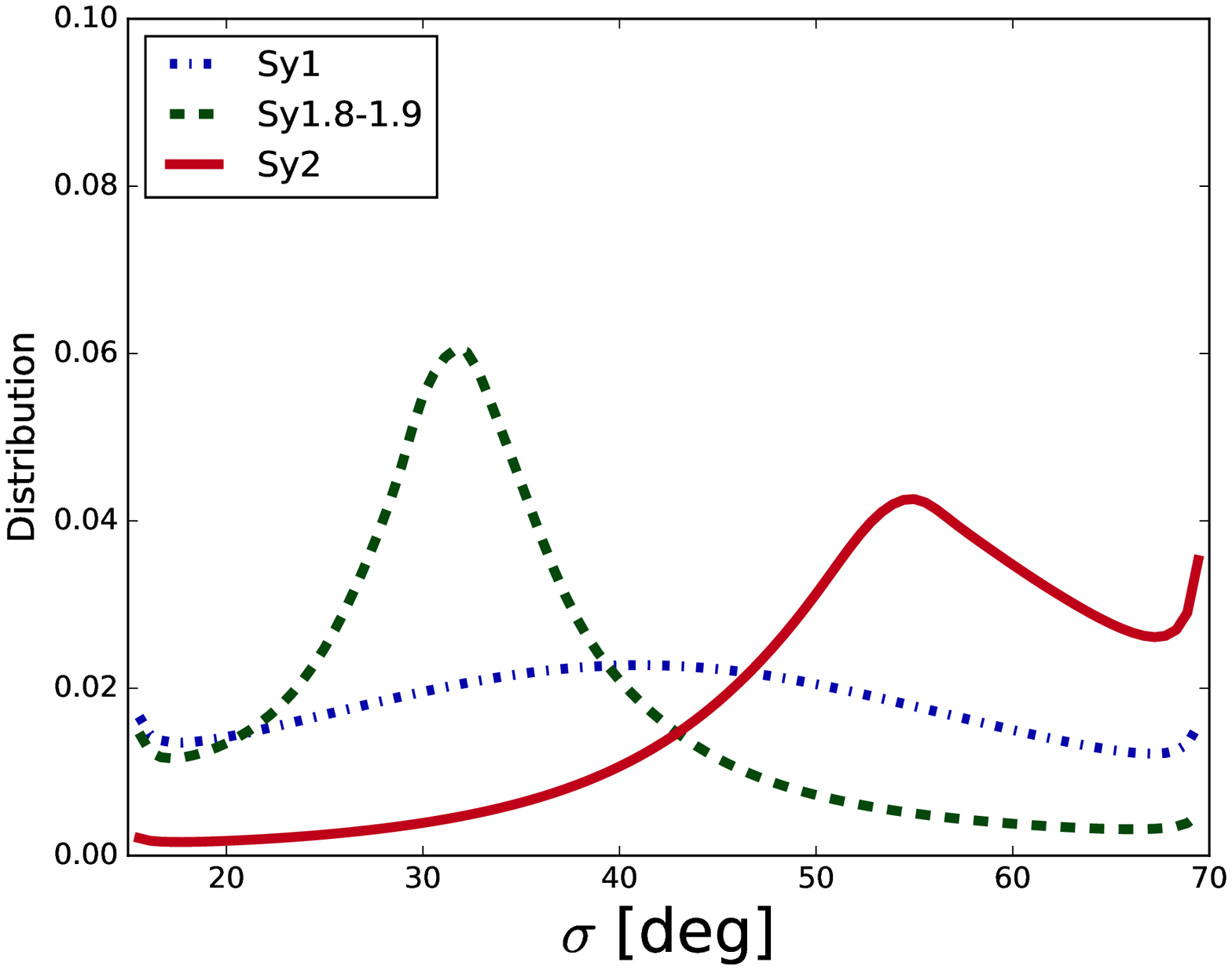}
\includegraphics[width=7.82cm]{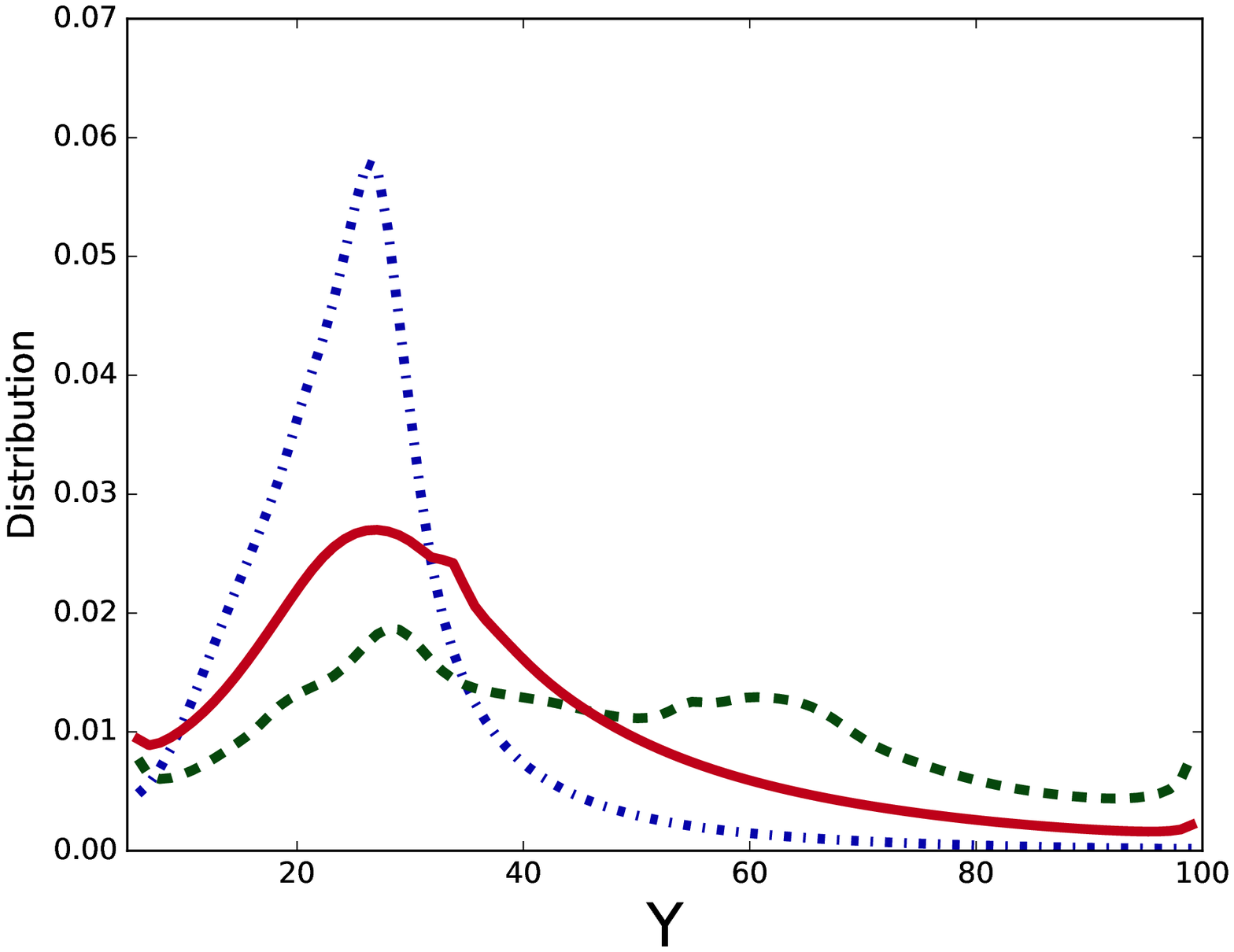}
\includegraphics[width=7.82cm]{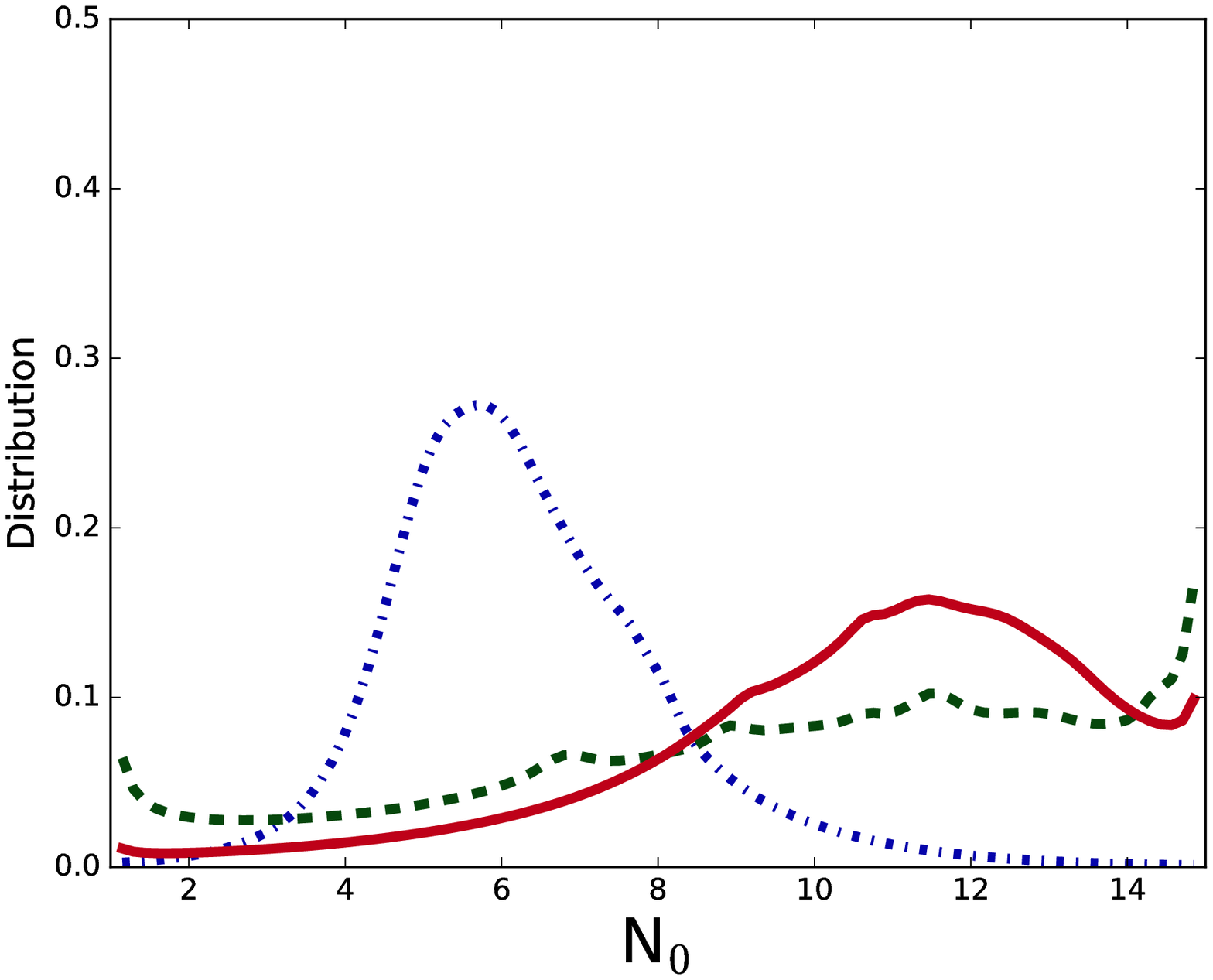}
\includegraphics[width=7.82cm]{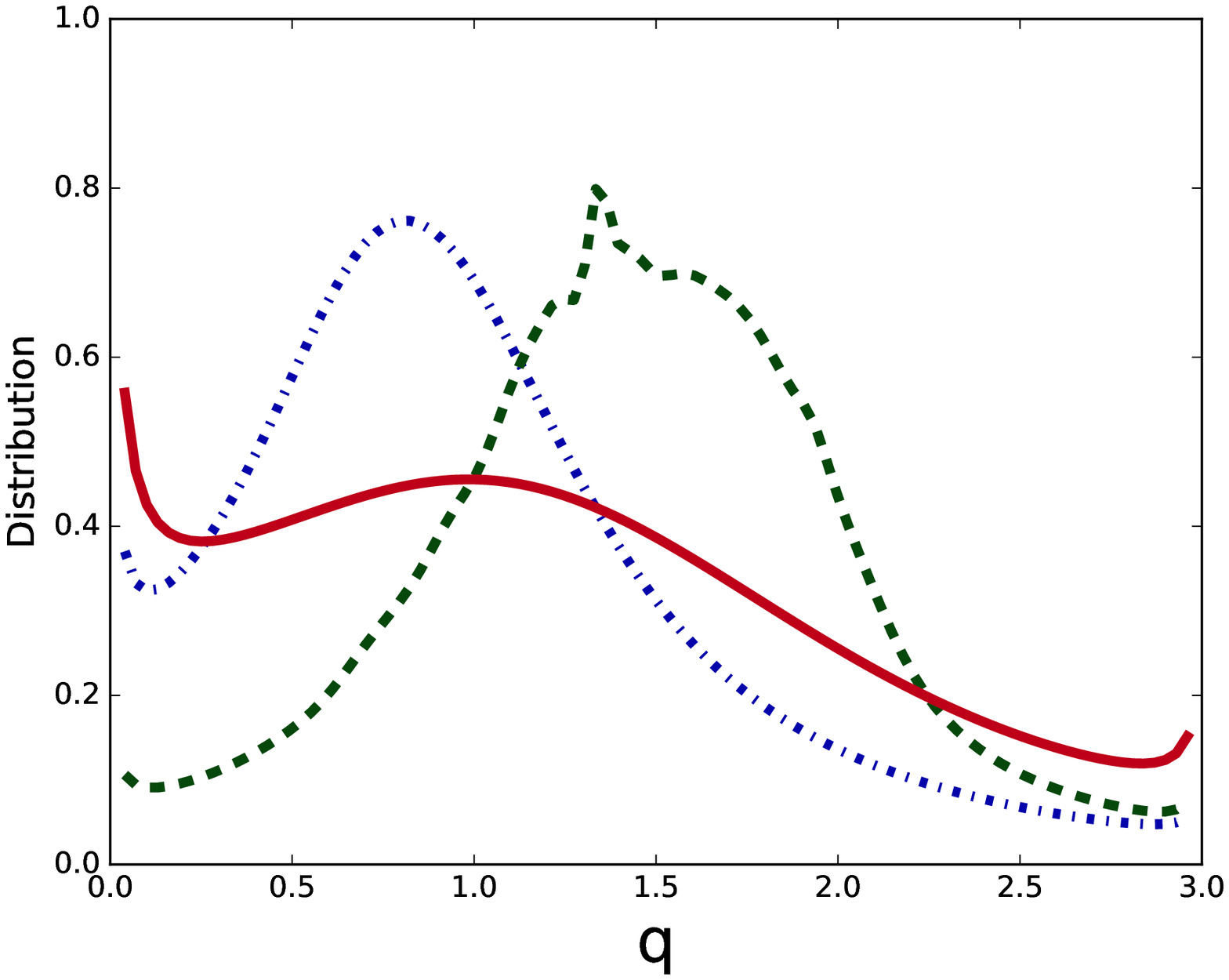}
\includegraphics[width=7.82cm]{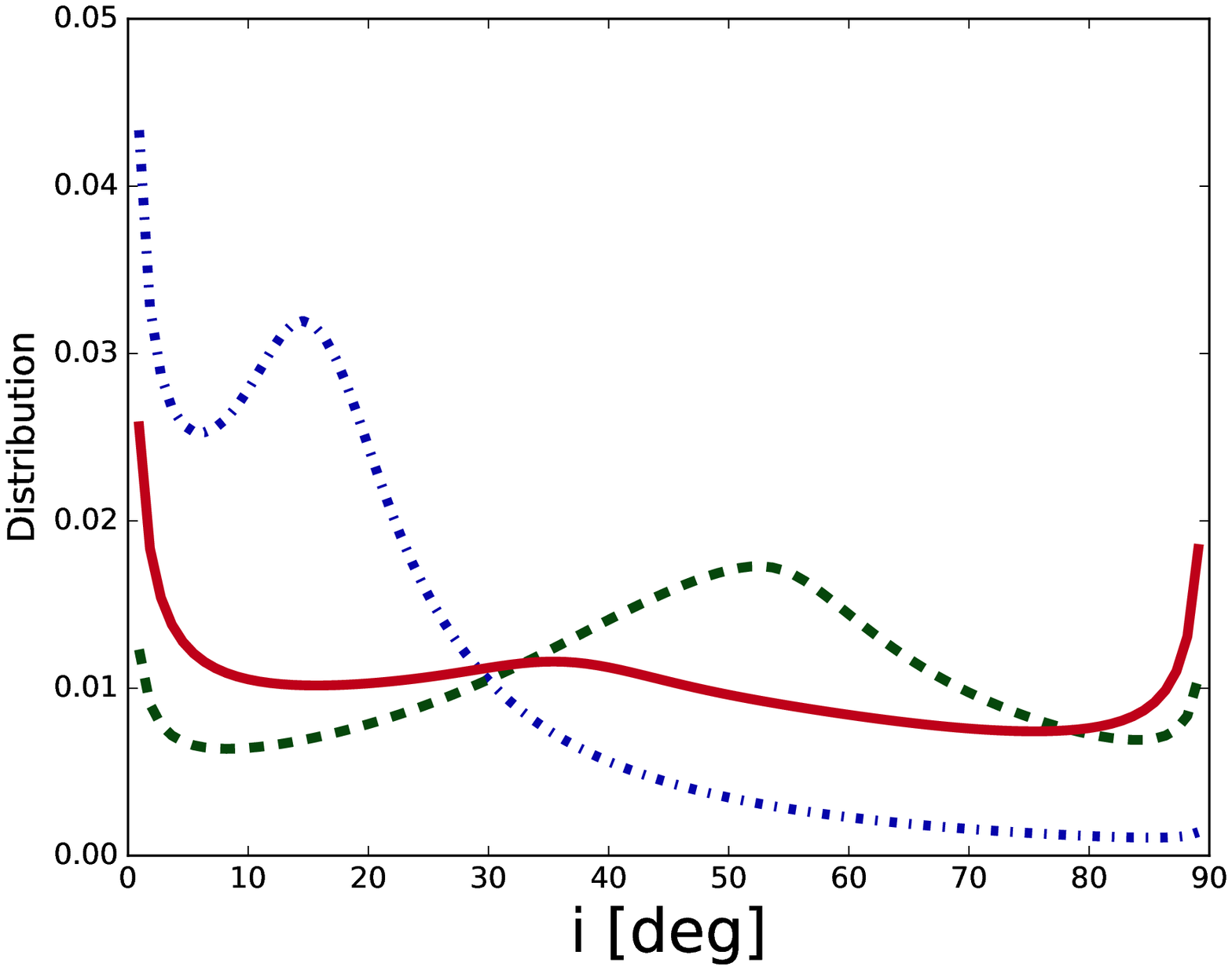}
\includegraphics[width=7.82cm]{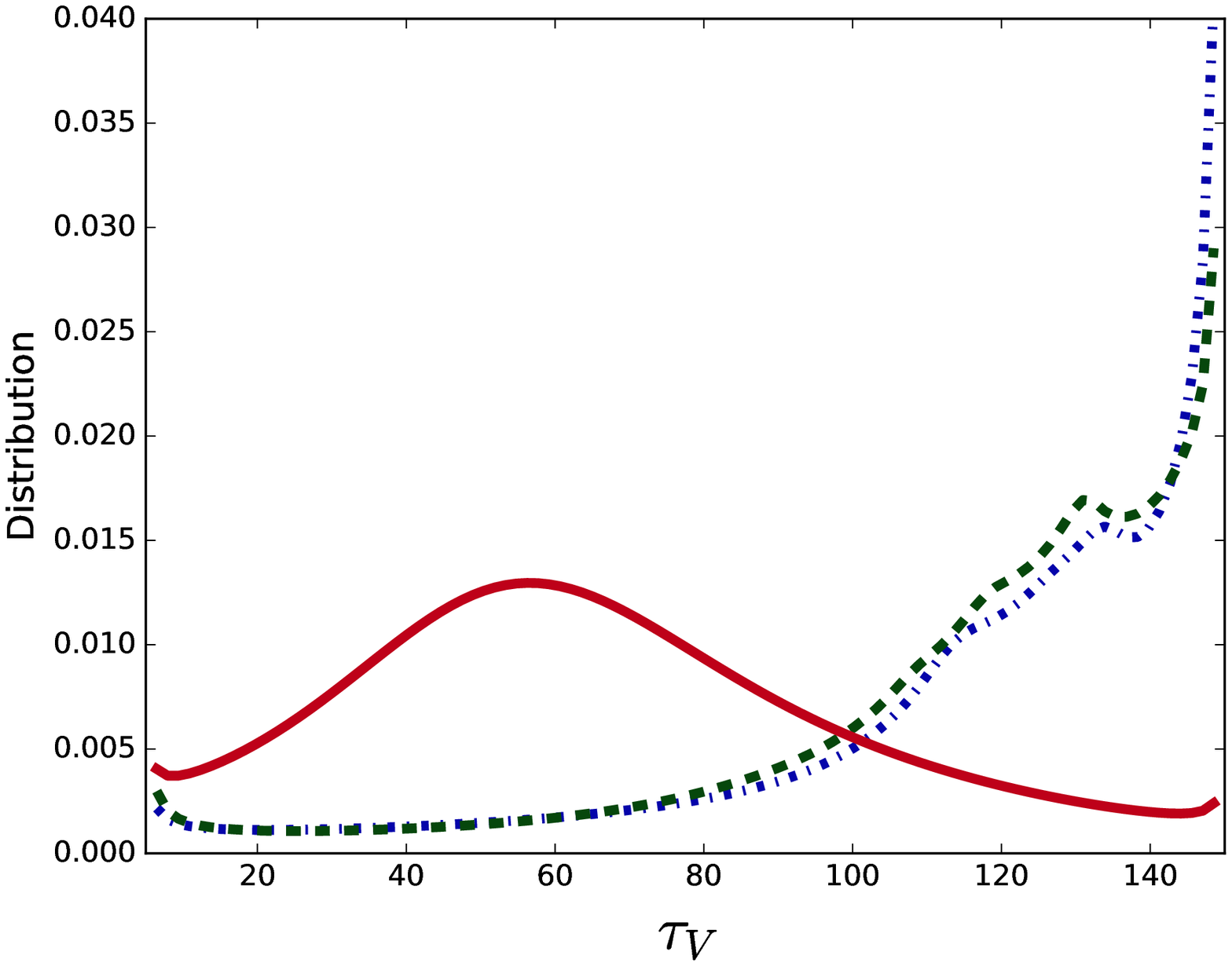}
\par}
\caption{Comparison between the clumpy torus model parameter global posterior distributions for the optical classification using only good fits ($\chi^2/dof<$2.0). Blue dot-dashed, green dashed and red solid lines represent the parameter global posterior distributions of Sy1, Sy1.8/1.9, and Sy2 galaxies, respectively.}
\label{figA2}
\end{figure*}

\begin{figure}
\centering
\includegraphics[width=8.2cm]{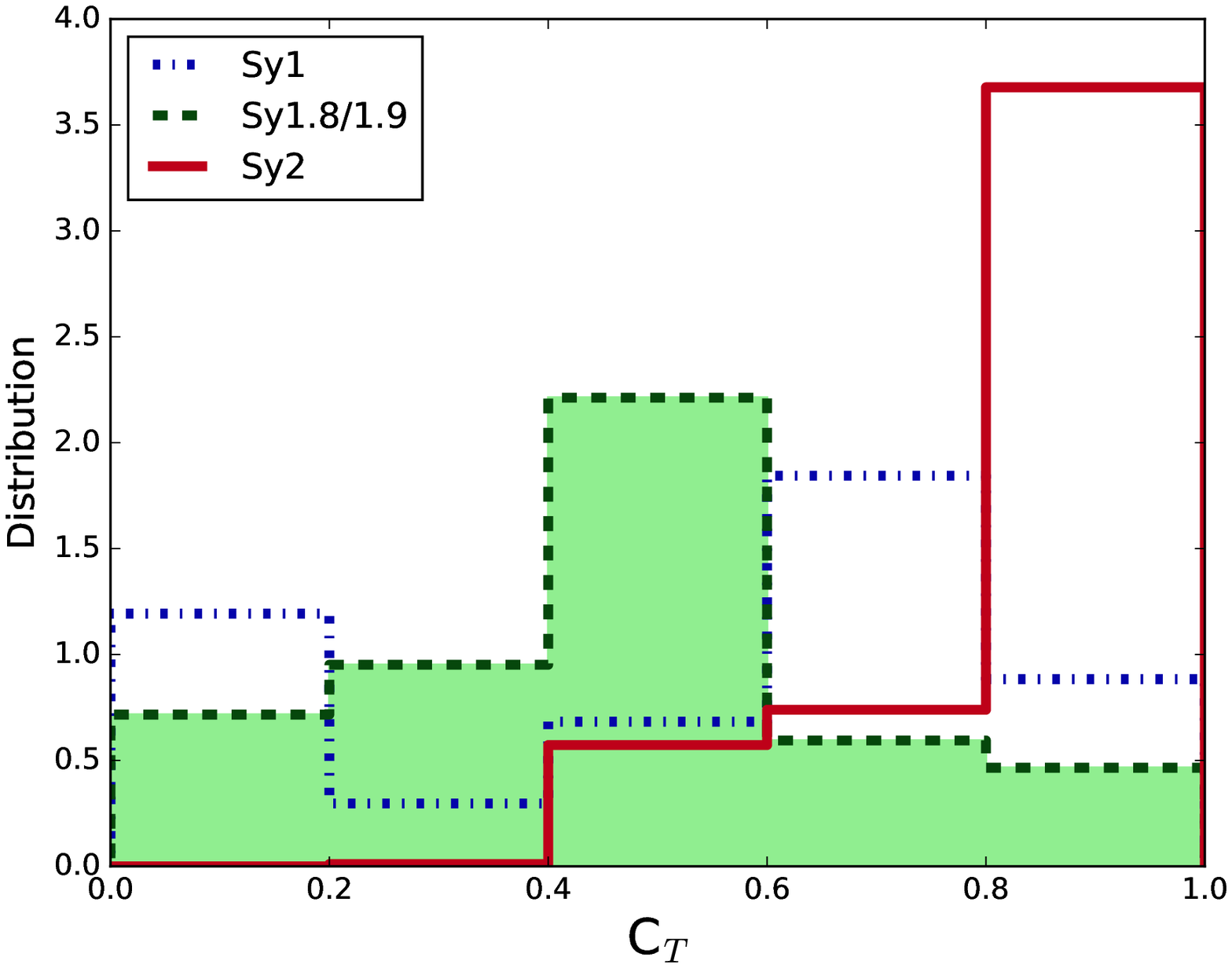}
\caption{Comparison between the torus covering factor (C$_T$) combined probability distributions for each Seyfert galaxy subgroup using only good fits ($\chi^2/dof<$2.0). Blue dot-dashed, green dashed and red solid lines represent the parameter distributions of Sy1, Sy1.8/1.9, and Sy2, respectively.}
\label{figA3}
\end{figure}

We note that while the MIR photometry and spectroscopy are well fitted in the majority of the cases, we found for 5/24 galaxies (i.e. NGC\,3783, NGC\,4395, NGC\,5506, NGC\,7172 \& NGC\,7314; see Figs. A1-A4) a NIR excess that the CLUMPY models cannot reproduce. This suggests that an extra component of very hot dust is needs to reproduce their IR SEDs (see also \citealt{Mor09}). Therefore, we repeat the global posterior distribution of each subgroup considered here for only the best ($\chi^2/dof<$1.0; $\sim$63\% of the sample) and good ($\chi^2/dof<$2.0; $\sim$79\% of the sample) fits and we find the same results within 1$\sigma$ (see Fig. \ref{figA2} and Fig. \ref{figA3}).

\section{AGN Luminosity}
\label{C}

\begin{figure}
\centering
\includegraphics[width=7.2cm, angle=90]{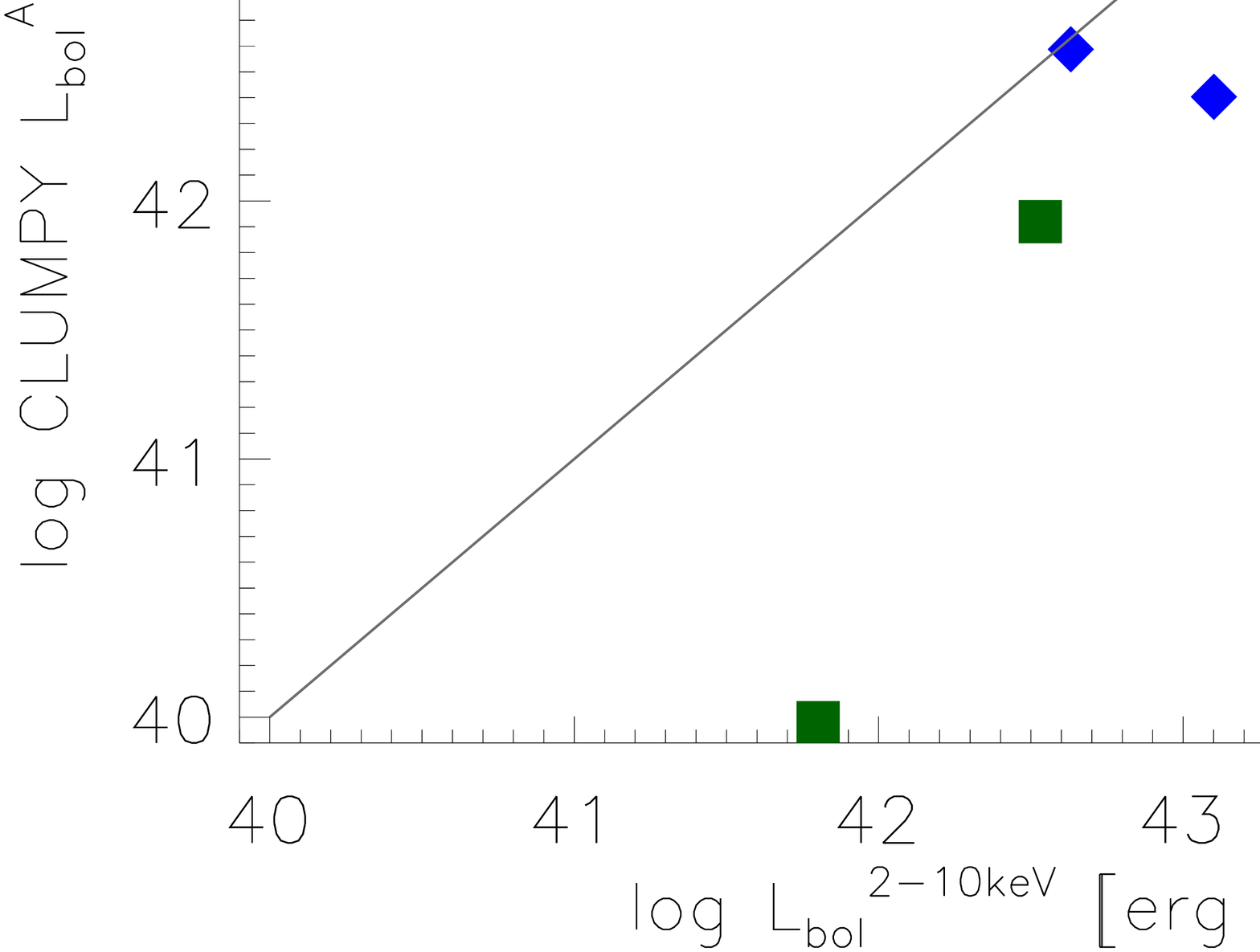}
\includegraphics[width=7.2cm, angle=90]{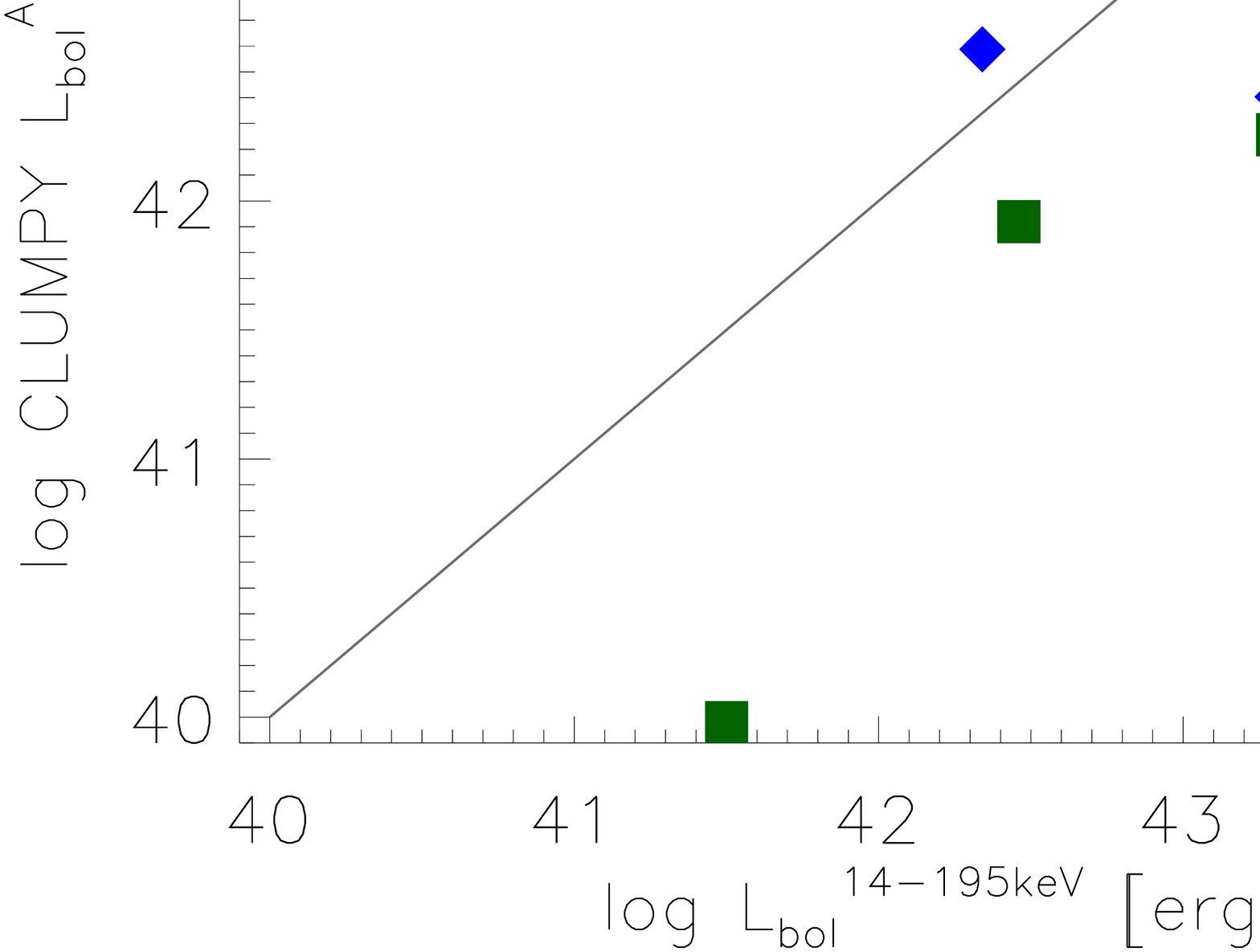}
\caption{Bolometric luminosities derived from the torus models versus Swift/BAT bolometric X-ray luminosities
 derived from the 2--10~keV and 14--195~keV X-ray band. We plot the 1:1 line for comparison. Blue diamonds, green squares and red circles are Sy1, Sy1.8/1.9 and Sy2 galaxies, respectively.}
\label{figC1}
\end{figure}

We can compare the AGN bolometric luminosities derived from the torus model fits with those derived from X-ray measurements. With this aim we compiled X-ray luminosities (2--10 and 14--195 keV; see Table \ref{tab9}) from \citet{Ricci17} and used the fixed bolometric correction factor of 20 \citep{Vasudevan09} and 7.42 for the 2--10~keV and 14--195 keV bands, respectively. The latter was obtained from the fixed bolometric factor at 2--10~keV by assuming a power-law of 1.8 as in \citet{Trakhtenbrot17}. We found that the relationship between bolometric luminosites derived from both X-ray bands and those derived from the clumpy torus models show an offset (see Fig. \ref{figC1}) which had not been found in previous works (e.g. AH11). As a sanity check, we compare the bolometric luminosities derived from the older and new version of CLUMPY models and we find that all the sources are in the 1:1 line. This new finding should be mainly related to the CLUMPY scaling factors. Therefore, in Section \ref{AGN_proper}, we used the bolometric luminosities derived from the harder X-ray band (14--195 keV) to obtaining tori sizes and masses.

\section{Black hole masses}
\label{D}
In this appendix, we estimate the black hole masses for the remaining sources in our sample not included in the work of \citet{Ricci17c}. The only exceptions are NGC\,7213 and ESO\,005-G004, for which we take their black hole masses from \citet{Vasudevan10}. To compare with the results reported by \citet{Ricci17c}, we follow the same methodology as in \citet{Koss17} to estimate the black hole masses. To do so, we use broad lines (e.g. Pa$_{\beta}$, H$_{\alpha}$, and H$_{\beta}$) and the stellar velocity dispersion ($\sigma_*$), from the 0.85~$\mu$m calcium triplet (hereafter, CaT) and CO (H- and K-bands) absorption features, for Sy1s and Sy1.8/1.9/2s, respectively. See Table \ref{tabB1}, \ref{tabB2} and \ref{tabB3}.

To calculate black hole masses from the broad Pa$_{\beta}$ line, we used the formula from  \citet{LaFranca15} (equation B1), which is based on the FWHM and luminosity of Pa$_{\beta}$. In the case of broad H$_{\alpha}$ and H$_{\beta}$ lines, we used the formulas from \citet{Greene05} (equation B2) and \citet{Trakhtenbrot12} (equation B3), respectively. For Sy1.8/1.9 and Sy2 galaxies, we used the formula from \citet{Kormendy13} (equation B4) that is based on the stellar velocity dispersion. We note that the relation from La Franca et al. (2015) was calibrated using a virial factor f = 4.31. We prefer to use the stellar velocity dispersion measurements derived from the CaT band, rather than those from the CO bands, when possible. Although the extinction at 0.85~$\mu$m (CaT) should be larger ($\sim$5 times) than at 2.3~$\mu$m (CO$^{K-band}$), the CaT feature trace an old stellar population that is more representative of the galaxy dynamical mass (e.g. \citealt{Riffel15}). CO absorption features may have a stronger contribution from young stars than the CaT feature. Indeed, the effect of the young stellar population on the stellar velocity dispersion measured from the CaT feature is practically insignificant (see e.g. \citealt{Riffel15}).

\begin{equation}
\frac{M_{BH}}{M\textsubscript{\(\odot\)}} = 10^{7.83\pm0.03} \times \left( \frac{L_{Pa_\beta}}{10^{40}~ergs^{-1}} \right)^{0.436\pm0.02} \times \left( \frac{FWHM_{Pa_\beta}}{10^4~kms^{-1}} \right)^{1.74\pm0.08}
\end{equation}

\begin{equation}
\frac{M_{BH}}{M\textsubscript{\(\odot\)}} = 1.3 \times 10^6 \times \left( \frac{L_{H_\alpha}}{10^{42}~ergs^{-1}} \right)^{0.57} \times \left( \frac{FWHM_{H_\alpha}}{10^3~kms^{-1}} \right)^{2.06}
\end{equation}

\begin{equation}
\frac{M_{BH}}{M\textsubscript{\(\odot\)}} = 1.05 \times 10^8 \times \left( \frac{L_{5100}}{10^{46}~ergs^{-1}} \right)^{0.65} \times \left( \frac{FWHM_{H_\beta}}{10^3~kms^{-1}} \right)^{2}
\end{equation}

\begin{equation}
log \left( \frac{M_{BH}}{M\textsubscript{\(\odot\)}} \right) = 4.38 \times log \left( \frac{\sigma_{*}}{200~kms^{-1}} \right)+8.49
\end{equation}

\begin{table*}
\scriptsize
\centering
\begin{tabular}{lccccccccc}
\hline
Name    & bPa$_{\beta}$ & L$_{bPa_{\beta}}$ &FWHM$_{bPa_{\beta}}$      & L$_{5100}$ & L$_{5100}^{bol}$ &FWHM$_{bH_{\beta}}$ & bH$_{\alpha}$ & L$_{bH_{\alpha}}$& FWHM$_{bH_{\alpha}}$ \\
&{\tiny (10$^{-15}$ erg~s$^{-1}$ cm$^{-2}$)}& {\tiny(erg~s$^{-1}$)}&{\tiny(km~s$^{-1}$)}& \multicolumn{2}{|c|}{{\tiny(erg~s$^{-1}$)}}& {\tiny(km~s$^{-1}$)}& {\tiny(10$^{-15}$ erg~s$^{-1}$ cm$^{-2}$)}& {\tiny(erg~s$^{-1}$)}&{\tiny(km~s$^{-1}$)}\\
 (1)&(2)&(3)&(4)&(5)&(6)&(7)&(8)&(9)&(10)\\
\hline
\multicolumn{10}{|c|}{Sy1 galaxies}\\
\hline
MCG-06-30-015	& $\cdots$        & $\cdots$ & $\cdots$            & $\cdots$ & $\cdots$ & $\cdots$                       &183.0$\pm$5.90 & 43.20 & 2007$\pm$154\\
NGC\,3783   	&  $\cdots$       & $\cdots$ & $\cdots$            & 42.96    & 43.92    & 3524$\pm$165                   &8525.1$\pm$12.7& 45.13 & 2880$\pm$5\\
NGC\,4151   	& 761.0$\pm$3.30  & 40.56    & 4535$\pm$16         & 43.21    & 44.22    & 3828$\pm$133                   &$\cdots$      &$\cdots$&  $\cdots$\\
\hline
\end{tabular}					 
\caption{Broad P$_{\beta}$, H$_{\alpha}$ and H$_{\beta}$ lines. The P$_{\beta}$ measurements were taken from \citet{Lamperti17}, and the H$_{\alpha}$ and H$_{\beta}$ values from \citet{Koss17}. Note that all the luminosities from the literature are rescaled to the luminosity distances used in this work.}
\label{tabB1}
\end{table*}

\begin{table*}
\scriptsize
\centering
\begin{tabular}{lccc}
\hline
Name    &       $\sigma_{CaT}$   & $\sigma_{CO}^{H-band}$   & $\sigma_{CO}^{K-band}$\\
		& (km~s$^{-1}$)	& (km~s$^{-1}$)& (km~s$^{-1}$)\\
	(1) & (2) & (3) & (4) \\
\hline
\multicolumn{4}{|c|}{Sy1.8/1.9 galaxies}\\
\hline 
NGC\,1365		& $\cdots$         & 141$\pm$20$^a$        & 154$\pm$20$^b$        \\
NGC\,4395		& 30$\pm$5$^h$    & $\cdots$               & $\cdots$              \\
NGC\,5506		& $\cdots$         & 180$\pm$20$^a$        & $\cdots$              \\
\hline
\multicolumn{4}{|c|}{Sy2 galaxies}\\
\hline
MCG-05-23-016	& $\cdots$        & 153$\pm$20$^a$       & $\cdots$               \\
NGC\,4945		& $\cdots$         & 151.0$\pm$20$^b$      & 117$\pm$20$^b$        \\ 
NGC\,5128		& $\cdots$         & $\cdots$              & 150$\pm$4$^i$         \\
NGC\,6300		& 92$\pm$5$^c$    & $\cdots$               & $\cdots$              \\
NGC\,7582		& 120$\pm$7$^c$  & 156$\pm$20$^b$          & 155$\pm$20$^b$        \\ 
\hline
\end{tabular}					 
\caption{Stellar velocity dispersion ($\sigma_*$) measurements. Columns 2, 3 and 4 correspond to the stellar velocity dispersion derived from CaT, CO$^{H-band}$ and CO$^{K-band}$ bands, respectively. References: a) \citet{Oliva99}; b) \citet{Oliva95}; c) \citet{Garcia-Rissmann05}; d) \citet{Riffel13}; e) \citet{Nelson95}; f) \citet{Lamperti17}; g) \citet{Riffel15}; h) \citet{Greene06}; i) \citet{Cappellari09}.}
\label{tabB2}
\end{table*}

\begin{table*}
\scriptsize
\centering
\begin{tabular}{lccc}
\hline
Name            & log(M$_{BH}^{P\beta}$/M$_\odot$)   & log(M$_{BH}^{H\beta}$/M$_\odot$) & log(M$_{BH}^{H\alpha}$/M$_\odot$) \\ 
	(1) & (2) & (3) & (4) \\

\hline
\multicolumn{4}{|c|}{Sy1 galaxies}\\
\hline
MCG-06-30-015	& $\cdots$          &   $\cdots$     & 7.42                    \\    
NGC\,3783   	&  $\cdots$       &   7.14              &  8.84                \\
NGC\,4151   	& 7.48&            7.38     &   $\cdots$                       \\
\hline
   &       log($M_{BH}^{CaT}$/M$_\odot$) & log($M_{BH}^{{CO}^{H-band}}$/M$_\odot$) & log($M_{BH}^{{CO}^{K-band}}$/M$_\odot$)\\ 
\hline
\multicolumn{4}{|c|}{Sy1.8/1.9 galaxies}\\
\hline
NGC\,1365		& $\cdots$      & 7.83                   & 7.99               \\
NGC\,4395		& 4.88          & $\cdots$               & $\cdots$           \\
NGC\,5506		& $\cdots$      & 8.29                   & $\cdots$           \\
\hline
\multicolumn{4}{|c|}{Sy2 galaxies}\\
\hline
MCG-05-23-016	& $\cdots$      & 7.98                   & $\cdots$           \\
NGC\,4945		& $\cdots$      & 7.96                   & 7.47               \\ 
NGC\,5128		& $\cdots$      & $\cdots$               & 7.94               \\
NGC\,6300		& 7.01          & $\cdots$               & $\cdots$           \\
NGC\,7582		& 7.52          & 8.02                   & 8.01               \\ 
\hline
\end{tabular}					 
\caption{BH masses. Different columns correspond to the various methods used to estimate these masses. }
\label{tabB3}
\end{table*}

\label{lastpage}

\end{document}